\documentclass[submission, Phys]{SciPost}
\usepackage{amsmath,amssymb,mathtools,xspace}
\usepackage{booktabs,multirow,graphicx,tabularx,slashed}
\usepackage{hyperref}
\usepackage{color,xcolor}
\usepackage[normalem]{ulem}
\usepackage{enumitem}
\usepackage{braket}
\usepackage{stackrel}
\usepackage{tikz}
\usetikzlibrary{arrows}
\usetikzlibrary{shapes.geometric}
\usepackage{float}
\usepackage{subcaption}

\usepackage{epstopdf}
\graphicspath{{./figs/}}

\makeatletter
\def\BState{\State\hskip-\ALG@thistlm}
\makeatother

\makeatletter
\@ifundefined{pdfoutput}{}{\DeclareGraphicsRule{*}{mps}{*}{}}
\makeatother

\makeatletter
\DeclareRobustCommand*{\bfseries}{%
   \not@math@alphabet\bfseries\mathbf
   \fontseries\bfdefault\selectfont
   \boldmath
}
\makeatother

\setitemize{itemsep=2pt,topsep=2pt,parsep=0pt,partopsep=0pt,leftmargin=*}
\setenumerate{itemsep=0pt,topsep=2pt,parsep=0pt,partopsep=0pt,labelindent=3pt,leftmargin=*}
\definecolor{Gcolor}{HTML}{3b528b}
\definecolor{Dcolor}{HTML}{e41a1c}

\tikzstyle{generator} = [rectangle, rounded corners, minimum width=3cm, minimum height=1cm,text centered, draw=Gcolor]
\tikzstyle{discriminator} = [rectangle, rounded corners, minimum width=3cm, minimum height=1cm,text centered, draw=Dcolor]
\tikzstyle{io} = [circle, trapezium left angle=70, trapezium right angle=110, minimum width=1cm, minimum height=1cm, text centered, draw=black]

\tikzstyle{process} = [rectangle, minimum width=1cm, minimum height=1cm, text centered, draw=black]
\tikzstyle{decision} = [rectangle, minimum width=1cm, minimum height=1cm, text centered, draw=black]

\tikzstyle{arrow} = [thick,->,>=stealth]
\usepackage{xcolor}

\def\impc{Mpc$^{-1}$}

\newcommand\one{\leavevmode\hbox{\small1\normalsize\kern-.33em1}}

\newcommand{\qqquad}{\qquad \qquad}
\newcommand{\qqqquad}{\qquad \qquad \qquad}

\def\slashchar#1{\setbox0=\hbox{$#1$}           
   \dimen0=\wd0                                 
   \setbox1=\hbox{/} \dimen1=\wd1               
   \ifdim\dimen0>\dimen1                        
      \rlap{\hbox to \dimen0{\hfil/\hfil}}      
      #1                                        
   \else                                        
      \rlap{\hbox to \dimen1{\hfil$#1$\hfil}}   
      /                                         
   \fi}

\def\mpl{{m_{\rm{Pl}}}}

\newcommand{\ltsima}{$\; \buildrel < \over \sim \;$}
\newcommand{\lsim}{\lower.5ex\hbox{\ltsima}}
\newcommand{\gtsima}{$\; \buildrel > \over \sim \;$}
\newcommand{\gsim}{\lower.5ex\hbox{\gtsima}}

\newcommand{\dd}{\mathrm{d}}

\newcommand{\ci}{\mathrm{i}}

\begin{document}

\begin{center}{\Large \textbf{
      Probing the Inflaton Potential with SKA
}}\end{center}

\begin{center}
Tanmoy Modak\textsuperscript{1}, 
Tilman Plehn\textsuperscript{1}, 
Lennart R\"over\textsuperscript{1}, and
Bj\"orn Malte Sch\"afer\textsuperscript{2}
\end{center}

\begin{center}
{\bf 1} Institut f\"ur Theoretische Physik, Universit\"at Heidelberg, Germany\\
{\bf 2} Astronomisches Recheninstitut, Universit\"at Heidelberg, Germany\\
\end{center}

\begin{center}
\today
\end{center}

\section*{Abstract}
         {\bf SKA will be a major step forward not only in
           astrophysics, but also in precision cosmology. We show how
           the neutral hydrogen intensity map can be combined with the
           Planck measurements of the CMB power spectrum, to provide a
           precision test of the inflaton potential. For a
           conservative range of redshifts we find that SKA can
           significantly improve current constraints on the Hubble
           slow-roll parameters.}


\tikzstyle{int}=[thick,draw, minimum size=2em]


\vspace{10pt}
\noindent\rule{\textwidth}{1pt}
\tableofcontents\thispagestyle{fancy}
\noindent\rule{\textwidth}{1pt}
\vspace{10pt}

\newpage
\section{Introduction}
\label{sec:intro}

Cosmic inflation~\cite{Starobinsky:1980te,Sato:1980yn,Guth:1980zm},
originally designed as a solution to the flatness and horizon
problems, provides a viable mechanism for seeding cosmic
structures~\cite{Mukhanov:1981xt,Starobinsky:1982ee,Hawking:1982cz,Guth:1982ec}.
It can be studied through fluctuations in the cosmic microwave
background at early times and through the distribution of dark matter
and galaxies in the late Universe. The standard paradigm assumes a
spatially homogeneous scalar inflaton field $\varphi$ with a potential
$V(\varphi)$ and a coupling to gravity through its energy momentum
tensor.  If a single inflaton is initialised in a slow-roll
state~\cite{Linde:1981mu,Albrecht:1982wi,Linde:1983gd} it drives an
accelerated expansion of the background spacetime.  During this
exponential expansion the comoving horizon shrinks and sets the
amplitude of fluctuations at the moment of their horizon exit. We can
then compute the spectrum of scalar and tensor perturbations from the
time evolution of the Hubble function, which in turn is completely
determined by the inflaton potential and initial conditions.

Planck's observations of the cosmic microwave background (CMB)
temperature and polarisation anisotropies have advanced our
understanding of inflation
tremendously~\cite{Planck:2018vyg,Planck:2018jri}.  The natural next
step would be to include observations on a larger range of scales, to
further constrain the spectral shape and with that the inflationary
potential. Here, structures in the late Universe ideally complement
the CMB, but one needs to proceed with caution as non-linear structure
formation adds significant amounts of power in the spectra on small
scales. In addition, depending on the observational channel, the
relationship between the fundamental fields, density or gravitational
potential, and the actual observable might be nonlinear, as is the
case of the galaxy distribution. In this article we study the
distribution of neutral hydrogen, mapped out by the Square Kilometer
Array
(SKA)~\cite{Cosmology-SWG:2015tjb, Furlanetto:2005ax, Furlanetto:2006jb, Loeb:2008dp, Mellema:2012ht,Morales:2009gs, Natarajan:2014rra, Pritchard:2011xb, Pritchard:2015fia,Bull:2018lat,SKA:2018ckk,Weltman:2018zrl, Zaroubi:2012in},
as a probe of inflation models. The density field on these scales is,
to good approximation, in a linear stage of structure formation with
Gaussian statistics and can potentially take cosmology to the next level of precision \cite{Furlanetto:2019wsj, Feix:2019lpo, McQuinn:2005hk, Oyama:2015gma, Xu:2020uws}, even for non perfectly Gaussian fields \cite{Shaw:2019qin, Mondal:2016hmf}, along with a determination of the astrophysical parameters \cite{Doussot:2019rdm}. Combined with with comparatively low systematic
influences due to $X$-ray or UV-sources \cite{Watkinson:2015vla, Pacucci:2014wwa, Pritchard:2006sq, Warszawski:2008pz, Ma:2018ltb} or due to baryonic feedback processes \cite{Kim:2012jx, Geil:2009ee, Barkana:2004zy}, this should allow us to probe inflationary parameters through 21cm tomography in this window. While we will use idealising assumptions in constraining inflationary parameters in this work, modelling of the reionisation process at high redshift has reached a high degree of sophistication \cite{Gnedin:2006uz, Miralda-Escude:1998adl, Furlanetto:2004ha, Iliev:2015aia, Trac:2009bt, Shapiro:2008zf} and takes care of astrophysical processes, which are likewise modelled in machine learning approaches \cite{Villanueva-Domingo:2020wpt, Hassan:2019cal}.

Several studies have looked at 21cm neutral hydrogen tomography as a
tool for precision measurements of the spectral index $n_s$ and its
running~\cite{Mao:2008ug,Barger:2008ii,Cooray:2008eb,Kohri:2013mxa,Munoz:2016owz,Pourtsidou:2016ctq,Sekiguchi:2018kqe},
to constrain the inflationary potential. Focusing on redshift
$z=8~$--$~10$ we analyze the SKA potential in constraining the Hubble
slow-roll (HSR) parameters~\cite{Liddle:2003py}, which are defined either as
logarithmic derivatives of the Hubble function.  A wide range of scales enters this measurement, ideally
from $10^{-2}$ to $1$ \impc, and the systematics related to nonlinear
structure formation on small scales or nonlinear relation between
observable and the density perturbations can be controlled. This will
allow SKA to derive tight bounds on the Hubble slow-roll parameters,
in combination with the Planck measurements of the CMB spectrum.

In Sec.~\ref{sec:cosmology} we discuss the required formalism for
inflation assuming a single field $\varphi$ driving the inflation. We
validate our approach using the constraints on slow-roll parameters
from the Planck 2018 measurements in Sec.~\ref{sec:statistics}. We
discuss about forecast on the slow-roll parameters from SKA and,
combined Planck and SKA in Sec.~\ref{sec:ska_likelihood}.  Finally, we
summarize our results.

\section{Cosmic inflation and cosmic structures}
\label{sec:cosmology}

The Planck measurements of the CMB temperature and polarisation
anisotropies have been the first systematic probe of inflationary
parameters.  The spectral index $n_s$, its running $\dd
n_s/\dd\ln k$, and the scalar-to-tensor ratio $r$ have been measured
with high precision, and the impact of cosmological parameters and the
optical depth $\tau$ has been investigated in detail. Assuming single
field inflation, these measurements can be translated into slow-roll
parameters, as we will briefly review below.

\subsubsection*{Slow-roll inflation}

During cosmic inflation, the evolution of the Universe is dominated by
the gravitational effect of a single field $\varphi$, whose
energy-momentum content acts as a source of gravity. As $\varphi$ is
assumed to conform to the FLRW-symmetries, it can only depend on time.
Because its action
\begin{align}
	S = 
	\int\dd^4x\sqrt{-\text{g}}\:\left(\frac{1}{2}g^{\mu\nu}\nabla_\mu\varphi\nabla_\nu\varphi - V(\varphi)\right)
\end{align}
does not contain any dissipative terms or couplings to other fields,
it acts as an ideal fluid with density $\rho$, pressure $p$, and a
covariantly conserved energy momentum tensor.  Variation with respect
to $\varphi$ and imposing the FLRW-symmetries gives us the Klein-Gordon equation
\begin{align}
	\ddot{\varphi} + 3\frac{\dot{a}}{a}\dot{\varphi} = -\frac{\dd V(\varphi)}{\dd\varphi} \; .
\end{align}
The gravitational effect of $\varphi$ on a FLRW-spacetime with scale
factor $a(t)$ is given by the Friedmann equations
\begin{align}
  \left(\frac{\dot{a}}{a}\right)^2
  = \frac{8\pi}{3\mpl^2}\left(\frac{\dot{\varphi}^2}{2} + V(\varphi)\right)
  \qquad \text{and} \qquad 
  \frac{\ddot{a}}{a}
  = -\frac{8\pi}{3\mpl^2}\left(\frac{\dot{\varphi}^2}{2} - V(\varphi)\right) \; .
\end{align}
The limit $\dot{\varphi}^2\ll 2V(\varphi)$ is referred to as the
slow-roll phase. The Klein-Gordon equation together with the first
Friedmann equation allow us to write the evolution of the
FLRW-universe in terms of the Hubble function $H = \dot{a}/a = \dd\ln
a/\dd t$,
\begin{align}
	\dot{\varphi} &= -\frac{\mpl^2}{4\pi} H^\prime(\varphi) \notag \\
	V(\varphi) &= - \frac{\mpl^4}{32\pi^2} [H^\prime(\varphi)]^2 + \frac{3 \mpl^2}{8\pi} H^2(\varphi) \; ,
\end{align}
with $H^\prime = \dd H/\dd\varphi$. In ideal slow roll, the
expansion of the Universe is exponential with a constant
Hubble function. Deviations are parametrized by 
\begin{align}
	\epsilon_H = \frac{\mpl^2}{4\pi}\left(\frac{H^\prime}{H}\right)^2 \; .
\end{align}
which reflects the equation-of-state parameter $w = p/(\rho c^2)
\approx -1$, as required by an exponential expansion.  Analogously, a
small value of
\begin{align}
	\eta_H = \frac{\mpl^2}{4\pi}\left(\frac{H^{\prime\prime}}{H}\right)
\end{align}
makes sure that slow roll is maintained for a sufficiently long time.

Starting from these two intuitive parameters 
one defines a full hierarchy of Hubble slow-roll parameters that
quantify logarithmic changes to the Hubble function,
\begin{align}
  \lambda^{(n)}_H = \left(\frac{\mpl^2}{4\pi}\right)^n
  \left(\frac{(H^\prime)^{n-1}}{H^n}\frac{\dd^{n+1}H}{\dd\varphi^{n+1}}\right)
  \qqquad  n\geq 1 \; ,
\end{align}
with the usual correspondence $\eta_H =
\lambda^{(1)}$, $\xi^2_H = \lambda^{(2)}$, and $\omega^3_H =
\lambda^{(3)}$. Expanding around the inflaton field value $\varphi_*$
at the horizon crossing with the pivot scale $k_* = 0.05$ \impc, the
Hubble function can be reconstructed in the observable window defined by the range of observationally accessible spatial scales as
\begin{align}
  H(\varphi)
  = \sum_{n = 0}^{N} \frac{1}{n!} \;
  \frac{\dd^n H}{\dd\varphi^n} \Bigg|_{\varphi_*} (\varphi-\varphi_*)^n \; ,
  \label{eq:recohubble}
\end{align}
expressed in terms of the $\lambda^{(n)}_H$.

\subsubsection*{Perturbation theory}

A suitable coordinate choice for perturbation theory is comoving
gauge, where spatial hyper-surfaces are orthogonal to the worldlines
of FLRW-observers, and the corresponding gauge invariant quantity is
the Mukhanov potential
\begin{align}
	u = a\delta\varphi - \frac{\mathcal{R}}{H}\frac{\partial\varphi}{\partial\eta} \; .
\end{align}
It is constructed from the curvature perturbation $\mathcal{R}$ and
the inflationary field perturbation $\delta\varphi$, and $\eta$ is
conformal time. Its Fourier modes $u(k)$ evolve according to
\begin{align}
	\frac{\dd^2}{\dd \eta^2}u(k) + \left[k^2 - \frac{1}{z}\frac{\dd^2 z}{\dd \eta^2}\right]u(k) = 0\label{eq:perturb}
\end{align}
where $z = a\dot{\varphi}/H$ and the initial conditions are formally set at 
\begin{align}
	u(k,\eta\rightarrow -\infty) = \frac{e^{-\ci k\eta}}{\sqrt{2k}}. 
\end{align}
%
This evolution equation is tackled by first solving the background
evolution of the FLRW-universe with the Hubble function as a Taylor
series and then solving the mode equation. This way we obtain a
scale-dependent prediction of $u(k)$ at the end of inflation, where
slow roll is violated and the Universe transitions away from exponential expansion.
The amplitudes of the Fourier modes define the spectrum of curvature
perturbations
\begin{align}
	\mathcal{P}_\mathcal{R}(k) = \frac{k^3}{2\pi^2}\left|\frac{u(k)}{z}\right|^2 \; .
\label{eq:spectrum}
\end{align}
While perfect slow roll would guarantee scale-independent curvature
perturbations and generate a perfect Harrison-Zel'dovich-spectrum, any
deviation leads to modulations. They can be computed by mapping the
slow-roll parameters onto a logarithmic Taylor expansion of the
potential of the type
\begin{align}
	\ln\mathcal{P}_\mathcal{R}(k) = 
	\ln A_s + \ln\frac{k}{k_*}\left[(n_s - 1) + \frac{\alpha}{2}\ln\frac{k}{k_*} + \frac{\beta}{3!}\ln^2\frac{k}{k_*}+\ldots\right] \; .
	\label{eq:running_spectrum}
\end{align}
where the expansion scale $k_*$ is exactly the pivot scale.

For $n_s = 1$ and $\alpha = \beta = 0$ we recover the
Harrison-Zel'dovich spectrum
$\mathcal{P}_\mathcal{R}(k)=\text{const}$.
The curvature perturbation spectrum $\mathcal{P}_\mathcal{R}(k)$
defined in Eq.~\eqref{eq:spectrum} serves as an input for the
computation of all observables, most notably the CMB temperature and
polarisation spectra, as well as for fluctuations in the 21cm
brightness.

To determine the constraints on the slow-roll parameters we
use the MCMC engine MontePython3~\cite{Brinckmann:2018cvx,Audren:2012wb}, 
interfaced with the Boltzmann code CLASS~III~\cite{Lesgourgues:2011rg,Blas:2011rf} to solve the background and
perturbation equations and find the power spectrum. We 
truncate the series in Eq.\eqref{eq:recohubble} at $N=4$,
such that our parameter space is spanned by
\begin{align}
  \left\{ \; \tilde{A}_s, \epsilon_H, \eta_H, \xi^2_H, \omega^3_H \; \right\} \; ,
\label{eq:paras}
\end{align}
also denoted as the Hubble slow-roll (HSR) parameters. The parameter
$\text{HSR}_0 \equiv \tilde{A}_s$ is defined in
Ref.~\cite{Lesgourgues:2011rg}. 
The primary reason behind this choice of parameterization is to validate our results
as well as to compare the potential of SKA to that of Planck 2018~\cite{Planck:2018vyg}.
Note that this parameterization does not depend on the slow-roll approximation. 
Here, we chose to parameterize $H$ in Eq.~\eqref{eq:perturb} as a Taylor expansion
with respect to $(\varphi-\varphi^*)$ as given in Eq.~\eqref{eq:recohubble}. 
Hence the HSR parameters are not constant in the observable window and evaluated at the pivot scale $k_* = 0.05$ Mpc$^{-1}$ corresponding to the comoving horizon size. However, for the 
$\tilde{A}_s$, $n_s$, $\alpha$ and $\beta$  parameterization we have truncated the
Taylor expansion in Eq.~\eqref{eq:spectrum} at $\beta$ and assumed the primordial spectrum $\mathcal{P}_\mathcal{R}(k)$ is well captured
in the observable window of comoving wave numbers.

For the reference cosmological models through out
the paper we assume spatially flat $\Lambda$CDM-cosmology, with
specific fixed parameters choices
\begin{alignat}{8}
  \omega_b & = 2.242\times 10^{-2}
  &\qqqquad 
  \omega_c & = 0.12 \notag \\
  \tau_\text{reio} &=0.05678
  &\qqqquad 
  h & =0.6724 \, ,
\label{eq:cosmo}
\end{alignat}
corresponding to our reproduced Planck 2018
measurements~\cite{Planck:2018vyg}, as discussed in the next section.

\section{Planck validation}
\label{sec:statistics}

As illustrated above, the combined system of differential equations
for the slow-roll parameters and the mode equation for the amplitudes
$u(k)$ predict the spectrum $\mathcal{P}_\mathcal{R}(k)$. Any
deviation from perfect slow roll induces a scale dependence and a
deviation away from the idealised Harrison-Zel'dovich shape.  A
measurement which is sensitive to slight variations from a pure power
law necessarily encompasses a wide range of scales, ideally from the
horizon $ck = aH$ to as small scales as possible. Evading nonlinear
structure formation on the smallest scales, the requirement of a
linear relationship between observable and potential fluctuations
conserving all statistical properties and access to a wide range of
scales starting at the pivot scale $k_*$ suggests a combination of the
CMB at a redshift around $10^3$ and the neutral hydrogen density at a
redshift around 10 as a powerful probe of inflationary dynamics.

The established window to inflationary fluctuations are observations
of the CMB temperature and polarisation
anisotropies~\cite{Planck:2018jri}. Perturbations on the spectral
distribution of photons along a line of sight incorporate baryonic
acoustic oscillations and Sachs-Wolfe-type effects and can be cast
into the angular spectra $C^{TT}(\ell)$, $C^{EE}(\ell)$ and
$C^{TE}(\ell)$. As long as we neglect the mode equation associated
with gravitational waves, we set the spectrum of primordial tensor
mode to zero and compute the $E$-mode polarisation from the curvature
perturbation alone.  Combining the three measured spectra to a
likelihood with Planck's noise model and a suitable covariance allows
us to constrain Hubble slow-role parameters from simulated Planck data
and check our results with the conventional
$(\alpha,\beta)$-parametrization defined in
Eq.\eqref{eq:running_spectrum}.

\begin{figure}[t]
  \centering
  \includegraphics[width = 0.80 \textwidth]{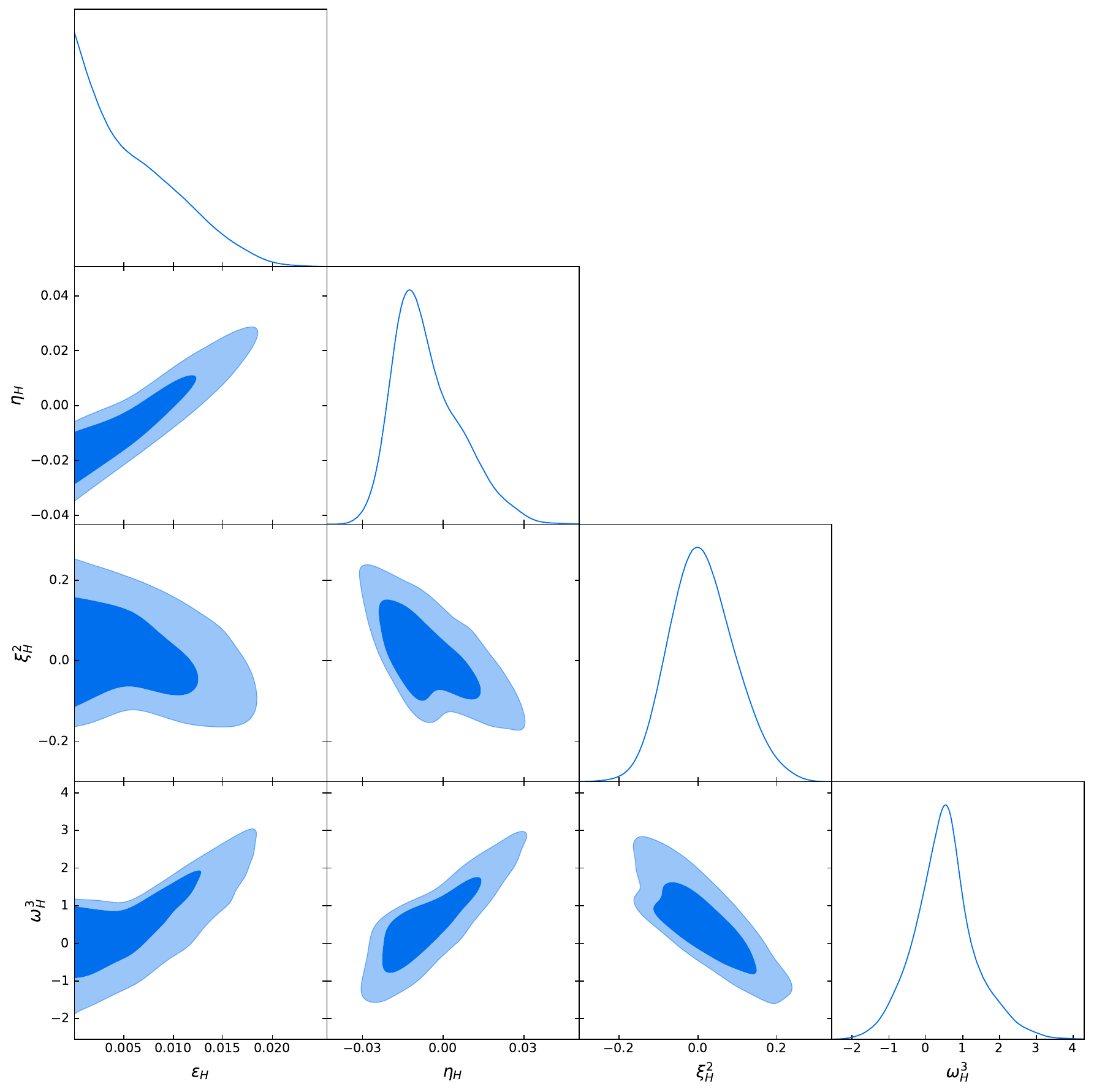}
  \caption{Marginalized joint distributions for parameter pairs at 68\% and 95\% confidence and marginalized distributions for individual parameters from the
    5-dimensional likelihood of the slow-roll parameters in
    Eq.\eqref{eq:paras}, also marginalized over the cosmological
    parameters in Eq.\eqref{eq:cosmo}.  We use the joint TT,TE,EE+lowE
    Planck data, these results should be compared with Fig.~13 of
    Ref.~\cite{Planck:2018jri}.}
  \label{fig:HSR_planck}
\end{figure}

\begin{table}[b!]
  \centering
  \begin{small}
  \begin{tabular}{c|lc} 
    \toprule
    Parameters & mean& 95\% CL \\ 
    \midrule
    $\tilde{A}_s \times 10^9$ & $\phantom{-}2.084$ & $[1.978, 2.197]$ \\
    $\epsilon_H$ & $\phantom{-}0.006095$ & $< 0.01518$ \\ 
    $\eta_H$ & $-0.005849$ & $[-0.02804, 0.02104]$ \\ 
    $\xi^2_H$ & $\phantom{-}0.01133$ & $[-0.1498, 0.1797]$ \\ 
    $\omega^3_H$ & $\phantom{-}0.5182$ & $[-1.213, 2.309]$ \\ 
    \bottomrule
  \end{tabular}
  \end{small}
  \caption{Mean values and error bars (95\% CL) for the slow-roll
    parameters shown in Fig~\ref{fig:HSR_planck}. }
  \label{tab:HSR_planck}
\end{table}

\begin{figure}[t]
  \includegraphics[width = 0.495\textwidth]{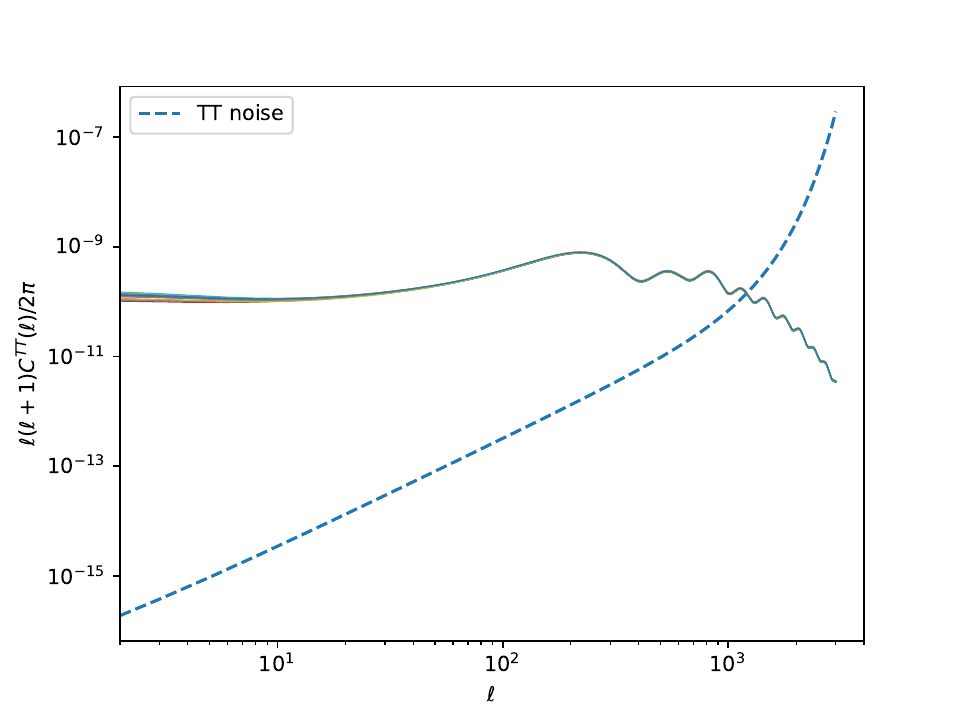}
  \includegraphics[width = 0.495\textwidth]{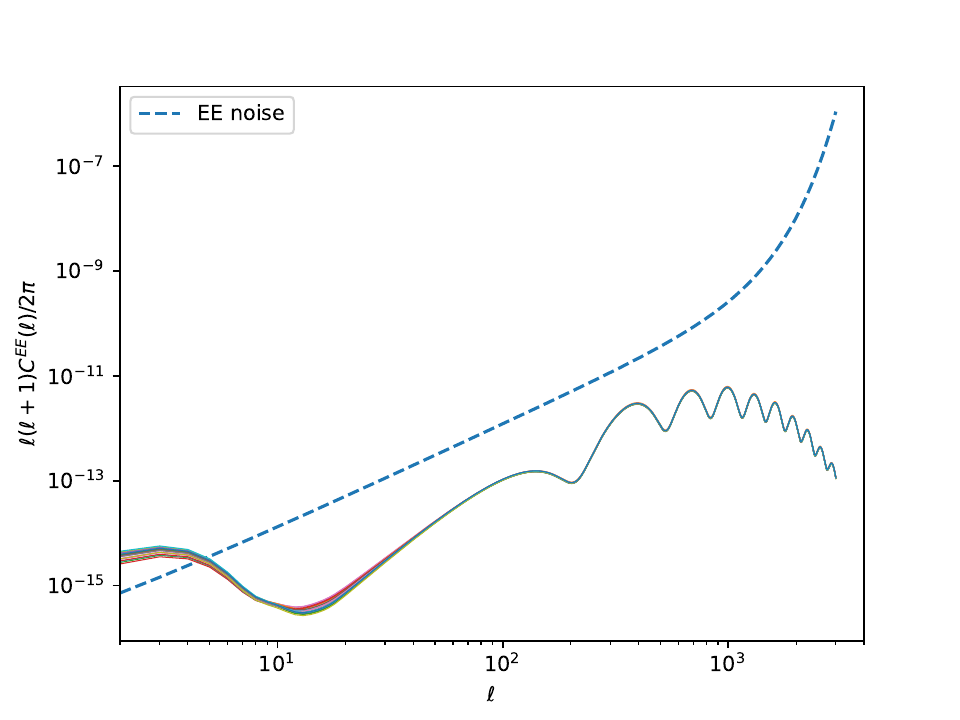}
  \caption{Angular power spectra for the TT (left) and EE (right)
    correlations for representative samples of the slow-roll
    parameters along with respective noise power spectrum.}
  \label{fig:TTEE_noise}
\end{figure}

For a first test of our method we fix the background cosmology to a
conventional $\Lambda$CDM-model.  The noise model uses Gaussian beam
shapes and the typical noise levels as specified for Planck.  We
restrict ourselves up to $N=4$ in Eq.\eqref{eq:recohubble}, as done in
Ref.~\cite{Planck:2018jri}. We sample the slow-roll parameters with flat
priors~\cite{Lesgourgues:2007aa,Planck:2015sxf,Planck:2018jri}.  The
reconstructed inflation parameters from TT,TE,EE+lowE data are shown
in Fig.~\ref{fig:HSR_planck} and in Tab.~\ref{tab:HSR_planck}. Our
results are in good agreement with the dashed contours of Fig.~13 of
Planck 2018~\cite{Planck:2018jri}.  Our values shown in
Tab.~\ref{tab:HSR_planck} are mildly weaker than the values shown in
Tab.~7 of Ref.~\cite{Planck:2018jri}, as we to not include BK15 and
lensing data.  For illustration, we also show the angular power
spectra for the TT and EE correlations with the respective noise for
few representative samples from our parameter scan in
Fig.~\ref{fig:TTEE_noise}.

\begin{figure}[b!]
  \includegraphics[width = 0.245 \textwidth]{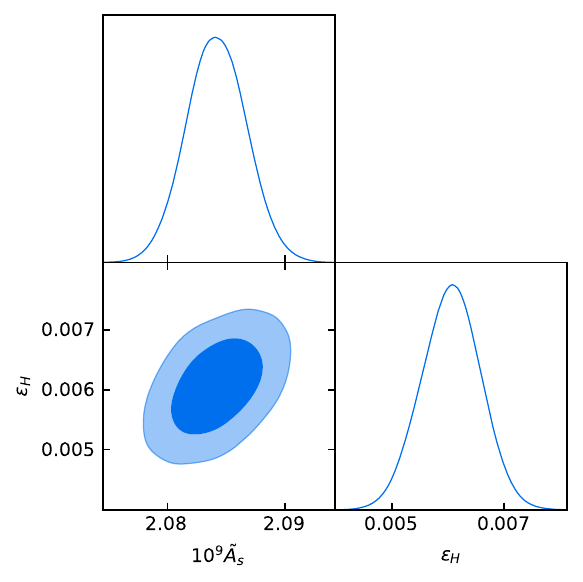}
  \includegraphics[width = 0.245 \textwidth]{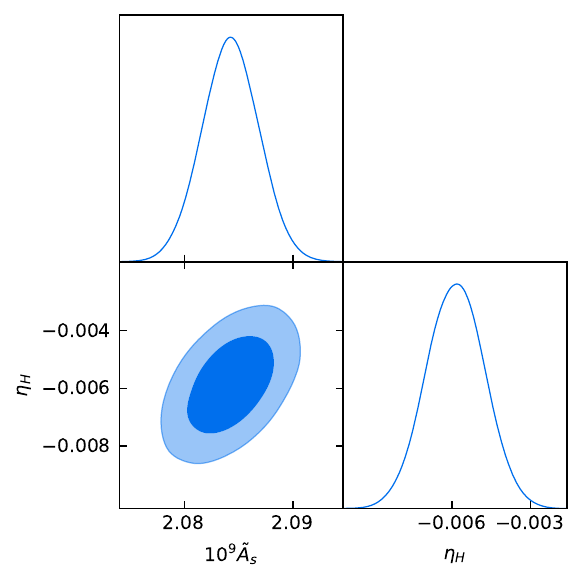}
  \includegraphics[width = 0.245 \textwidth]{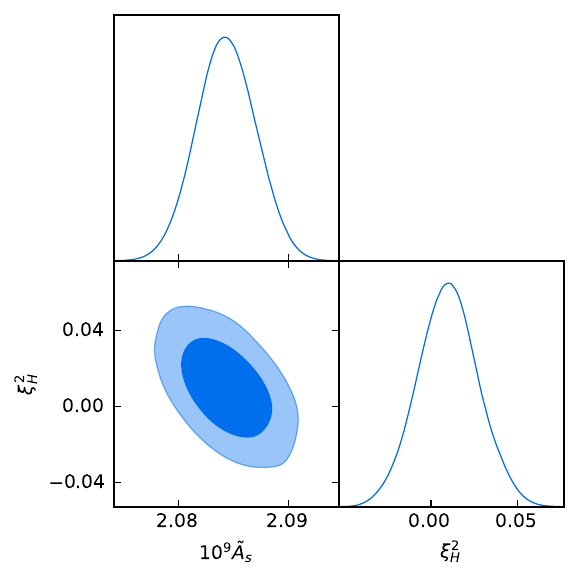} 
  \includegraphics[width = 0.245 \textwidth]{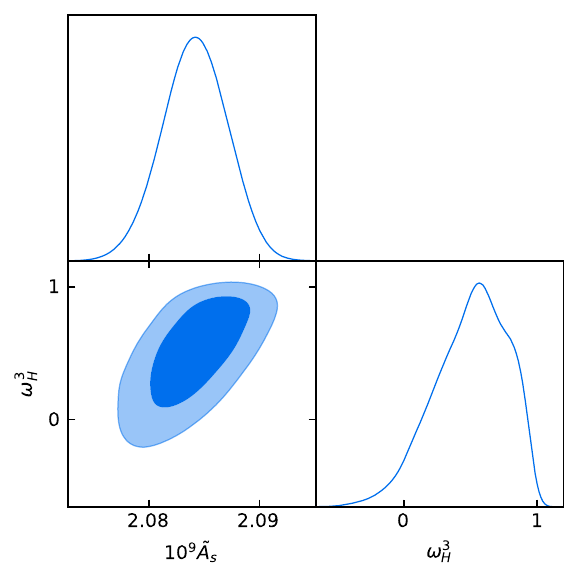} \\
  \includegraphics[width = 0.245 \textwidth]{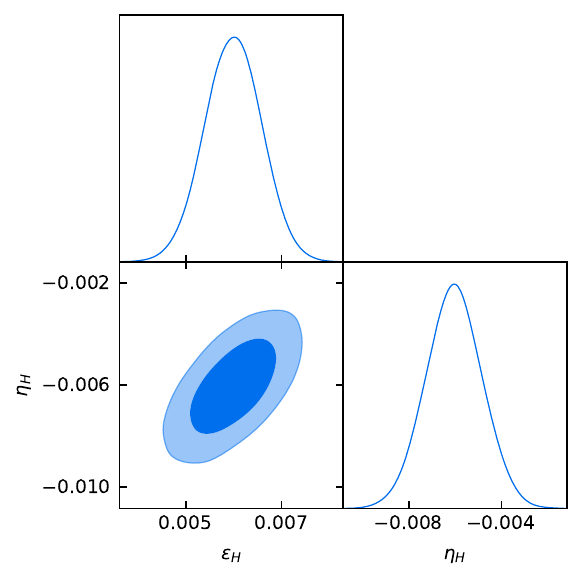}
  \includegraphics[width = 0.245 \textwidth]{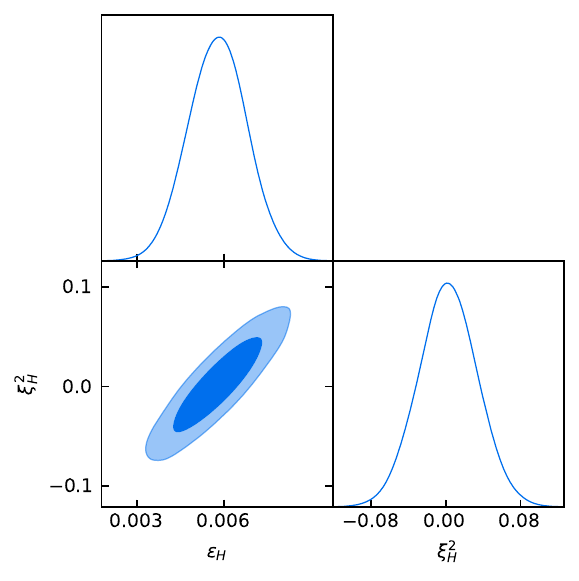} 
  \includegraphics[width = 0.245 \textwidth]{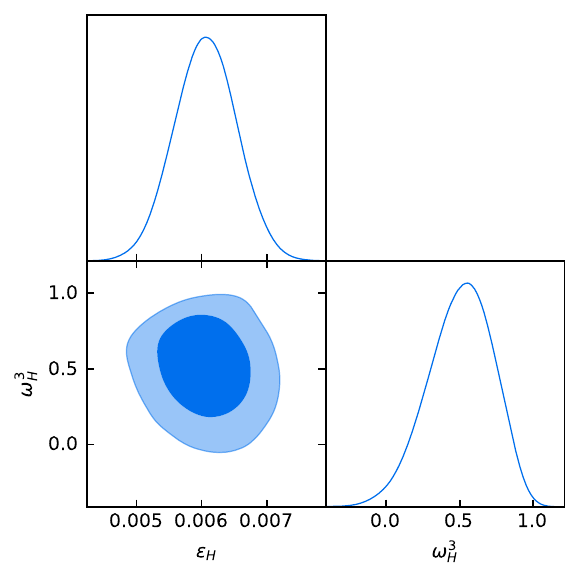}
  \includegraphics[width = 0.245 \textwidth]{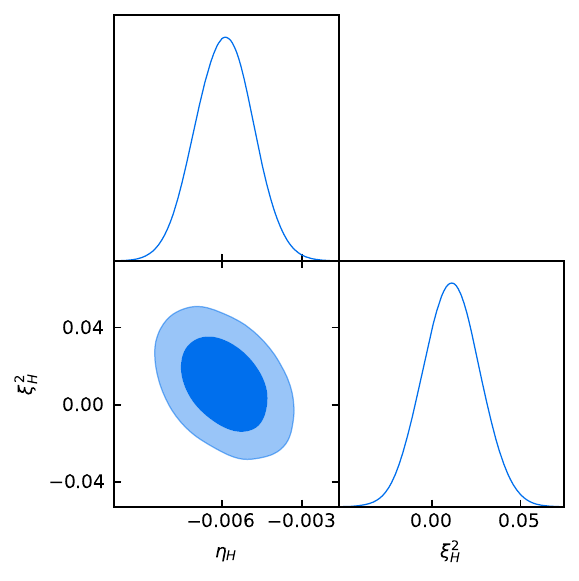} \\
  \includegraphics[width = 0.245 \textwidth]{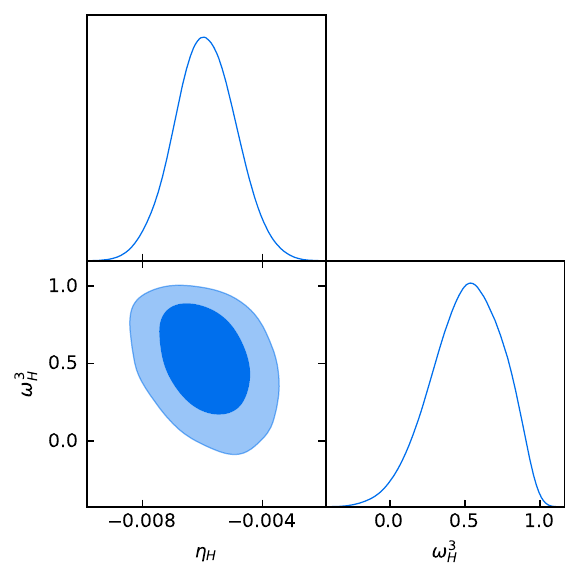} 
  \includegraphics[width = 0.245 \textwidth]{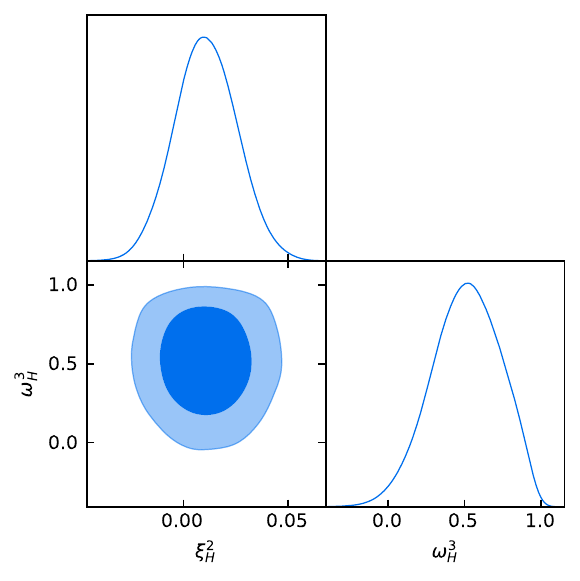} \\
  \caption{Sliced 2-dimensional likelihoods for the slow-roll
    parameters from CMB data.}
  \label{fig:Planck2D}
\end{figure}

While the primary aim of this paper is to estimate the potential of
SKA and 21cm tomography in measuring the inflaton potential, we need
to keep in mind that any SKA measurement will be combined with the
Planck CMB constraints. This means we first need to understand the way
this correlated set of fundamental parameters affects the CMB power
spectra. To illustrate the relation between the different model
parameters, we start with a set of ten 2-dimensional parameter scans,
fixing three parameters of the 5-dimensional model space defined in
Eq.\eqref{eq:paras}.  For each 2-dimensional scan we set the remaining
three parameters to the mean values given in
Tab.~\ref{fig:HSR_planck}. We can then assume that the maximum in the
2-dimensional scan should also reproduce the mean values in
Tab.~\ref{fig:HSR_planck}, but with a correlated uncertainty. In
Fig.~\ref{fig:Planck2D} we show these 2-dimensional parameter planes
and confirm that for the combined Planck measurements there do not
exist especially strong correlations.  In Tab.~\ref{tab:Planck2D} we
give the mean values and the 95\% confidence level limits for the
2-dimensional parameter planes shown in Fig.~\ref{fig:Planck2D}.

\begin{table}[t]
  \centering
  \resizebox{\textwidth}{!}{
  \begin{tabular}{l|lc|lc|lc} 
    \toprule 
    Parameter & \multicolumn{2}{c|}{Planck} & \multicolumn{2}{c}{SKA} & \multicolumn{2}{c}{SKA+Planck} \\
    & mean & 95\% CL & mean & 95\% CL & mean & 95\% CL\\
    \midrule
    \multirow{2}{*}{$\tilde{A}_s \times 10^9$ vs $\epsilon_H$}
    & $\phantom{-}2.0842$&$[2.0792,2.0892]$ 
    & $\phantom{-}2.08401$&$[2.08314,2.08489]$ 
    & $\phantom{-}2.08403$ & $[2.08318,2.08491]$\\
    & $\phantom{-}0.0061$&$[0.0050,0.0071]$ 
    & $\phantom{-}0.006096$&$[0.006050,0.006141]$ 
    & $\phantom{-}0.006096$&$[0.006052,0.006142]$\\
    \multirow{2}{*}{$\tilde{A}_s \times 10^9$ vs $\eta_H$}
    & $\phantom{-}2.0842$&$[2.0791,2.0894]$ 
    & $\phantom{-}2.08400$&$[2.08331,2.08469]$
    & $\phantom{-}2.08401$&$[2.08333,2.08468]$\\
    & $-0.0059$&$[-0.0081,-0.0037]$ 
    & $-0.005849$&$[-0.005941,-0.005756]$ 
    &$-0.005850$&$[-0.005940,-0.005758]$\\
    \multirow{2}{*}{$\tilde{A}_s \times 10^9$ vs $\xi_H^2$}
    & $\phantom{-}2.0843$&$[2.0791,2.0897]$ 
    & $\phantom{-}2.08400$&$[2.08365,2.08436]$ 
    & $\phantom{-}2.08400$&$[2.08365,2.08436]$ \\
    & $\phantom{-}0.010$&$[-0.024,0.044]$ 
    & $\phantom{-}0.01134$&$[0.01052,0.01218]$
    & $\phantom{-}0.01134$&$[0.01055,0.01215]$\\
    \multirow{2}{*}{$\tilde{A}_s \times 10^9$ vs $\omega^3_H$}
    & $\phantom{-}2.0841$&$[2.0782,2.0898]$ 
    & $\phantom{-}2.08400$&$[2.08379,2.08420]$ 
    & $\phantom{-}2.08400$&$[2.08381,2.08420]$\\
    & $\phantom{-}0.51$&$[-0.02,0.97]$ 
    & $\phantom{-}0.518$&$[0.498,0.537]$ 
    & $\phantom{-}0.518$&$[0.498,0.537]$\\
    \multirow{2}{*}{$\epsilon_H$ vs $\eta_H$}
    & $\phantom{-}0.0060$&$[0.0048,0.0072]$ 
    & $\phantom{-}0.00609$&$[0.00593,0.00625]$
    & $\phantom{-}0.00609$&$[0.00593,0.00624]$\\
    & $-0.0061$&$[-0.0085,-0.0037]$ 
    & $-0.00586$&$[-0.00627,-0.00545]$ 
    & $-0.00587$&$[-0.00628,-0.00549]$ \\
    \multirow{2}{*}{$\epsilon_H$ vs $\xi_H^2$}
    & $\phantom{-}0.0058$&$[0.0038,0.0078]$ 
    & $\phantom{-}0.006095$&$[0.006066,0.006123]$ 
    & $\phantom{-}0.006095$&$[0.006066,0.006123]$\\
    & $\phantom{-}0.002$&$[-0.059,0.065]$ 
    & $\phantom{-}0.0113$&$[0.0101,0.0126]$
    & $\phantom{-}0.0113$&$[0.0101,0.0126]$\\ 
    \multirow{2}{*}{$\epsilon_H$ vs $\omega^3_H$}
    & $\phantom{-}0.00606$&$[0.00513,0.00699]$ 
    & $\phantom{-}0.006095$&$[0.006085,0.006105]$ 
    & $\phantom{-}0.006095$&$[0.006085,0.006105]$\\
    & $\phantom{-}0.51$&$[0.07,0.92]$ 
    & $\phantom{-}0.518$&$[0.499,0.536]$ 
    & $\phantom{-}0.517$&$[0.498,0.536]$\\
    \multirow{2}{*}{$\eta_H$ vs $\xi_H^2$}
    & $-0.0059$&$[-0.0080,-0.0038]$ 
    & $-0.005848$&$[-0.005933,-0.005762]$
    & $-0.005848$&$[-0.005935,-0.005760]$\\
    & $\phantom{-}0.011$&$[-0.020,0.043]$     
    & $\phantom{-}0.0114$&$[0.0099,0.0128]$ 
    & $\phantom{-}0.0113$&$[0.0099,0.0128]$\\
    \multirow{2}{*}{$\eta_H$ vs $\omega^3_H$}
    & $-0.0059$&$[-0.0079,-0.0039]$ 
    & $-0.005849$&$[-0.005875,-0.005822]$ 
    & $-0.005849$&$[-0.005875,-0.005822]$\\
    & $\phantom{-}0.52$&$[0.07,0.94]$ 
    & $\phantom{-}0.518$&$[0.498,0.537]$ 
    & $\phantom{-}0.517$&$[0.498,0.536]$ \\
    \multirow{2}{*}{$\xi_H^2$ vs $\omega^3_H$}
    & $\phantom{-}0.011$&$[-0.018,0.040]$ 
    & $\phantom{-}0.01133$&$[0.01083,0.01183]$ 
    & $\phantom{-}0.01132$&$[0.01082,0.01183]$ \\
    & $\phantom{-}0.51$&$[0.09,0.93]$ 
    & $\phantom{-}0.518$&$[0.497,0.538]$ 
    & $\phantom{-}0.517$&$[0.496,0.538]$\\
    \bottomrule
  \end{tabular}} 
  \caption{Mean values and error bars for 2-dimensional contours of
    the slow-roll parameters space for the CMB
    (Fig.~\ref{fig:Planck2D}), 21cm hydrogen spectrum
    (Fig.~\ref{fig:SKA2D}) and their combination
    (Fig.~\ref{fig:SKA+Planck2D}).}
  \label{tab:Planck2D}
\end{table}

\section{SKA projections}
\label{sec:ska_likelihood}

The second window to the matter spectrum at relatively high redshift
are intensity fluctuations of the 21cm hydrogen line. We focus on the
redshifts range between 8 and 10. The upper bound avoids the position
dependence of the spin temperature, since the spin temperature couples
to the gas temperature through the Wouthuysen-Field effect in this
redshift range~\cite{Mao:2008ug}. The lower bound allows us to avoid
position-dependent reionization, as there is still nearly no reionized
helium. Since this redshift regime probes patterns from before the
reionization started, the neutral hydrogen fraction is
$\overline{x}_{H} = 1$ and we can identify the power spectrum of the
neutral hydrogen perturbations $P_\text{HI}(k)$ with the matter power
spectrum $P_\delta(k,z)$.  This means the two-point temperature
correlations of the 21cm intensity can be expressed
as~\cite{Munoz:2015eqa, Munoz:2016owz}
\begin{align}
	\langle \Delta T_{21}(\mathbf{k}) \Delta T_{21}(\mathbf{k}')\rangle \equiv P_{21}(\mathbf{k},z)(2\pi)^3 \delta(\mathbf{k}-\mathbf{k}'),
\end{align}
where $\Delta T_{21}(\mathbf{k})$ is the Fourier transformation of the
difference between the 21cm temperature $T_{21}(\mathbf{x})$ with,
\begin{align}
	P_{21}(\mathbf{k}) = \left[\mathcal{A}(z) + \overline{T}_{21}(z) \mu^2\right]^2 P_\text{HI}(k,z). \label{P21_from_matter}
\end{align}
Here the parameter $\mu \equiv k_{\parallel}/k$ is the cosine
between the line of sight $k_{\parallel}$ and the absolute value $k$
and the $\overline{T}_{21}(z)$ is the average 21cm temperature at
redshift $z$ where the function $\mathcal{A}(z)$ can be found from
Refs.~\cite{Munoz:2015eqa,Bharadwaj:2004nr}.  The $P_{HI}(k,z)$ is the
spectrum of the neutral hydrogen density fluctuation which we assumed
to be equal to the matter spectrum $P_{\delta}(k,z)$, implying zero
(re)ionisation \cite{Gnedin:2004nj}.  Before the beginning of reionization the function
$\mathcal{A}(z)$ and the average temperature at a specific redshift
can be approximated as~\cite{Munoz:2016owz}
\begin{align}
	\mathcal{A}(z) = \overline{T}_{21}(z) = 27.3~\text{mK} \times \overline{x}_\text{H}
	\frac{T_s - T_\gamma}{T_s}\left(\frac{1+z}{10}\right)^{1/2}.
\end{align} 
During the epoch of recombination the spin temperature can be taken to
be much larger than the photon temperature due to the Wouthuysen-Field
effect. The gas temperature in the inter-galactic medium is heated by
$X$-ray photons up to hundreds of Kelvin~\cite{Mao:2008ug}.  This
allows us to drop the temperature factor, which reduces the previous
expression to
\begin{align}
	\mathcal{A}(z) = \overline{T}_{21}(z) =27.3~\text{mK} \times \overline{x}_\text{H} \left(\frac{1+z}{10}\right)^{1/2}.
\end{align}
This way, the 21cm-intensity and the matter distribution are linked in
the most straightforward way possible, with a uniform modelling of the
relationship between fundamental field and
observable~\cite{Choudhury:2016cyh, Gluscevic:2010iz}, ignoring
cross-correlations~\cite{Furlanetto:2006pg, Fialkov:2019jcx} and
taking into account velocities only~\cite{Wang:2005my}, while ignoring
structures beyond that of a continuous Gaussian random field such as
halo formation \cite{Schneider:2020xmf}.

The instrumental noise power spectrum in Fourier space can be
expressed as~\cite{Tegmark:2008au,Zaldarriaga:2003du}
\begin{align}
  P_{21}^N = \frac{\pi T_\text{sys}^2}{t_o f_\text{cover}^2}d_A^2(z) y_\nu(z) \frac{\lambda^2(z)}{D_\text{base}^2}
  \label{eq:noise_spectrum},
\end{align}
where $D_\text{base}$ is the baseline of the antenna array that is
uniformly covered up to a fraction $f_\text{cover}$ and $t_o$
is the observation time, with $\lambda(z)$ the 21cm-transition
wavelength at redshift $z$ (for optimisations of the design, please refer to \cite{Greig:2015zra}). The conversion function from frequency
$\nu$ to line of sight $k_\parallel$ is $y_\nu = 18.5
((1+z)/10)~\text{Mpc}/\text{MHz}$, while the system
temperature can be parameterized as~\cite{Munoz:2016owz}
\begin{align}
 T_\text{sys} = 180~\text{K} \times \left(\frac{\nu}{180~\text{MHz}}\right)^{-2.6}.
\end{align}
Here the frequency is the 21cm transition at redshift $z$, $\nu =
\nu_0/(1+z)$.  We take the observation time as $t_o = 10000$ hours
(hrs) for our analysis, however we also provide results for 1000 hrs
for comparison. The baseline $D_\text{base} = 1~\text{km}$ is taken to
be the baseline specified for SKA-LOW in Ref.~\cite{SKA:2018ckk}. The
coverage fraction in the nucleus of the antenna array can be computed
as~\cite{Tegmark:2008au}
\begin{align}
 f_\text{cover} = N_a \frac{D^2}{D_\text{base}^2},
\end{align}
where $N_a$ is the number of antennas while $D$ is their diameter.
For SKA-LOW~\cite{SKA:2018ckk} the coverage fraction is approximately
$f_\text{cover} \approx 0.0091$.

\begin{figure}[t]
  \includegraphics[width = 0.245\textwidth]{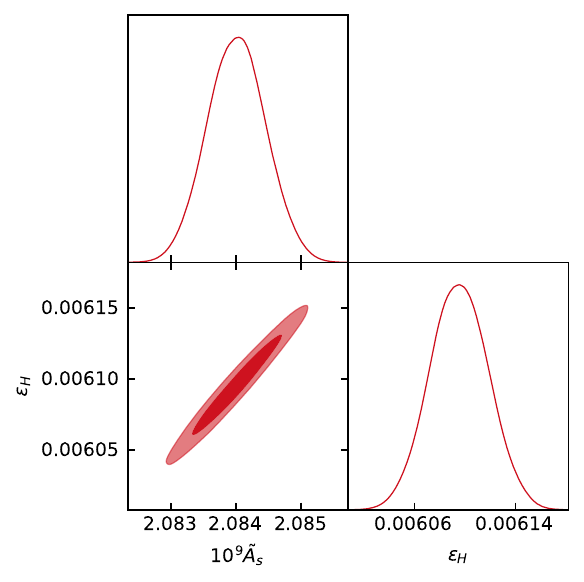}
  \includegraphics[width = 0.245\textwidth]{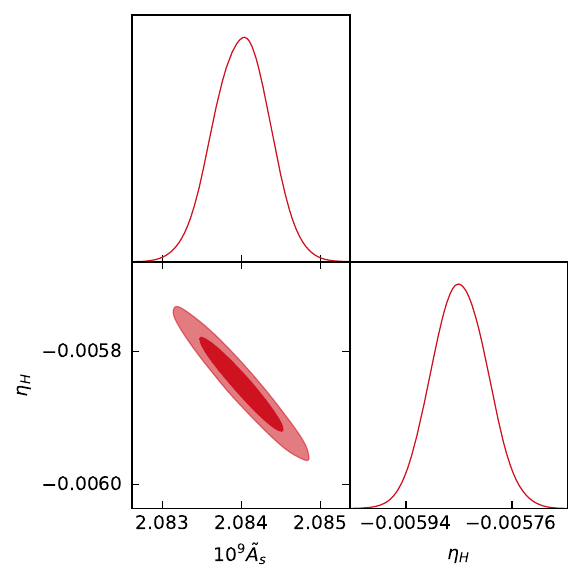}
  \includegraphics[width = 0.245\textwidth]{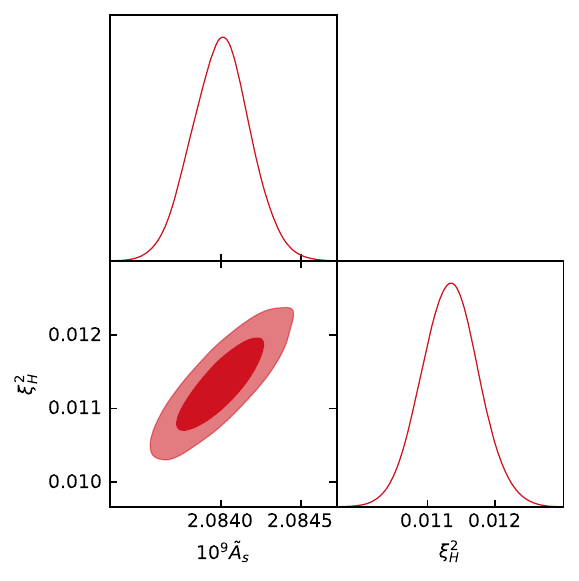} 
  \includegraphics[width = 0.245\textwidth]{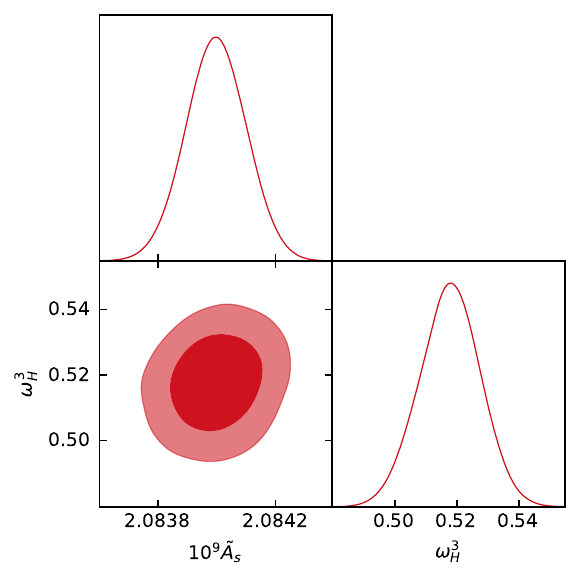} \\
  \includegraphics[width = 0.245\textwidth]{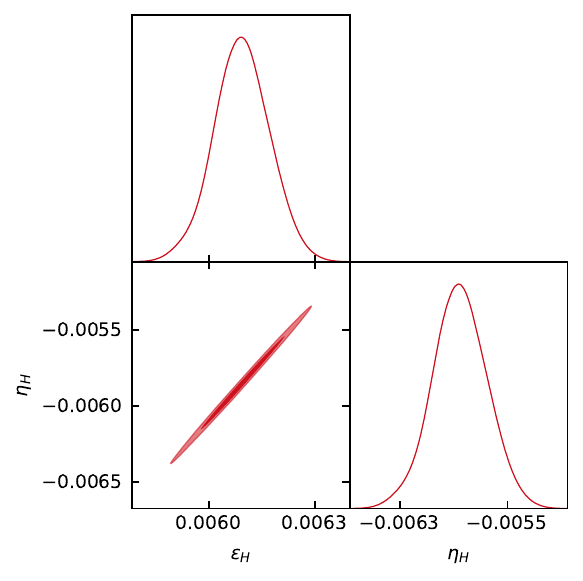}
  \includegraphics[width = 0.245\textwidth]{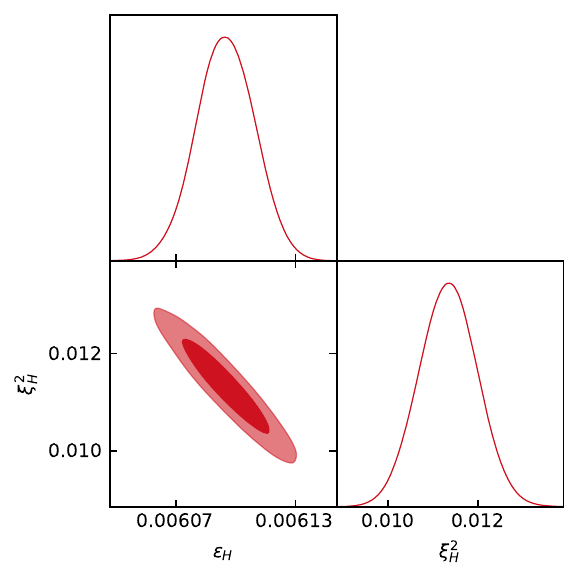} 
  \includegraphics[width = 0.245\textwidth]{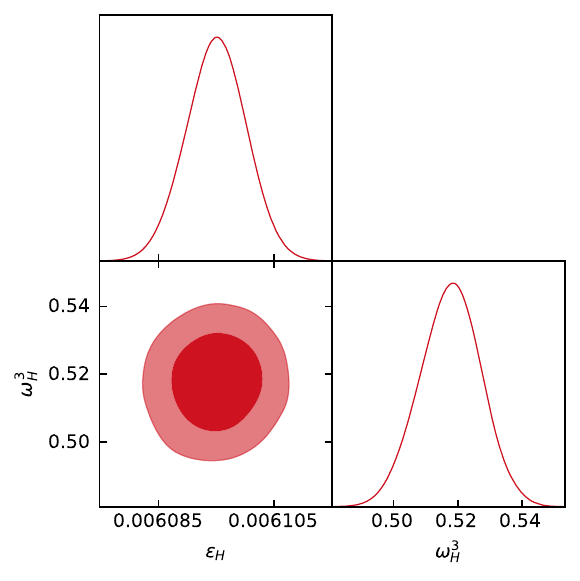}
  \includegraphics[width = 0.245\textwidth]{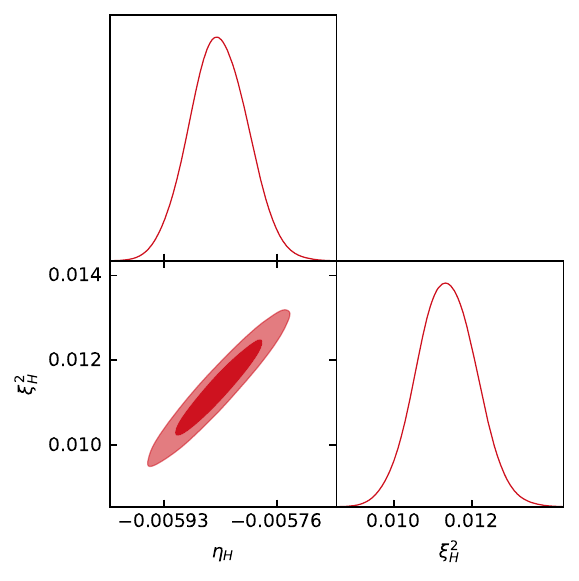} \\
  \includegraphics[width = 0.245\textwidth]{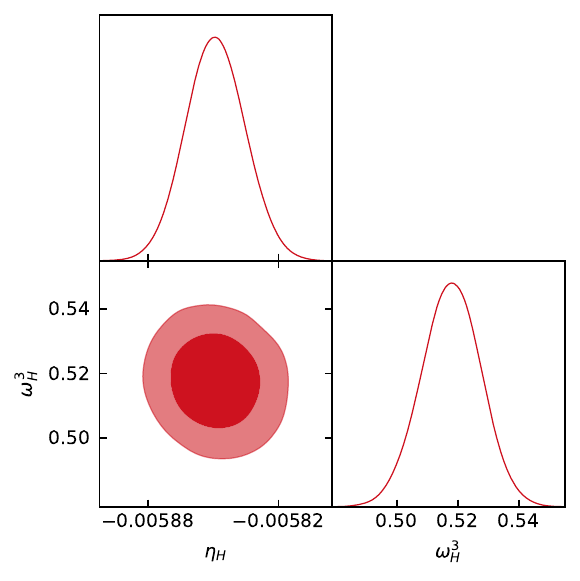} 
  \includegraphics[width = 0.245\textwidth]{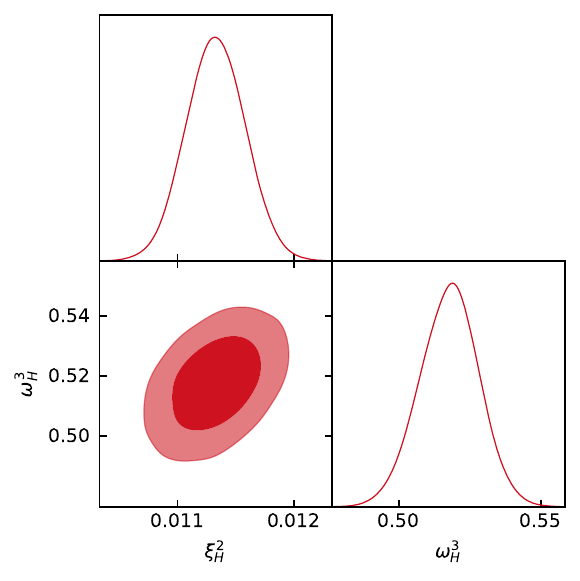}
  \caption{Sliced 2-dimensional likelihoods for the slow-roll
    parameters from SKA projections.}
  \label{fig:SKA2D}
\end{figure}

For a specific redshift bin centered at $z_i$ the $\chi^2$ functional
can be expressed as~\cite{Munoz:2016owz, Munoz:2020itp}
\begin{align}
	\chi^2_i = \frac{f_\text{sky}}{2}\frac{\text{Vol}_i}{(2\pi)^3} 
	\int_{k_\text{min}}^{k_\text{max}}dk(2\pi k^2) \int_{-1}^1d\mu \frac{[P_{21}
	(\mathbf{k},z,\mathbf{\theta})-P_{21}^{fid}(\mathbf{k},z,\mathbf{\theta}_\text{fid})]^2}{[P_{21}
	(\mathbf{k},z,\mathbf{\theta}) + P_{21}^N(z)]^2},
\end{align}
where subscripts $i$ denote the redshift bin and $\mathbf{\theta} =
\{\tilde{A}_s, \epsilon, \eta, \xi, \omega \}$. The comoving volume of the
redshift bin $\text{Vol}_i$ can be computed as a spherical shell in comoving
distance $r(z_i)$ and $r(z_{i-1})$, where $z_i$ and $z_{i-1}$ are the
edges of the redshift bin of interest. The expression to compute the
volume reads approximately as
\begin{align}
	\text{Vol}_i = \frac{4}{3}\pi \left(r(z_i)^3 - r(z_{i-1})^3\right),
\end{align} 
which is over the redshift range considered very accurate in
comparison to integration over the volume evolution, due to the fine
slicing in redshift.

For our analysis we take the 22 equally spaced redshift bins in the
region $z\in [8,10]$.  The comoving wave numbers are bounded from
above by the non linear scale which we set as $k_\text{NL} = 1$\impc,
as in Ref.~\cite{Munoz:2016owz}.  On the other hand, the astrophysical
foregrounds will cut off wave numbers smaller than $k^\text{min}
\approx 10^{-2}$ \impc \cite{Munoz:2016owz}.  Summing up all the
different $\chi^2_i$ we get the overall $\chi^2 = \sum_i \chi^2_i$.
The fiducial power spectrum $P_{21}^\text{fid}
(\mathbf{k},z,\mathbf{\theta}_\text{fid})$ is computed according to
Eq.\eqref{P21_from_matter} and $f_\text{sky}$ is set to $0.58$
according to Ref.~\cite{Sprenger:2018tdb}. The parameters
$\mathbf{\theta}$ are used to compute the matter power spectrum.  For
the computation of the fiducial power spectrum we choose the mean
values for these parameters based on the Planck likelihoods.

Whenever we combine the data sets, we assume that CMB and 21cm data
are uncorrelated. This assumption could be challenged if one takes
into account effects such as gravitational lensing on the radiation
backgrounds by the same structures or correlated secondary
anisotropies~\cite{Pourtsidou:2014pra, Zahn:2005ap}. We ignore such
effects also because they are expected to remain sub-leading compared
to the primary fluctuations of the two radiation backgrounds.

\subsubsection*{Slow-roll parameters from SKA}

\begin{figure}[t]
  \includegraphics[width = 0.32\textwidth]{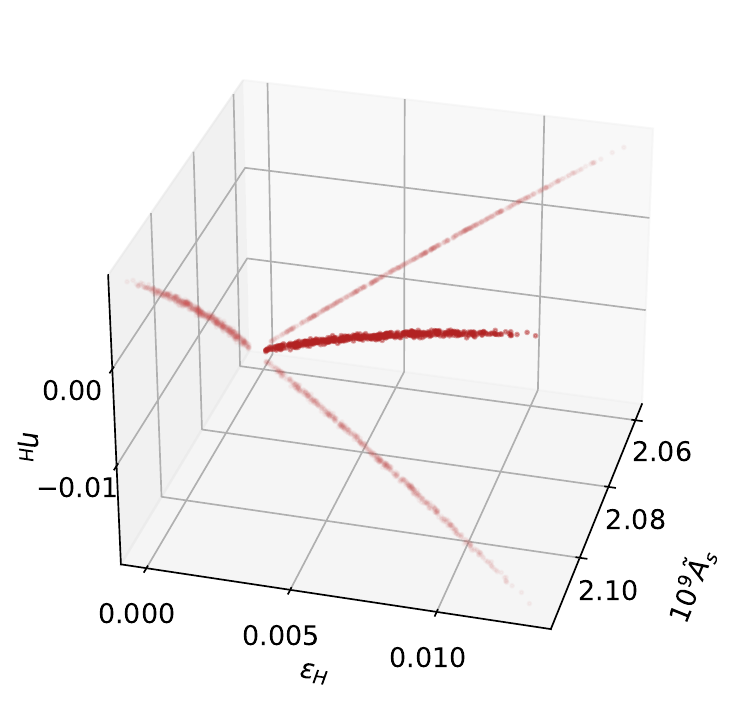}
  \includegraphics[width = 0.32\textwidth]{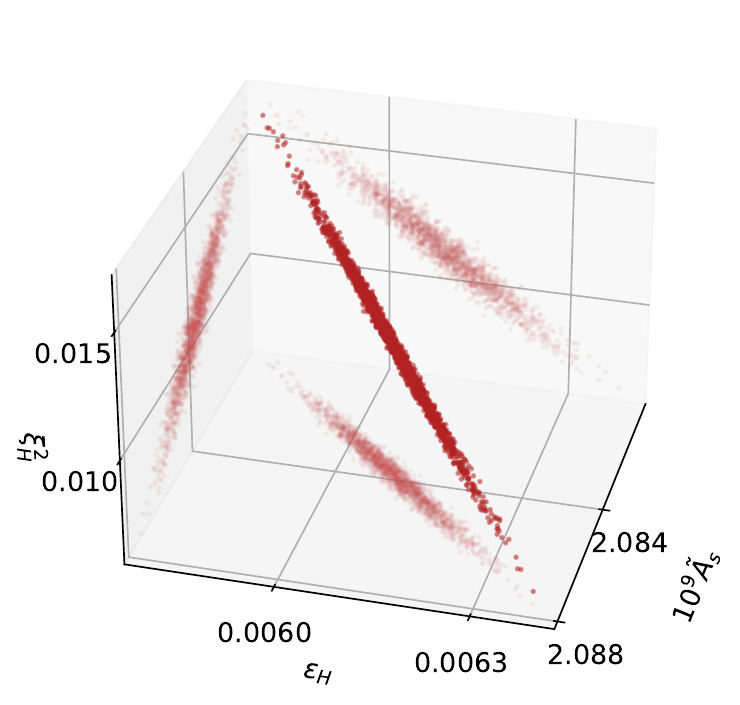}
  \includegraphics[width = 0.32\textwidth]{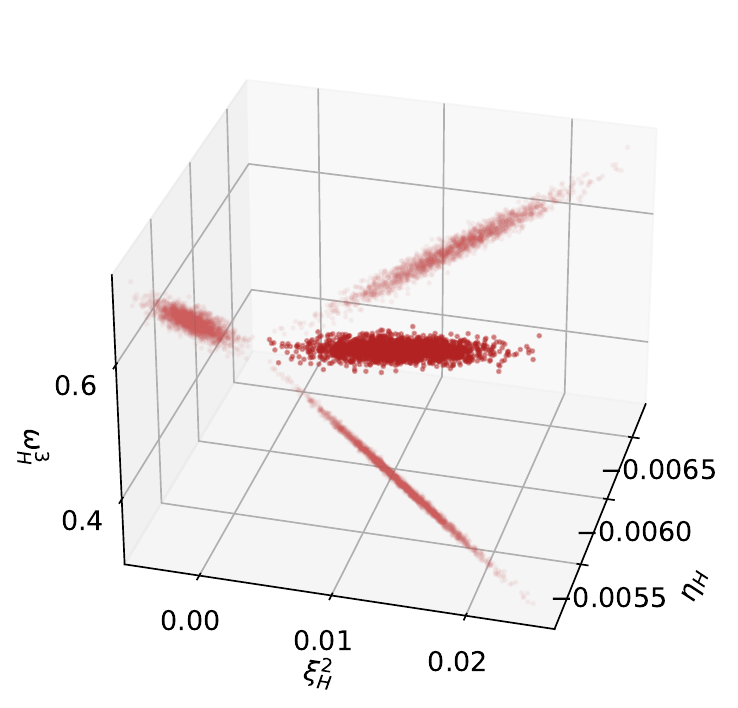}
  \caption{Sliced 3-dimensional likelihood ellipsoids for a selection
    of slow-roll parameters from SKA projections.}
  \label{fig:SKA3Dx}
\end{figure}

In this section we discuss the potential of 21cm tomography in
constraining the slow-roll parameters in detail.  As we shall see
shortly a clear hierarchy in sensitivity of observables on
cosmological parameters, we simply keep the background cosmology fixed
to the $\Lambda$CDM given in Eq.\eqref{eq:cosmo}. With this vanilla
parameter choice, the SKA data only constrains the inflationary
potential through the slow-roll parameters defined in
Eq.\eqref{eq:paras}.

\begin{figure}[t]
  \includegraphics[width = 0.32\textwidth]{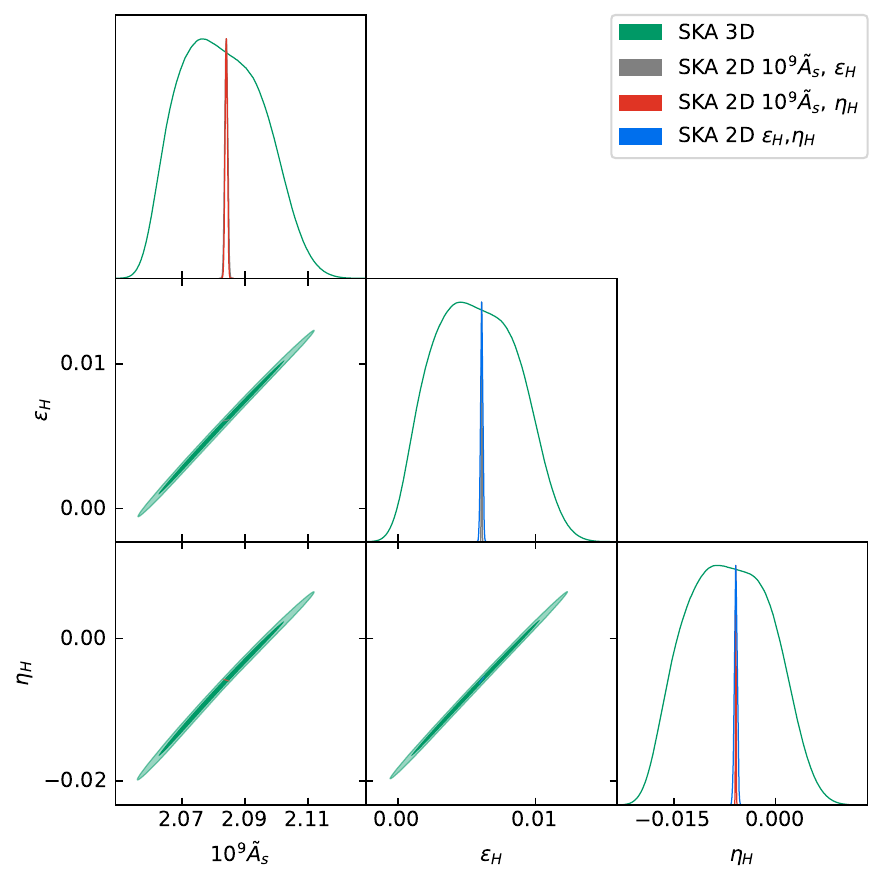}
  \includegraphics[width = 0.32\textwidth]{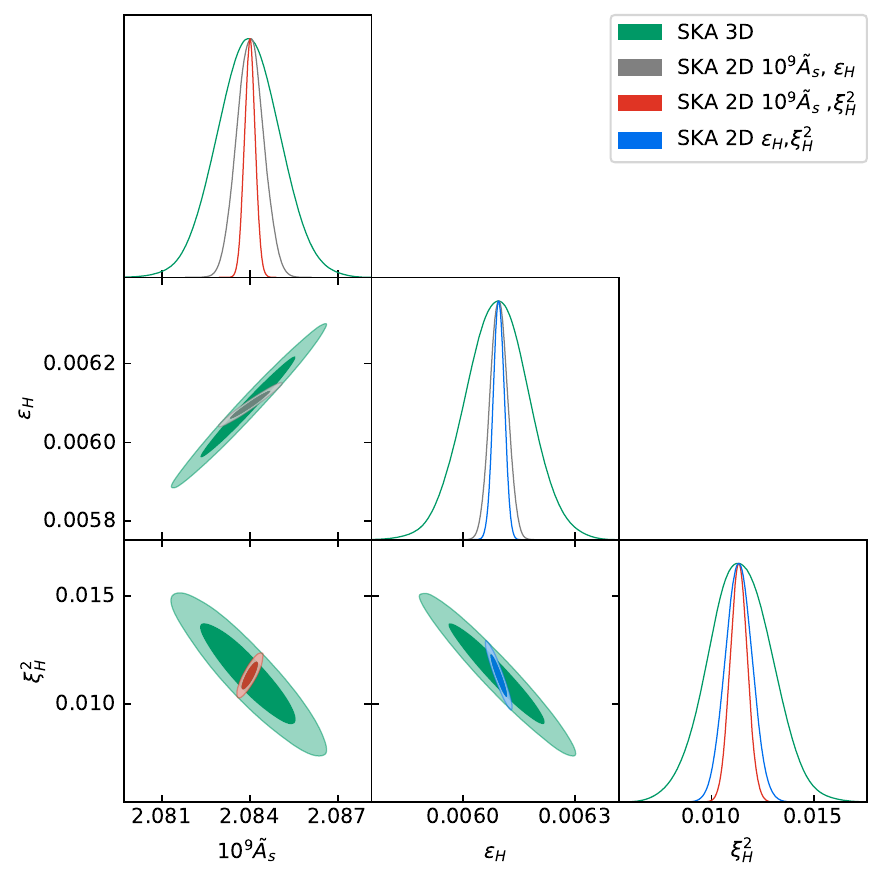}
  \includegraphics[width = 0.32\textwidth]{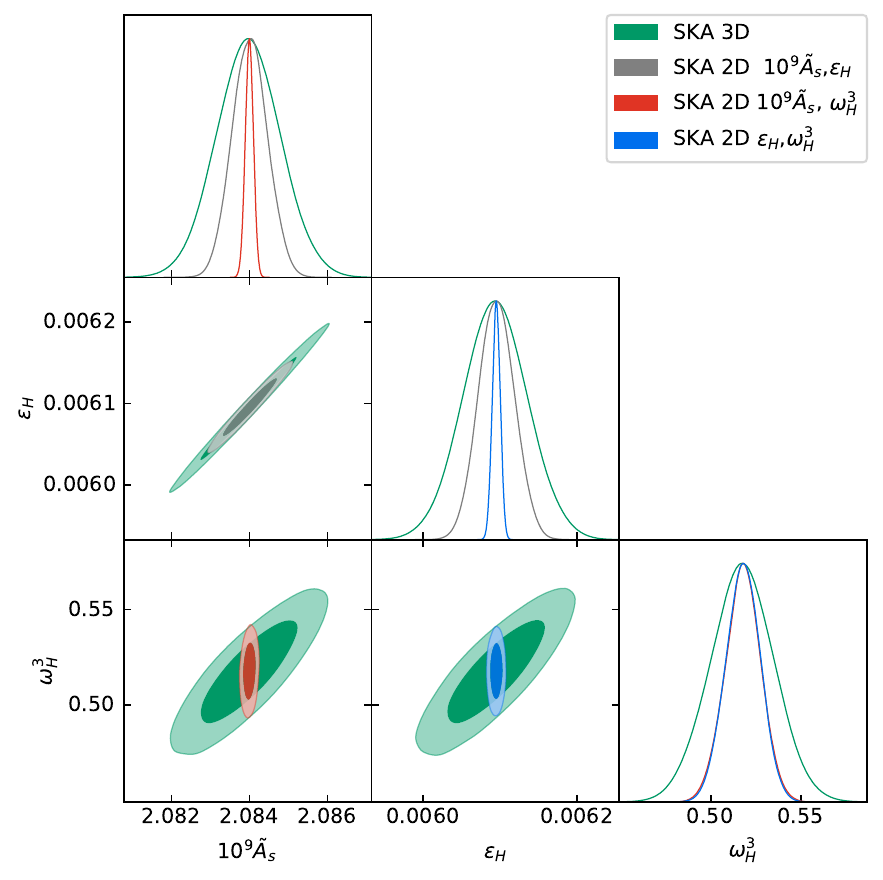} \\
  \includegraphics[width = 0.32\textwidth]{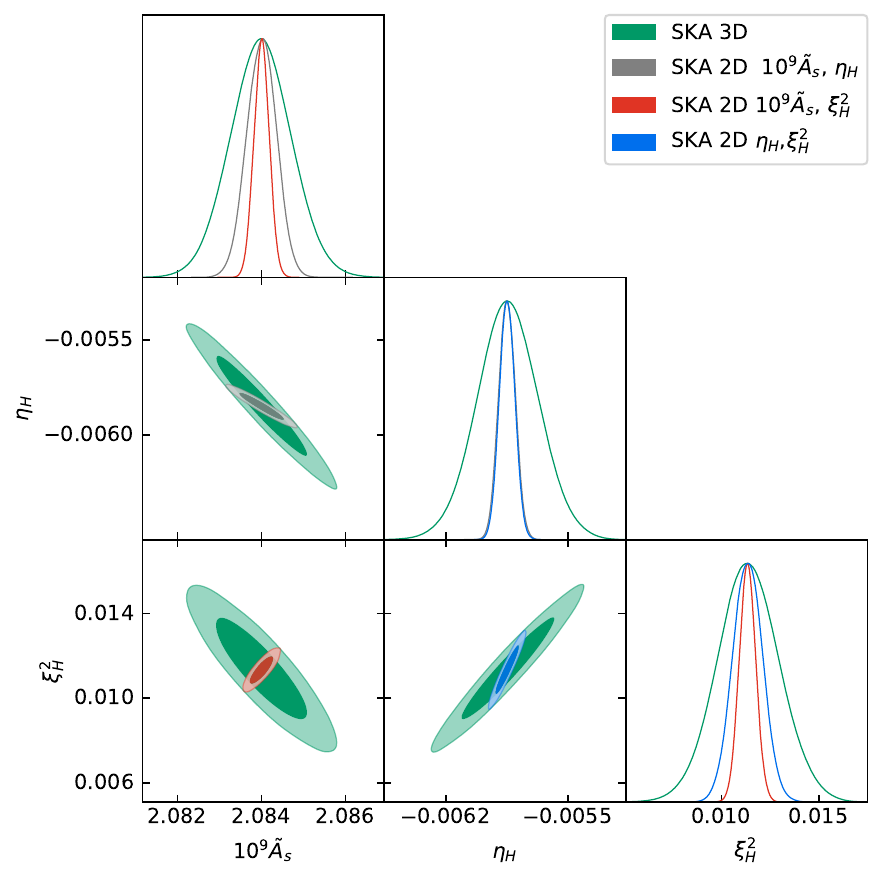}
  \includegraphics[width = 0.32\textwidth]{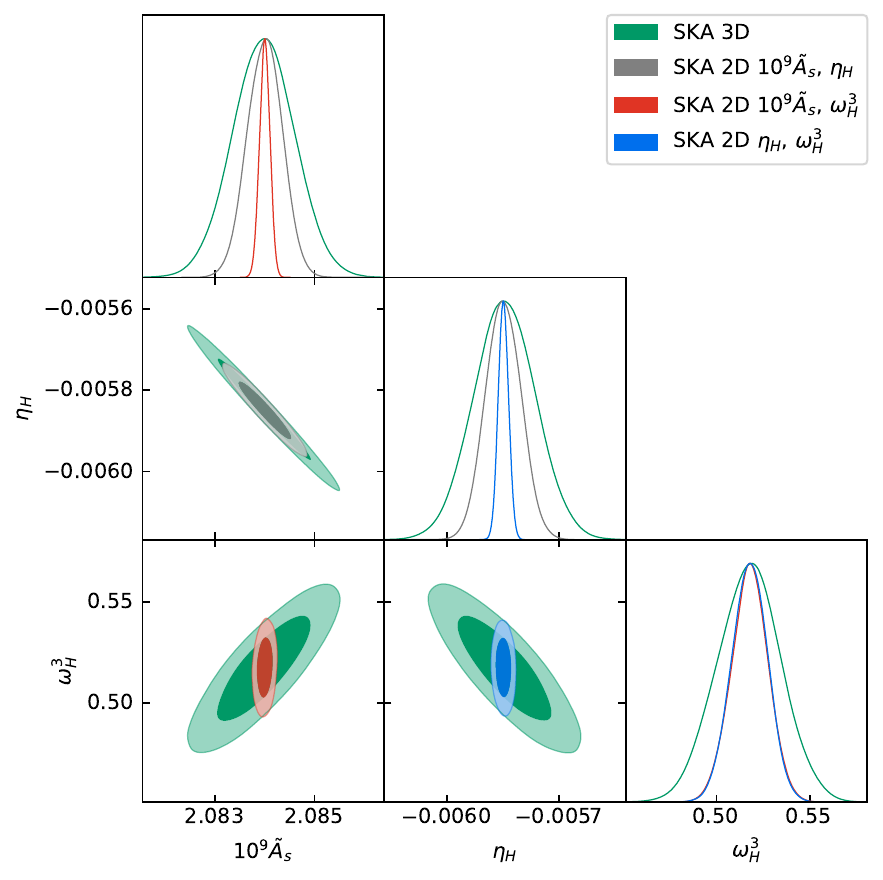}
  \includegraphics[width = 0.32\textwidth]{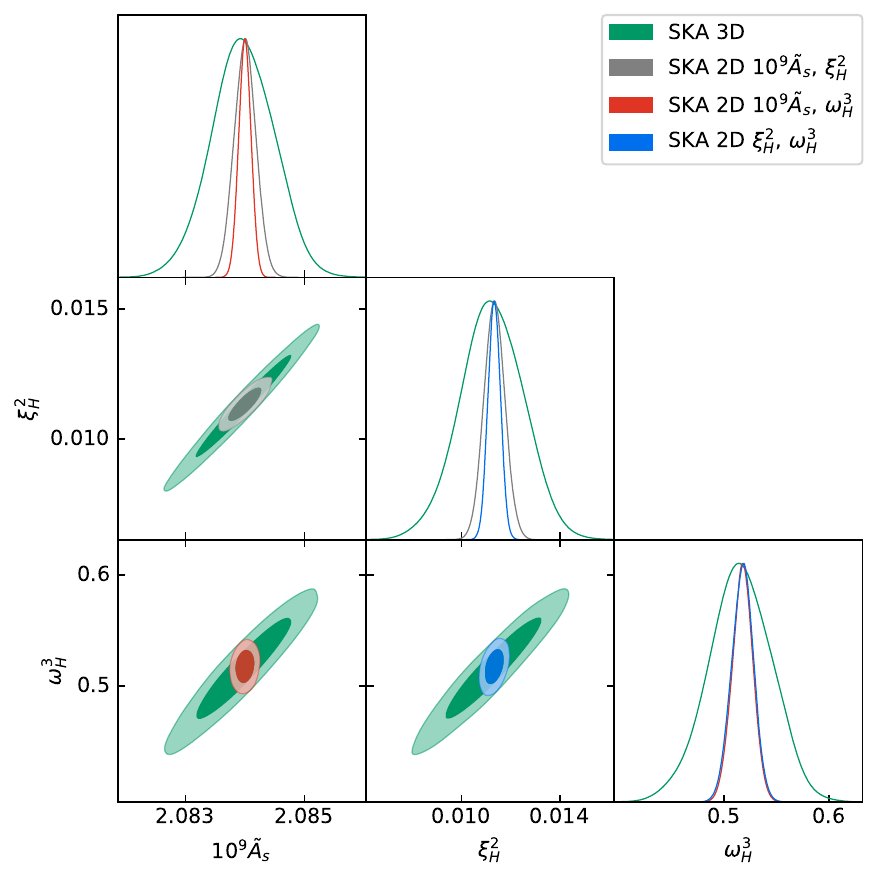} \\
  \includegraphics[width = 0.32\textwidth]{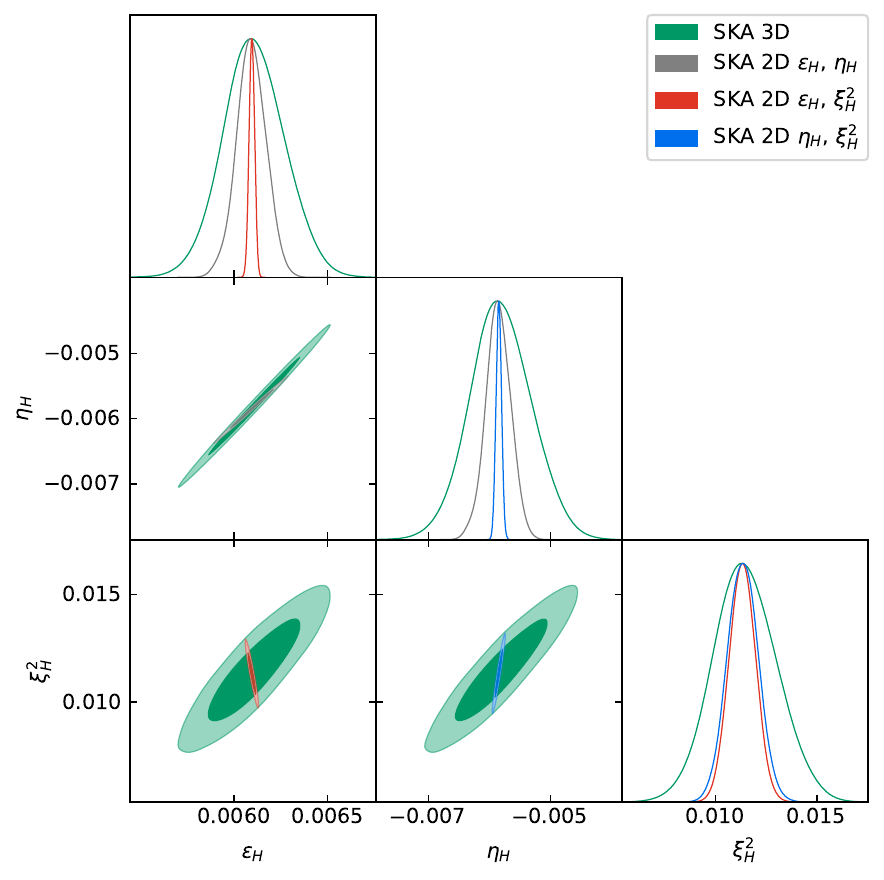}
  \includegraphics[width = 0.32\textwidth]{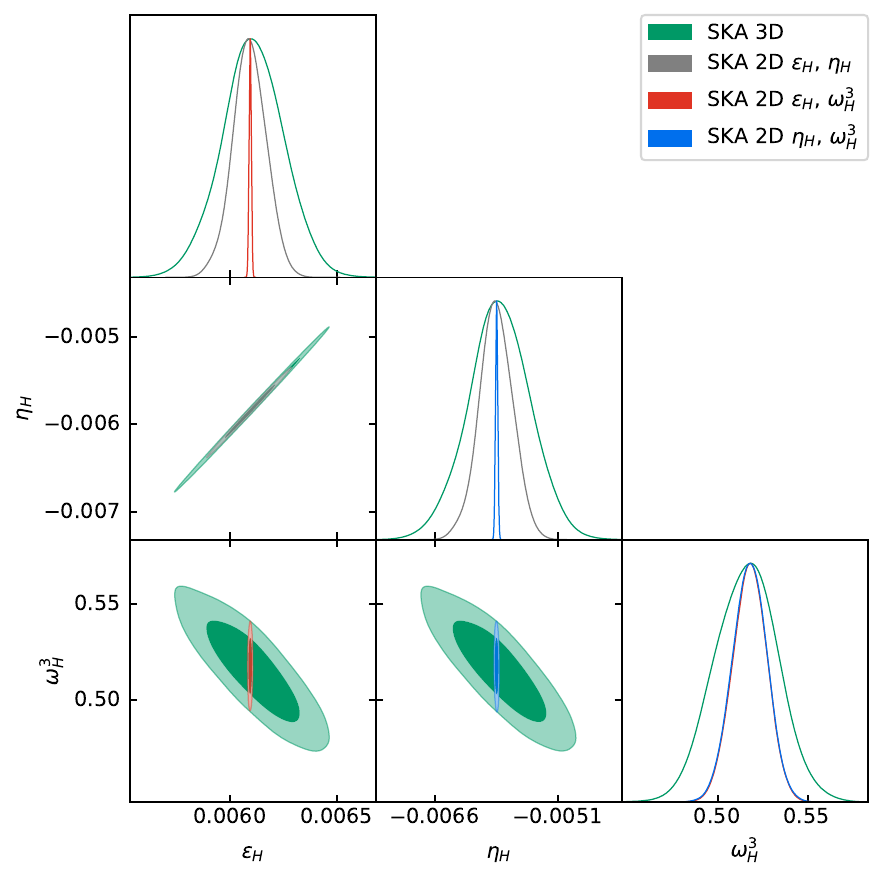}
  \includegraphics[width = 0.32\textwidth]{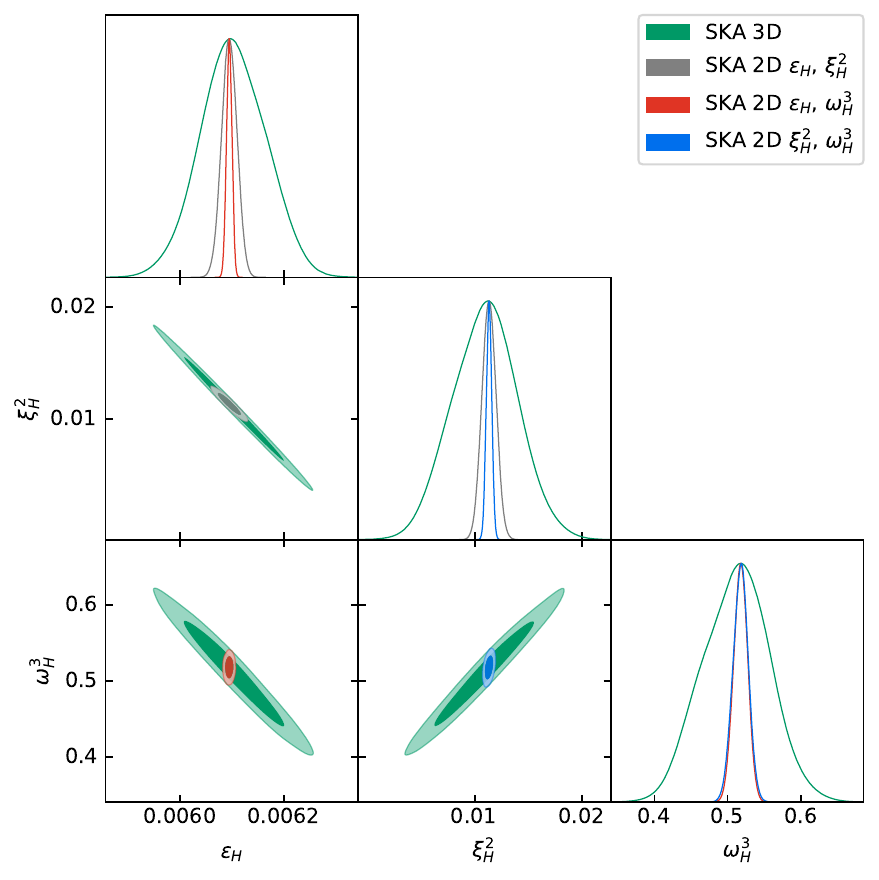} \\
  \includegraphics[width = 0.32\textwidth]{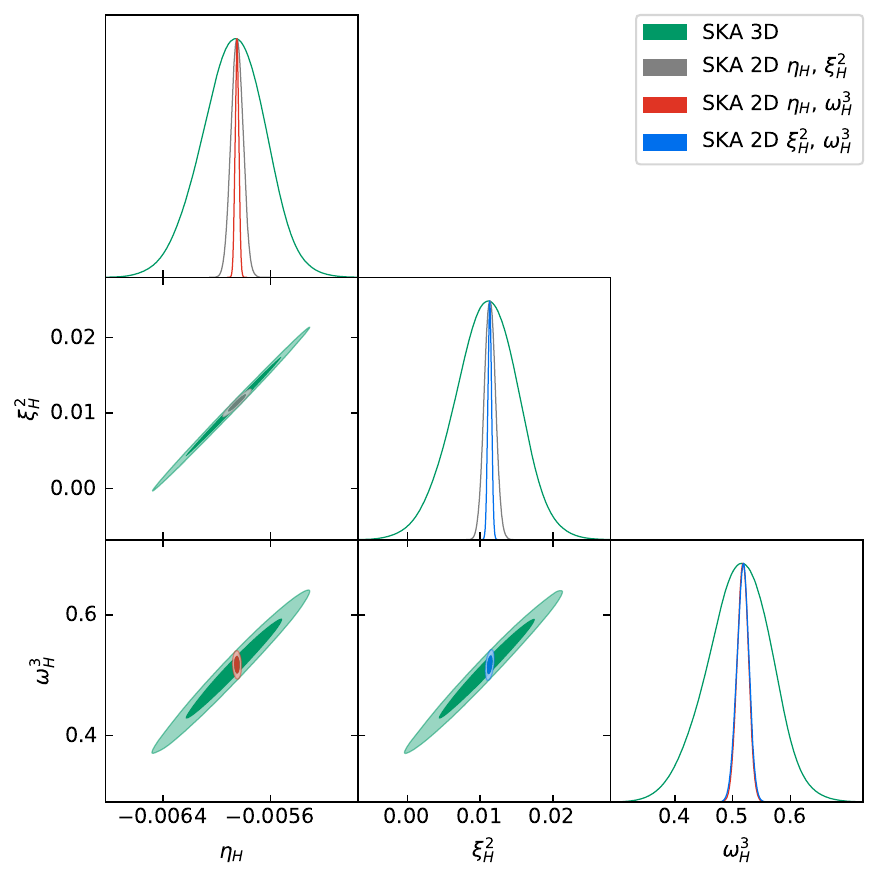}
  \caption{Marginalized 3-dimensional likelihoods from
    Fig.~\ref{fig:SKA3Dx}, compared with the 2-dimensional likelihoods
    in Fig.~\ref{fig:SKA2D}.}
  \label{fig:SKA3D}
\end{figure}

To understand the correlations between different slow-roll parameters
we first plot all possible 2-dimensional Markov chains in
Fig.~\ref{fig:SKA2D} based on the likelihood discussed in
Sec.~\ref{sec:ska_likelihood}. The figure should be compared with
Fig.~\ref{fig:Planck2D} where we show the 2-dimensional Markov
chains for Planck 2018 data. The corresponding mean values and errors
are presented in Tab.~\ref{tab:Planck2D} which can be compared with
similar one for Planck 2018 as given in Tab.~\ref{tab:Planck2D}. It is
clear that SKA offers more stringent constraints on slow-roll
parameters then Planck for all 2-dimensional combinations of slow-roll
parameters. As for Fig.~\ref{fig:Planck2D} we set the all remaining
parameters of our 2D analyses to the mean values given in the
Tab.~\ref{tab:HSR_planck}.

In a similar fashion we can study 3-dimensional Markov chains for the
slow-roll parameters, illustrating a few combinations in
Fig.~\ref{fig:SKA3Dx}. In Fig.~\ref{fig:SKA3D} we project the
3-dimensional error ellipses into two dimensions and superimpose the
2-dimensional chains from Fig.~\ref{fig:SKA2D}.  We also plot 2D
slices from the 3-dimensional Markov chains and superposed them with
2D Markov chains of
Figs.~\ref{fig:SKA2D_slices1} and \ref{fig:SKA2D_slices2} for all
slow-roll parameter combinations in App.~\ref{sec:2dslices}. These figures
illustrates that the 3-dimensional Markov chains still reflect the 
correlations observed in 2D chains. For the full set of slow-roll
parameters from Eq.\eqref{eq:paras} we rely on the combination with
the Planck data, where the numerics are less challenging than for SKA
alone.

\begin{figure}[t]
  \centering
  \includegraphics[width = 0.5\textwidth]{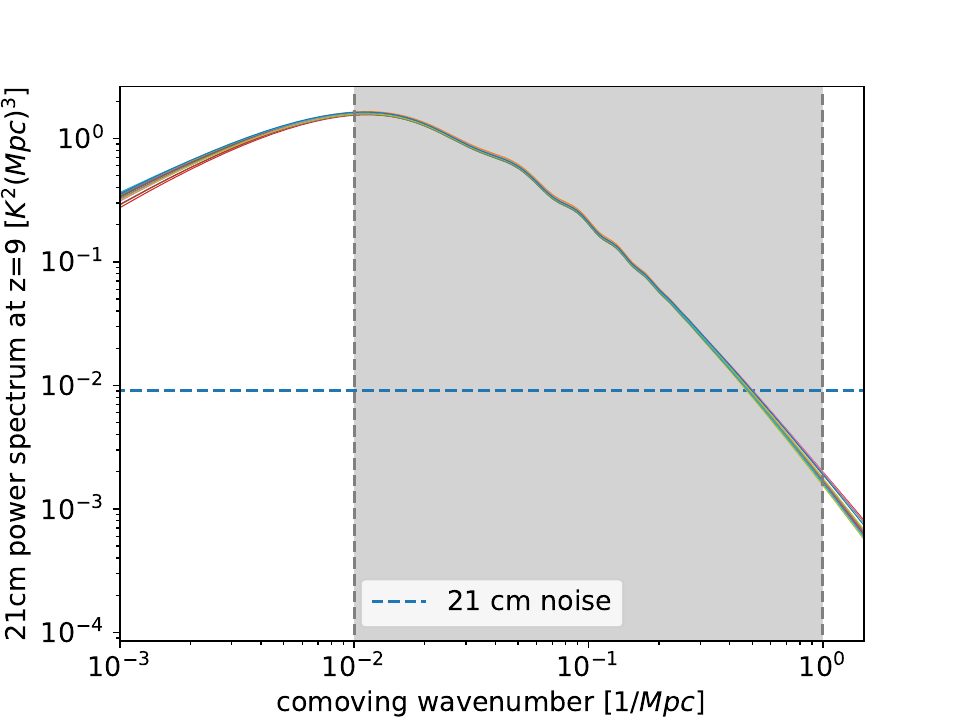}
  \caption{Comparison between 21cm power spectrum and the noise power
    spectrum computed through Eq.\eqref{eq:noise_spectrum}.  The gray
    lines denote the maximal and minimal scales
    considered.}
  \label{fig:SKA_noise}
\end{figure}

The observed improvement in sensitivity on slow-roll parameters
provided by SKA originates from the wide range of scales that are
probed and that the range of accessible scales stretches to small
spatial scales, too, giving SKA in comparison to Planck a better lever
to constrain the effect of slow-roll parameters on the spectrum. In
Fig.~\ref{fig:SKA_noise} we illustrate variations in the spectrum for
a representative selection of samples of the slow-roll parameters.  We
show the largest scales close to the pivot-scale, as probed by Planck,
and the smallest scales, where SKA plays out its unique sensitivity
and resolution. Comparing Fig.~\ref{fig:SKA_noise} and
Fig.~\ref{fig:TTEE_noise} one should keep in mind that on the largest
scales there is a significant cosmic variance, which is not included
in the shown noise levels, such that the constraining power of
CMB-spectra on the largest scales remains limited. We have chosen $h$ to be consistent with the comparatively low values from the CMB in all our forecasts, as both observations probe the high-redshift universe. The choice of a higher value of $h$ as reported from low-redshift observations would not have a strong influence on the constraints of slow-roll parameters, as preliminary tests suggest.

\begin{figure}[b!]
	\includegraphics[width = 0.245 \textwidth]{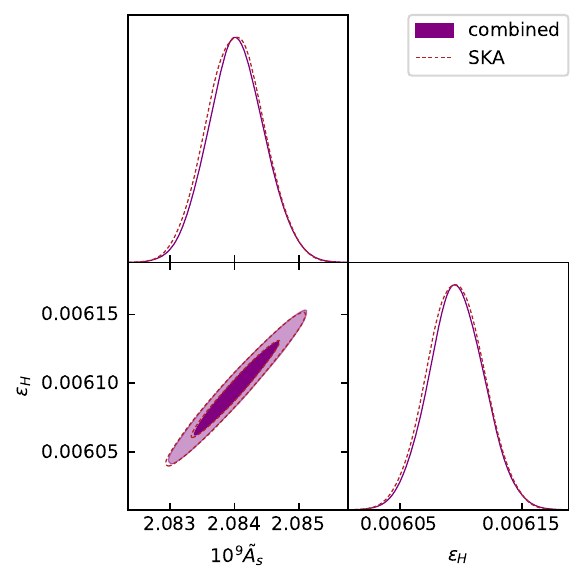}
	\includegraphics[width = 0.245 \textwidth]{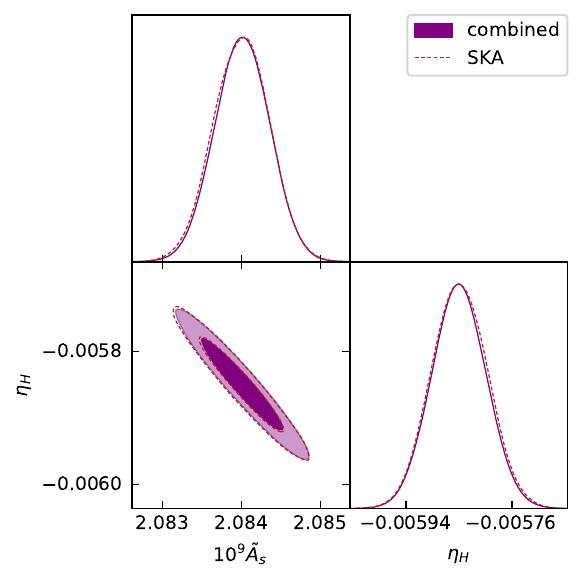}
	\includegraphics[width = 0.245 \textwidth]{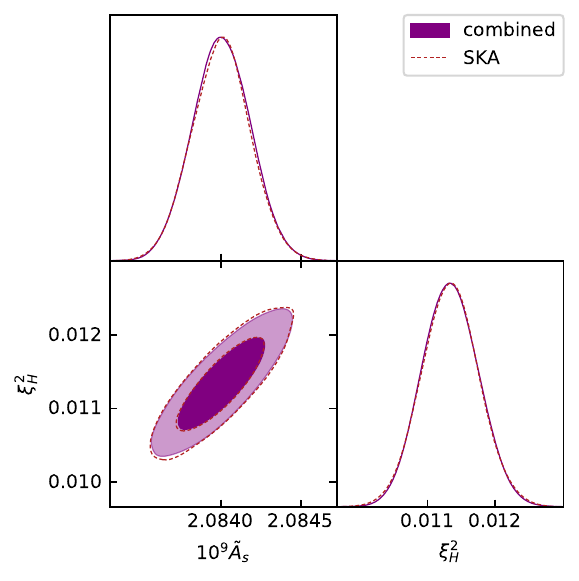} 
	\includegraphics[width = 0.245 \textwidth]{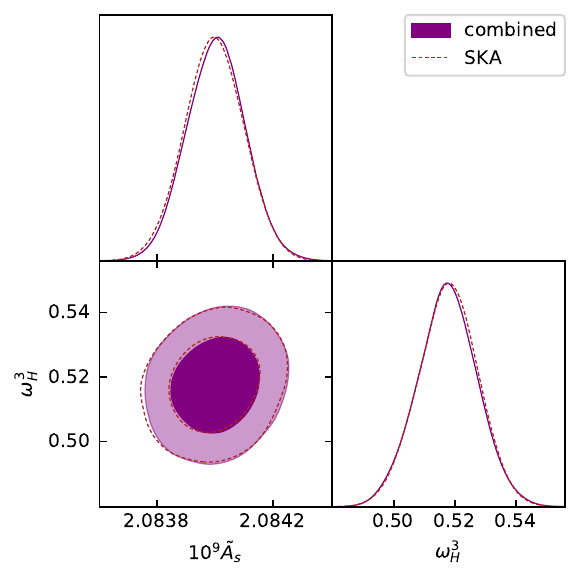} \\
	\includegraphics[width = 0.245 \textwidth]{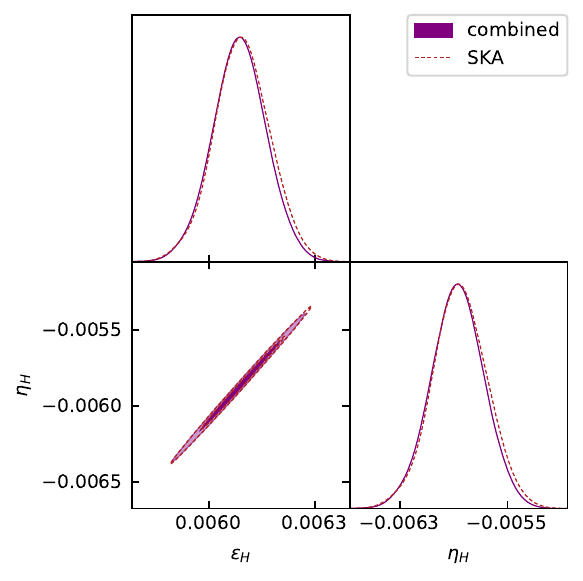} 
	\includegraphics[width = 0.245 \textwidth]{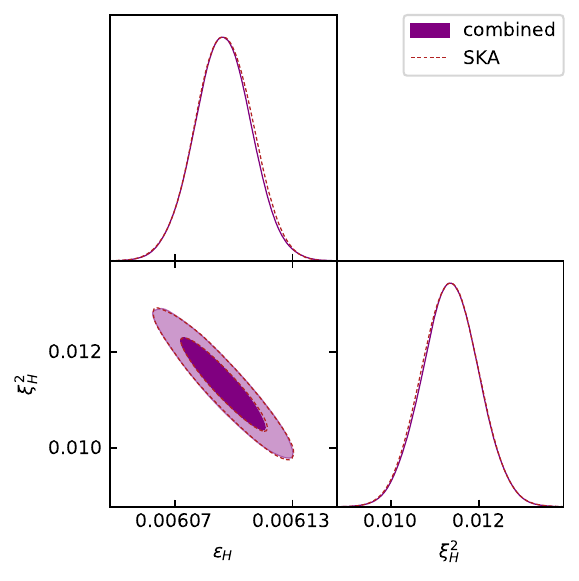} 
	\includegraphics[width = 0.245 \textwidth]{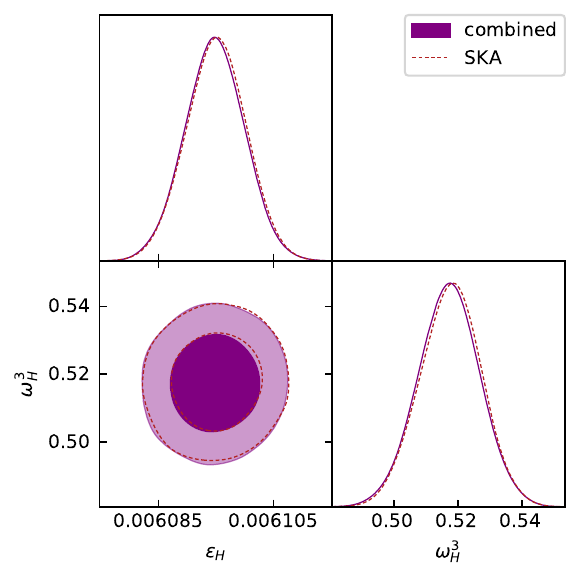}
	\includegraphics[width = 0.245 \textwidth]{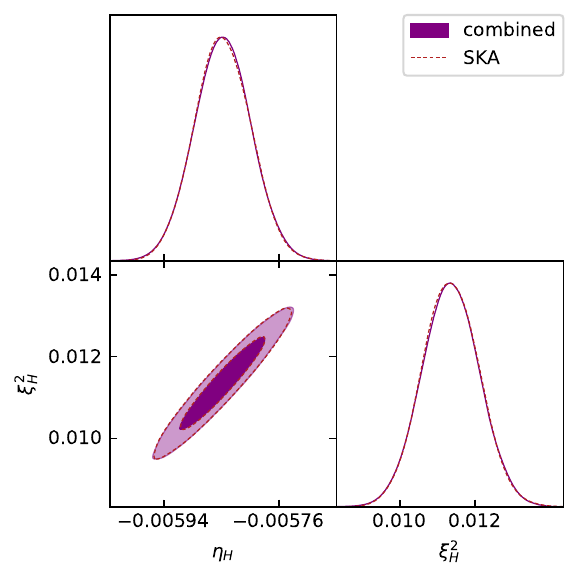} \\
	\includegraphics[width = 0.245 \textwidth]{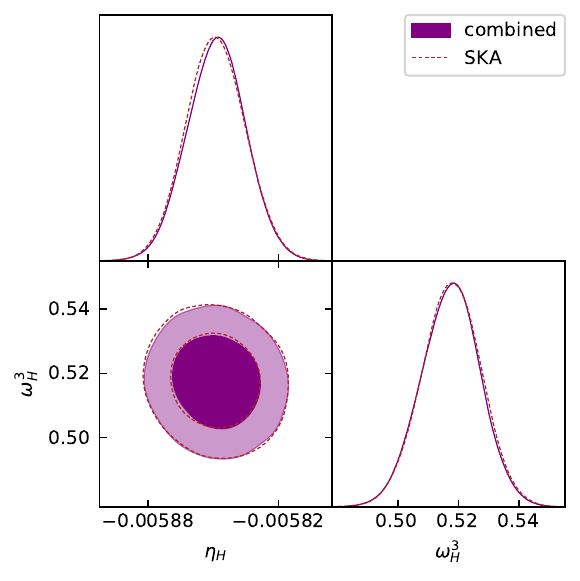} 
	\includegraphics[width = 0.245 \textwidth]{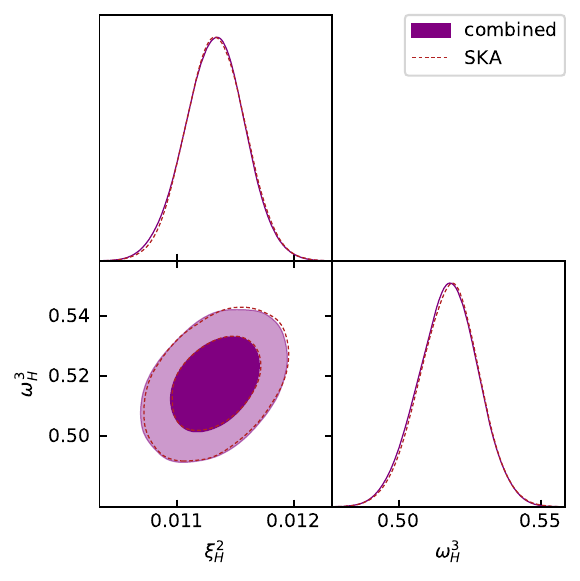}
	\caption{Sliced 2-dimensional likelihoods for the slow-roll
          parameters from a combination of Planck and SKA
          projections. We also show the SKA-only contours from
          Fig.~\ref{fig:SKA2D} as red dashed lines, the results should
          be compared to the Planck results shown in
          Fig.~\ref{fig:Planck2D}.}
	\label{fig:SKA+Planck2D}
\end{figure}

\subsubsection*{Slow-roll parameters from SKA and Planck}

\begin{figure}[t]
  \centering
  \includegraphics[width = 0.80\textwidth]{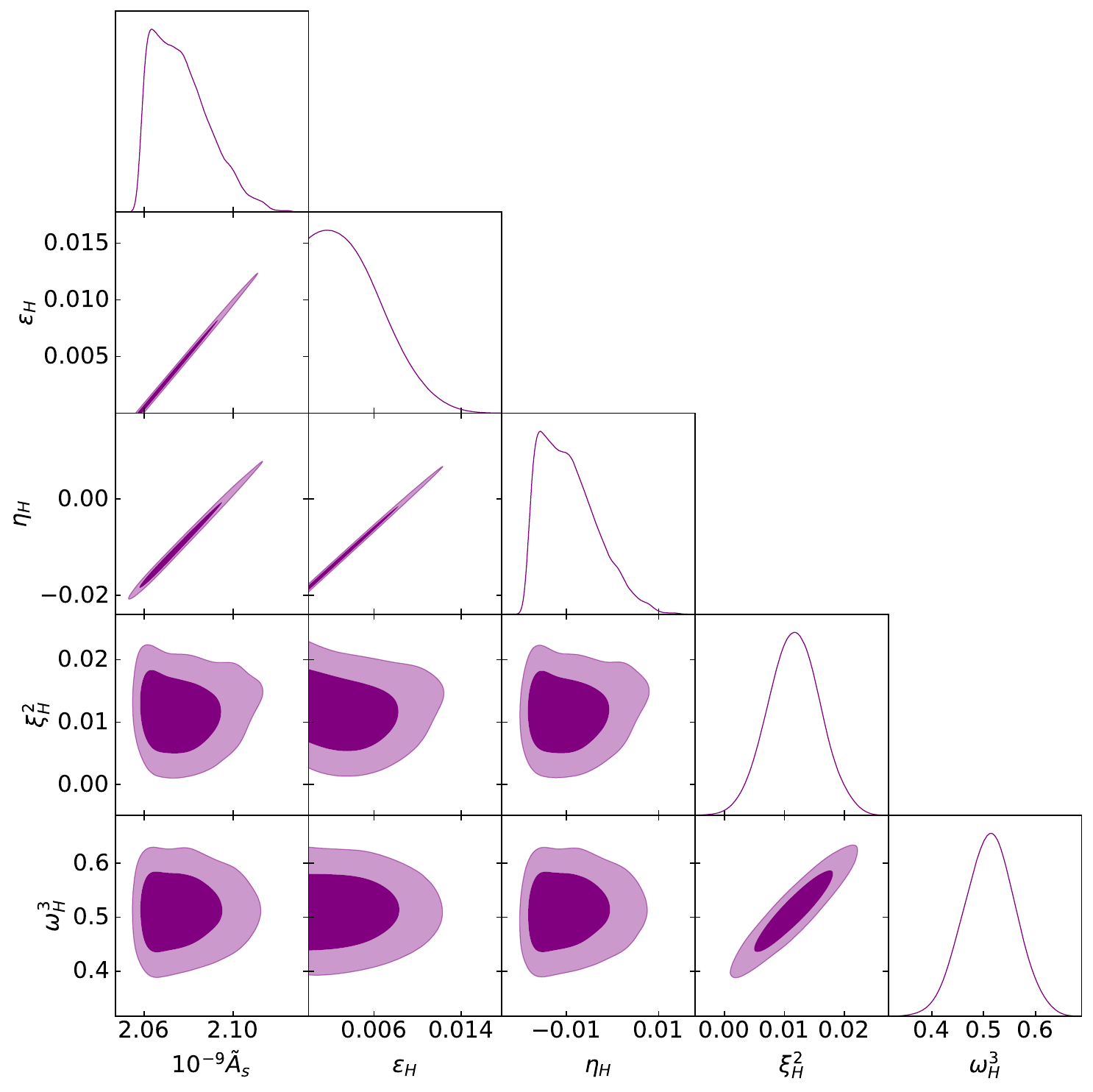}
  \caption{Marginalized 68\% and 95\% CL contours from the
    5-dimensional likelihood of the slow-roll parameters in
    Eq.\eqref{eq:paras}.  We combine the 2018 Planck results with SKA
    projections assuming 10000~hrs observation time.}
  \label{fig:SKA+Planck5D}
\end{figure}

To illustrate how the combined SKA and Planck likelihoods constrain
the slow-roll parameters, we again start with the 2-dimensional
contours in Fig.~\ref{fig:SKA+Planck2D}.  The corresponding mean
values and 95\% CL limits are included in Tab.~\ref{tab:Planck2D}. The
projected 2-dimensional constraints from the Planck and SKA
combination are extremely similar to the SKA limits alone, as expected
from the weaker Planck limits shown in Fig.~\ref{fig:Planck2D}. This
is expected since the constraints from SKA in the 2D parameters sets are
much stronger than the Planck as can be compared from
Tab.~\ref{tab:Planck2D}. We provide the 3-dimensional constraints from
the Planck and SKA combination in Fig.~\ref{fig:SKA+Planck3D} in the
Appendix. Unlike SKA likelihood alone as in previous section we find
that 5-dimensional slow-roll parameters converge well, since the
Planck likelihood cuts off approximately flat directions. The
marginalized 2D contours from the 5-dimensional Markov chains for the
slow-roll parameters given in Eq.\eqref{eq:paras} are shown in
Fig.~\ref{fig:SKA+Planck5D}.

\begin{figure}[t]
  \centering
  \includegraphics[width = 0.80\textwidth]{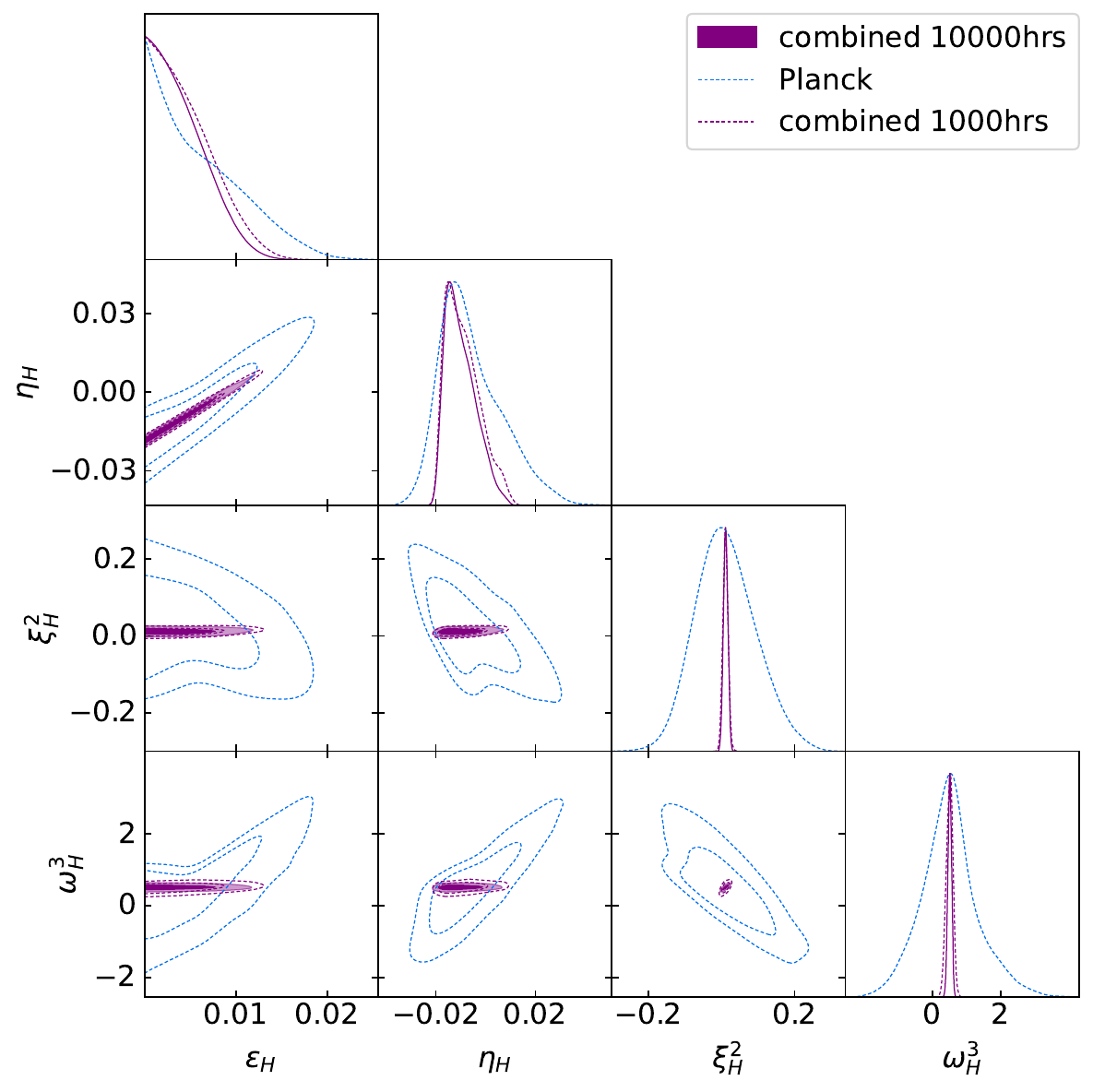}
  \caption{Marginalized 68\% and 95\% CL contours from the
    5-dimensional likelihood of the slow-roll parameters in
    Eq.\eqref{eq:paras}, also marginalized over the cosmological
    parameters in Eq.\eqref{eq:cosmo}.  We combine the 2018 Planck
    results with SKA projections assuming 10000~hrs (purple solid) and
    1000~hrs (purple dashed) observation time and overlay the Planck
    contours from Fig.~\ref{fig:HSR_planck}}
  \label{fig:HSR_combo}
\end{figure}

\begin{table}[b!]
  \centering
  \begin{small}
  \begin{tabular}{c|lc|lc|lc} 
    \toprule
    & \multicolumn{2}{c|}{Planck} & \multicolumn{2}{c}{SKA+Planck} (1000 hrs) & \multicolumn{2}{c}{SKA+Planck} (10000 hrs) \\
    Parameter & mean& 95\% CL & mean& 95\% CL & mean& 95\% CL \\ 
    \midrule
    $\tilde{A}_s \times 10^9$ & $\phantom{-}2.084$      & $[1.978, 2.197]$                      &$\phantom{-}2.075$&$[2.046,2.110]$ & $\phantom{-}2.075$&$[2.048,2.106]$    \\
    $\epsilon_H$  & $\phantom{-}0.006095$   & $\phantom{-}< 0.01518$                & $\phantom{-}0.0043$ &$< 0.0101$   & $\phantom{-}0.0041$ &$< 0.00951$  \\ 
    $\eta_H$      & $-0.005849$             & $[-0.02804, 0.02104]$                 & $-0.0097$ &$[-0.0198,0.0032]$     & $-0.0101$ &$[-0.0197,0.0022]$ \\ 
    $\xi^2_H$     & $\phantom{-}0.01133$    & $[-0.1498, 0.1797]$                   & $\phantom{-}0.011$ &$[-0.002,0.024]$ & $\phantom{-}0.011$   &$[0.000,0.023]$ \\ 
    $\omega^3_H$  & $\phantom{-}0.5182$     & $[-1.213, 2.309]$                     & $\phantom{-}0.51$ &$[0.33,0.67]$     & $\phantom{-}0.51$    &$[0.41,0.61]$  \\ 
    \bottomrule
  \end{tabular}
  \end{small}
  \caption{Mean values and 95\% CL error bars for the slow-roll
    parameters from Planck 2018 data and combined Planck plus SKA with
    1000~hrs or and 10000~hrs observation time, marginalized over
    cosmological paramers, and corresponding to
    Fig.~\ref{fig:HSR_combo}.}
  \label{tab:combo}
\end{table}

We finally provide constraints on the full set of slow-roll parameters
from Planck alone and from Planck and SKA combined in
Fig.~\ref{fig:HSR_combo}, also varying cosmological parameters
$\omega_b$, $\omega_c$, $\tau_\text{reio}$ and $h$.  These limits can
be compared directly to the final Planck results reproduced in
Fig.~\ref{fig:HSR_planck} and shown as dashed contours. We assume
total observation times of 1000 and 10000 hours for SKA.  The
corresponding mean and 95\% CL limits are summarized in
Tab.~\ref{tab:combo}. It is clear that the constraints from the
combined likelihoods are much stronger than the Planck 2018 data
alone. In particular, we find the combined constraints are about one
order of magnitude stronger than the Planck constraints for the
slow-roll parameters $\xi_H^2$ and $\omega_H^3$.  Moreover, the
combined data can constrain the slow-roll parameters $\xi^2_H$ and
$\omega^3_H$ more stringently than that of $\epsilon_H$ and $\eta_H$,
as can be easily seen from Fig.~\ref{fig:HSR_combo} and
Tab.~\ref{tab:combo}. The sensitivity gain for the higher-order
slow-roll parameters $\xi^2$ and $\omega^3$ is related to the fact
that those parameters impact the shape of the spectrum strongest far
away from the pivot scale $k_*$, similar to the parametrization given
in Ref.~\ref{eq:running_spectrum}.  As the pivot scale $k_*$ is chosen
to be the largest scale in the problem and covered by CMB-observations
(albeit with a limitation due to cosmic variance), significant
improvement is provided by the smallest scales, far away from
$k_*$.

A non-Harrison-Zel'dovich form of the spectrum in terms of $\alpha$
and $\beta$ can be measured with CMB-S4 experiments, as well as
precision 21cm-surveys~\cite{Munoz:2016owz}.  To allow for a
comparison we present our results in terms of these parameters in
Fig.~\ref{fig:SKA_alphabeta} and Tab.~\ref{tab:SKA_alphabeta} of the
Appendix. The mapping to slow-roll parameters is not unambiguous
because of its nonlinearity, so we prefer to work with the slow-roll
parameters directly.


\section{Outlook}

We have presented constraints on the single-field inflationary
potential in terms of the Hubble slow-roll parameters from the CMB
temperature and polarisation spectra combined with the 21cm brightness
fluctuations. We compute the spectrum of curvature perturbations for a
sample of initial values, parameterizing the Hubble function in terms
of the scalar field amplitude in a truncated Taylor-expansion. The
field itself, the background cosmology and the mode equations for
perturbations are evolved together to yield a curvature perturbation
spectrum at horizon-exit. We then evolve the modes using a
Boltzmann-code and link them to CMB temperature and polarisation
anisotropies, as well as the matter power spectrum at low redshift,
from which we model the fluctuations in the 21cm spectrum. All spectra
with their instrumental noise levels and covariances form a likelihood
for the slow-roll parameters, parameters of the $\Lambda$CDM background
cosmology, and parameters inherent to the observational channels such
as the optical depth.

Violation of slow roll causes scale-dependent variations of the
scale-invariant Harrison-Zel'dovich spectrum. It can be described by a
Taylor-expansion of the curvature perturbation spectrum in terms of
logarithmic wave number, relative to a pivot-scale close to the
horizon.
Both, primary CMB spectra and 21cm intensity fluctuations probe a wide
range of scales with a linear relation between observable and
fundamental field.  This range is key to the sensitivity to the
inflationary potential, as the variation of the shape of the spectrum
with logarithmic wave number is generically small.  In terms of the
Hubble slow-roll parameters especially the SKA limits showed strong
degeneracies and increasingly loose bounds on higher-order
parameters. We recovered a hierarchy in precision, where $\epsilon_H$
and $\eta_H$ are measured at a level of $\sim 10^{-2}$, followed by
$\xi^2_H$ at $10^{-1}$ and $\omega^3_H$ just slightly better than
order-one. The improvement of SKA over Planck, in particular on
$\xi^2_H$ and $\omega^3_H$, is driven by small scales, where
deviations from the Harrison-Zel'dovich shape far away from the pivot
scale $k_*$ become important.

Naturally, one would like to extend the scale range to higher wave
numbers and include low-redshift probes of the cosmic large-scale
structure such as weak cosmic shear or galaxy clustering at redshifts
around unity. However, on such scales nonlinear structure formation
starts to dominate.  Similarly, small scales at higher redshift can be
probed by Lyman-$\alpha$ measurements, which requires a detailed
understanding of baryonic dynamics. Additional constraints on the
spectral shape on small scales will eventually come from limits on
primordial black holes.  We leave these additional handles for future
analyses and instead follow a very conservative approach. Even with a
limited range of scales and a narrow redshift window we confirm that
SKA will provide excellent limits on the inflationary potential,
pushing precision cosmology significantly beyond the CMB measurements
by Planck.


\begin{center} \textbf{Acknowledgments} \end{center}

We thank Robert F. Reischke and Michel Luchmann for discussions. We
also thank Julien Lesgourgues for communication. TM is supported by
Postdoctoral Research Fellowship from Alexander von Humboldt
Foundation. The research of TP is supported by the Deutsche
Forschungsgemeinschaft (DFG, German Research Foundation) under grant
396021762 -- TRR~257 \textsl{Particle Physics Phenomenology after the
  Higgs Discovery}. This work was supported by the Deutsche
Forschungsgemeinschaft (DFG, German Research Foundation) under
Germany's Excellence Strategy EXC 2181/1 - 390900948 (the Heidelberg
STRUCTURES Excellence Cluster).

\clearpage
\appendix
\section{Appendix}

\subsubsection*{2D slices from 3D Markov Chains}
\label{sec:2dslices}

\begin{figure}[b!]
	\includegraphics[width = 0.245\textwidth]{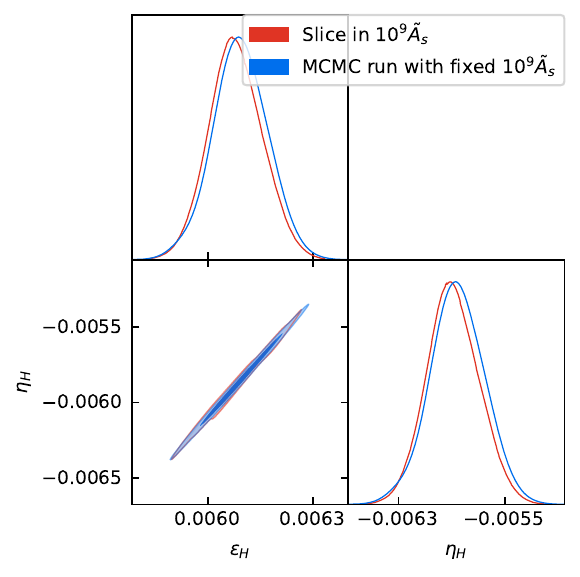}
	\includegraphics[width = 0.245\textwidth]{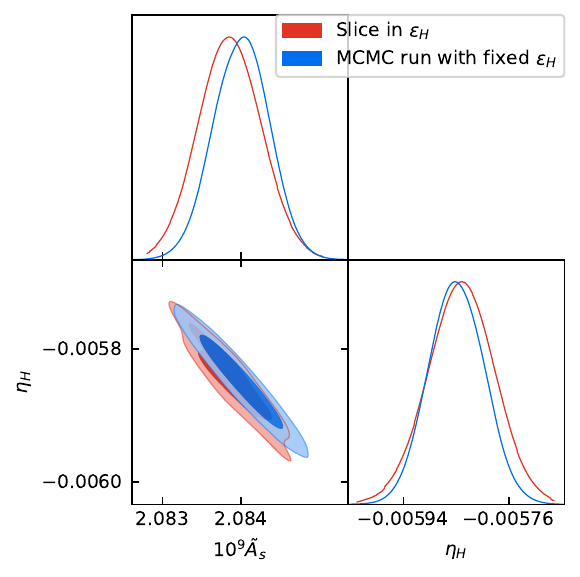}
	\includegraphics[width = 0.245\textwidth]{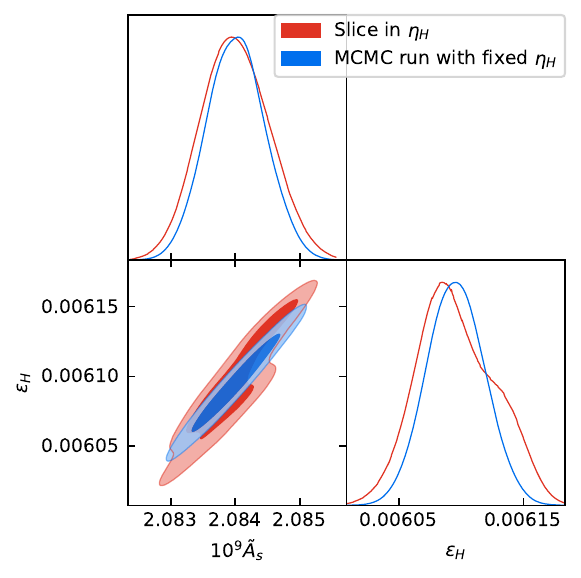} 
	\includegraphics[width = 0.245\textwidth]{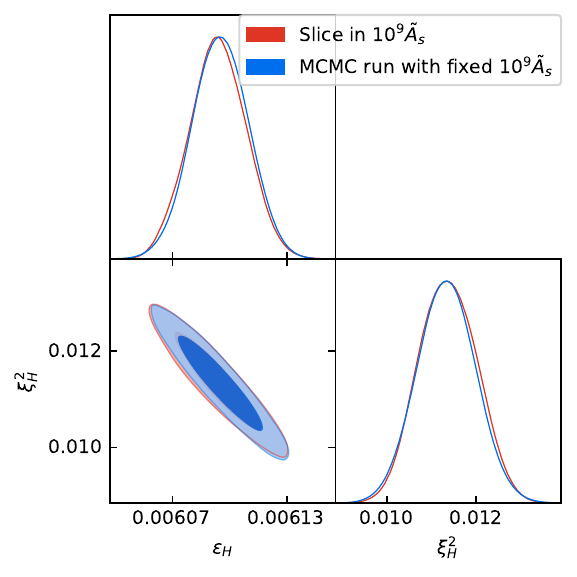} \\
	\includegraphics[width = 0.245\textwidth]{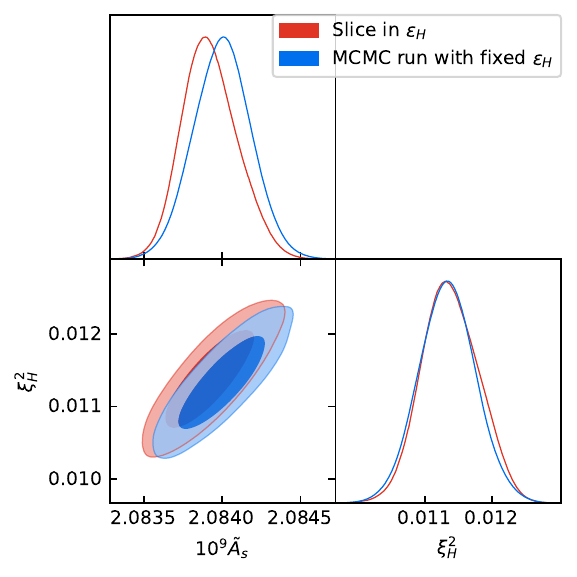}
	\includegraphics[width = 0.245\textwidth]{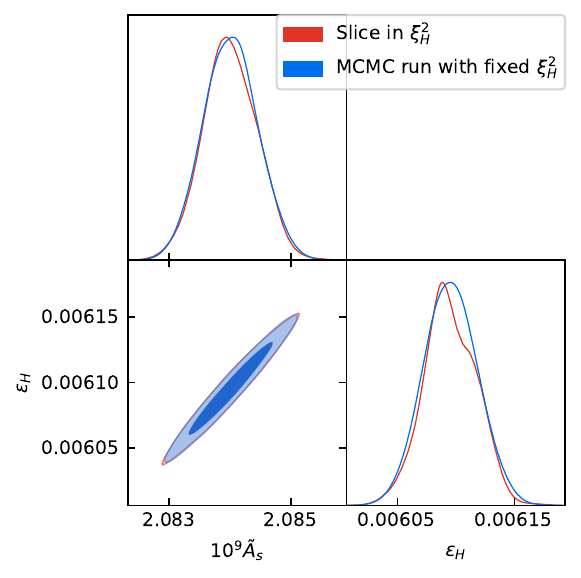} 
	\includegraphics[width = 0.245\textwidth]{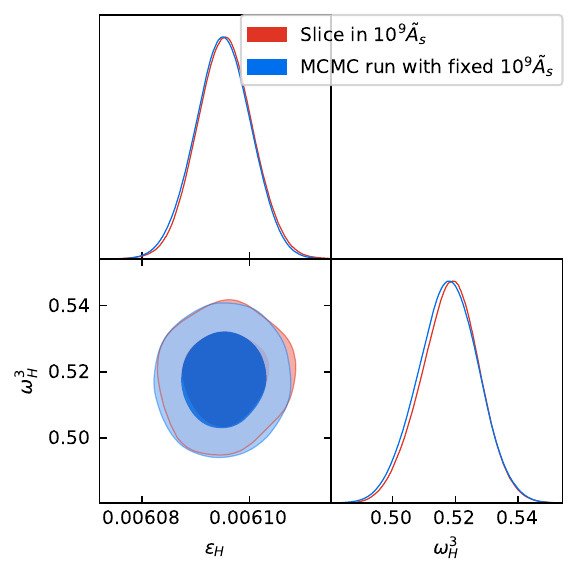}
	\includegraphics[width = 0.245\textwidth]{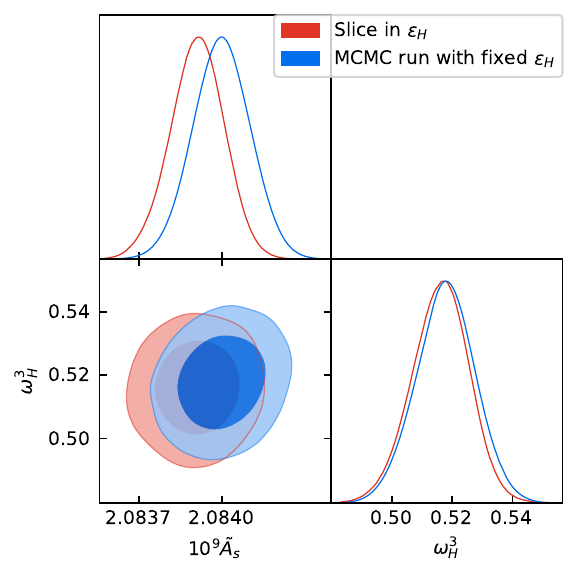} \\
	\includegraphics[width = 0.245\textwidth]{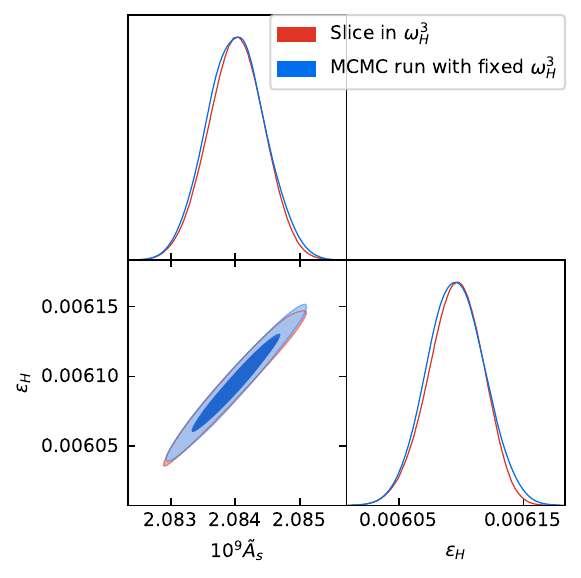} 
	\includegraphics[width = 0.245\textwidth]{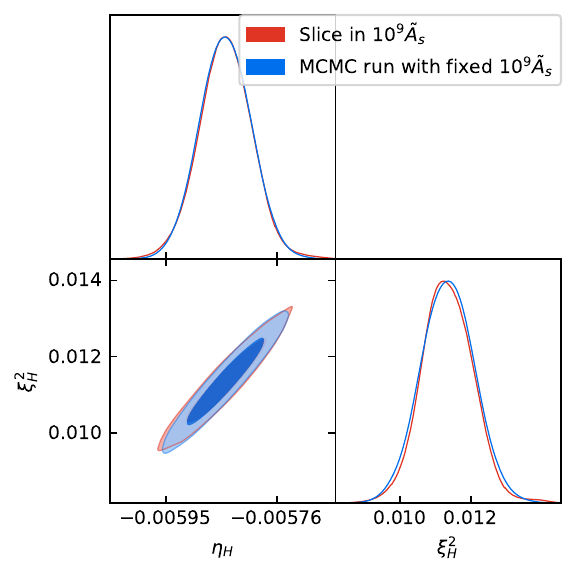}
	\includegraphics[width = 0.245\textwidth]{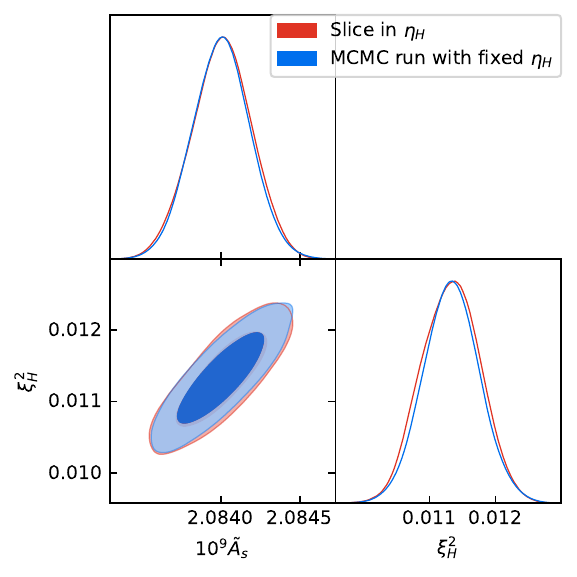}
	\includegraphics[width = 0.245\textwidth]{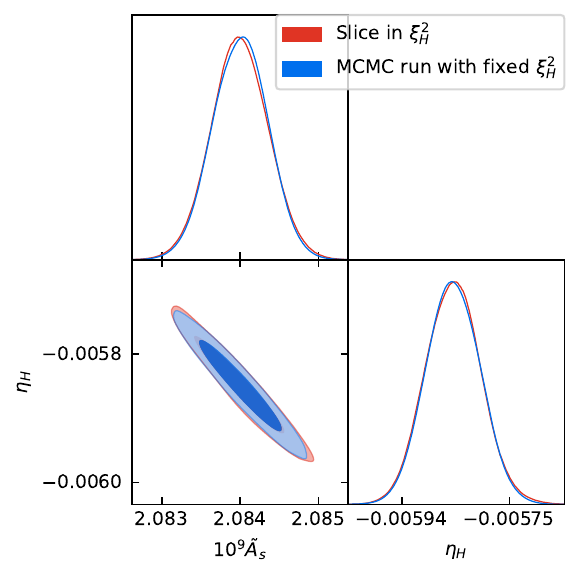} 
	\caption{Comparison between slices and 2D contours using the SKA projections. Slices are taken from the 3D chains in Fig. \ref{fig:SKA3D}.}
	\label{fig:SKA2D_slices1}
\end{figure}

\begin{figure}[t!]
	\includegraphics[width = 0.245\textwidth]{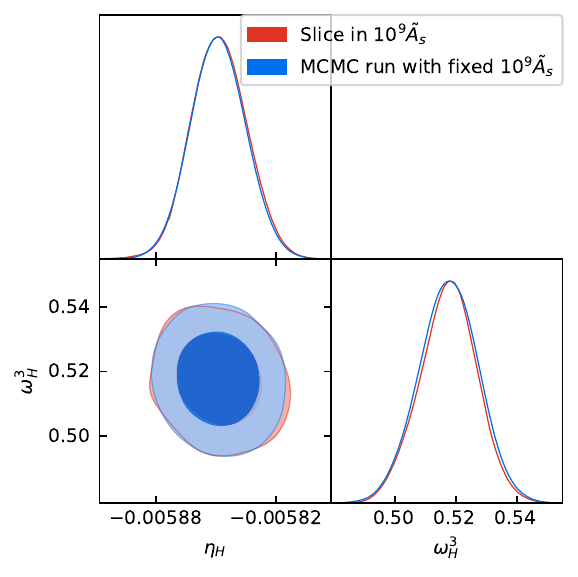}
	\includegraphics[width = 0.245\textwidth]{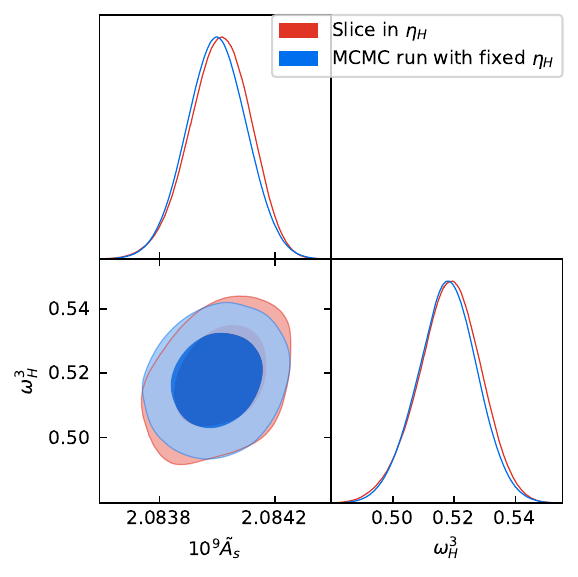}
	\includegraphics[width = 0.245\textwidth]{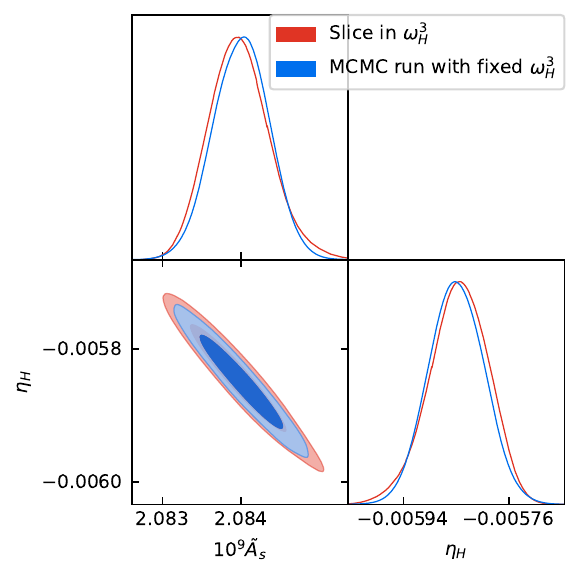} 
	\includegraphics[width = 0.245\textwidth]{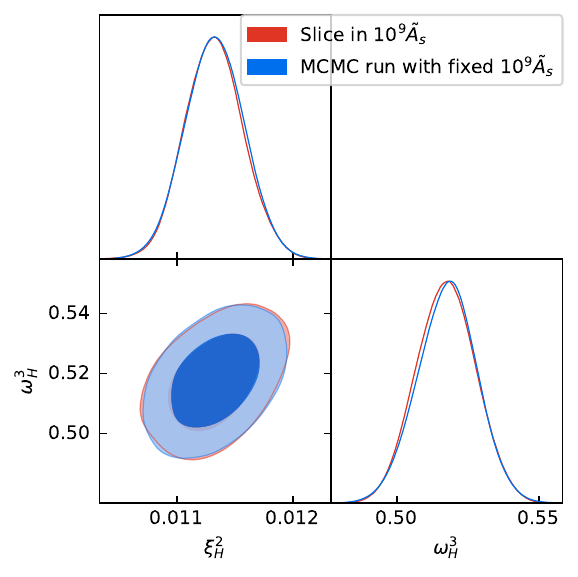} \\
	\includegraphics[width = 0.245\textwidth]{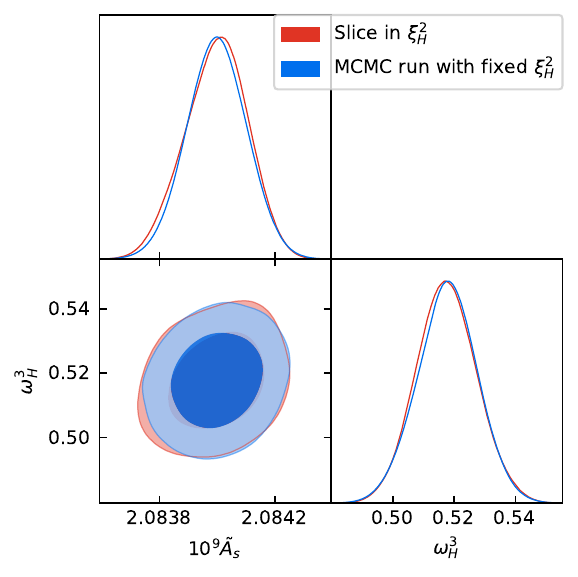}
	\includegraphics[width = 0.245\textwidth]{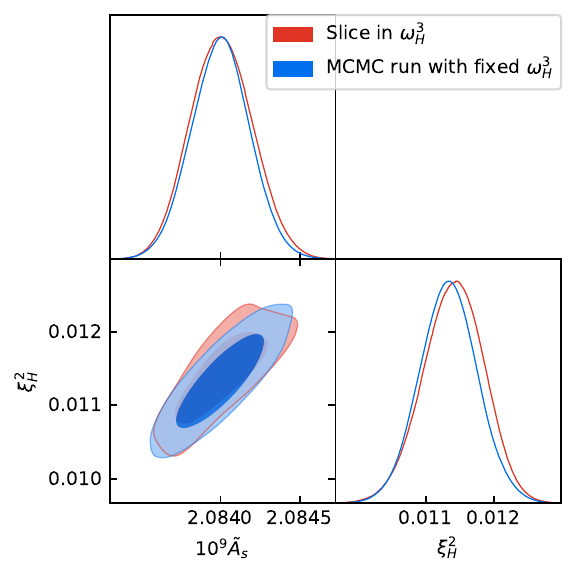} 
	\includegraphics[width = 0.245\textwidth]{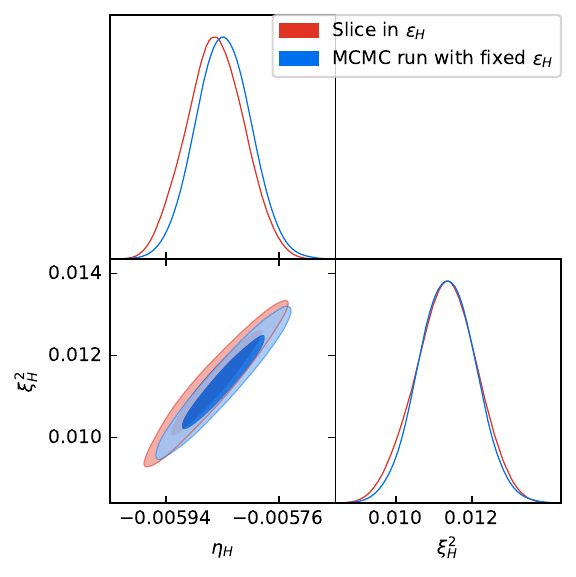}
	\includegraphics[width = 0.245\textwidth]{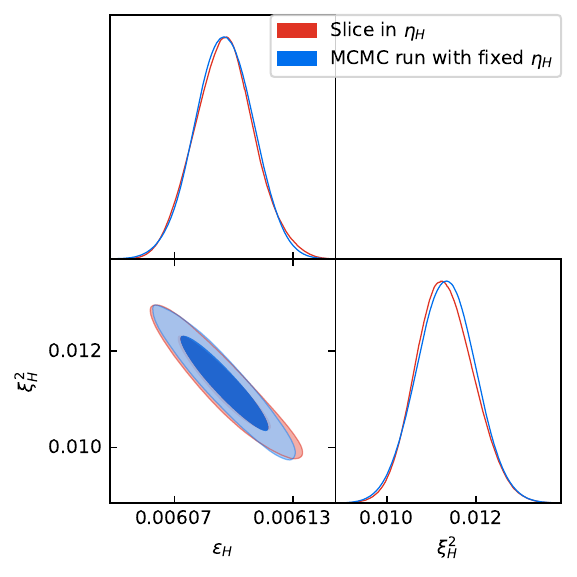}\\
	\includegraphics[width = 0.245\textwidth]{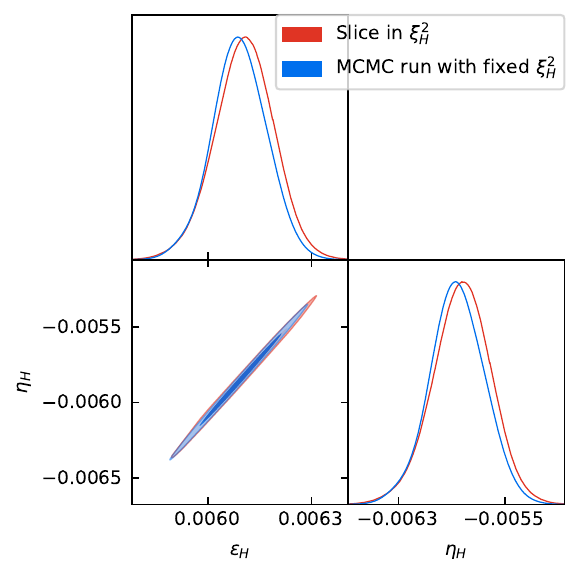} 
	\includegraphics[width = 0.245\textwidth]{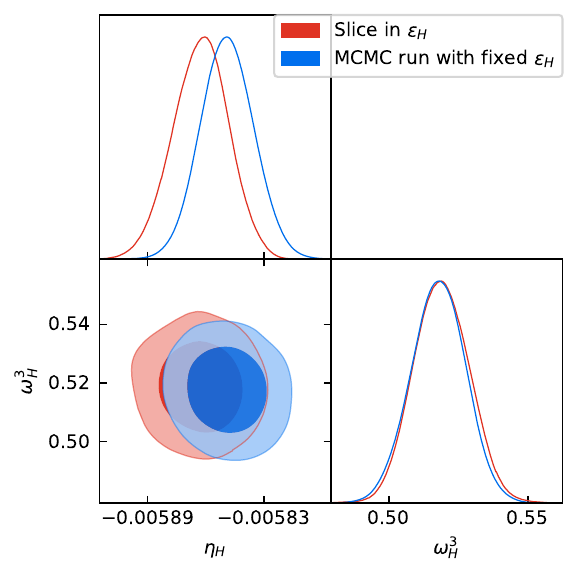}
	\includegraphics[width = 0.245\textwidth]{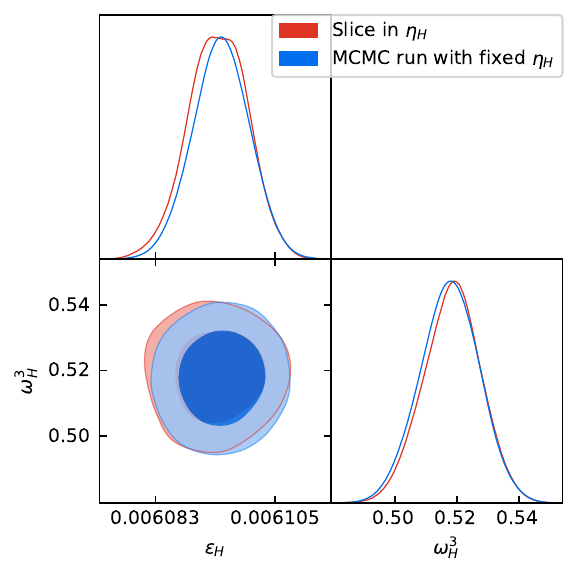}
	\includegraphics[width = 0.245\textwidth]{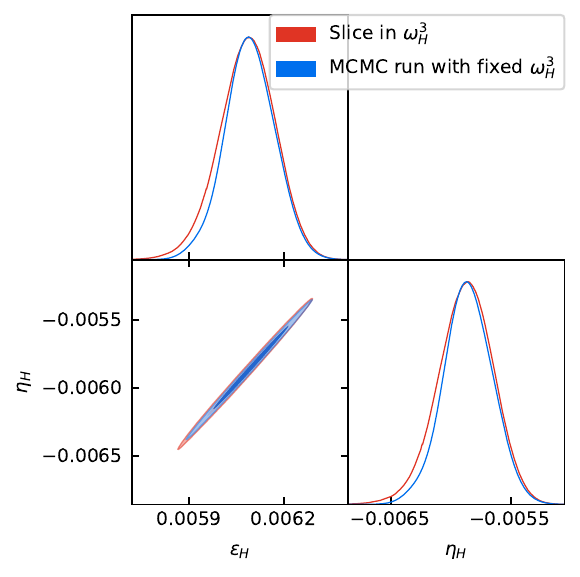}\\
	\includegraphics[width = 0.245\textwidth]{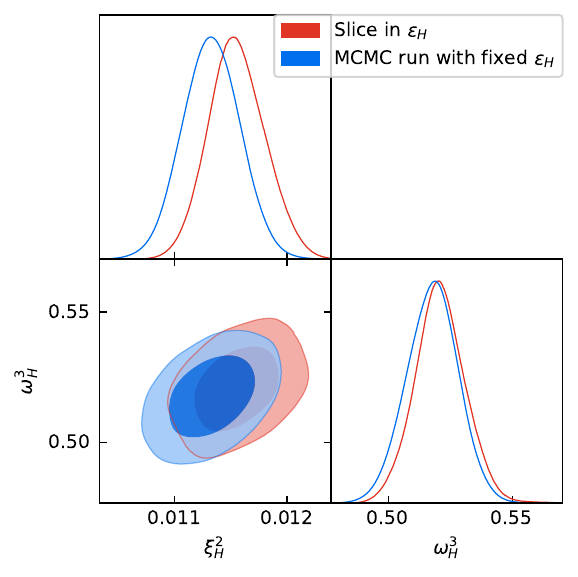}
	\includegraphics[width = 0.245\textwidth]{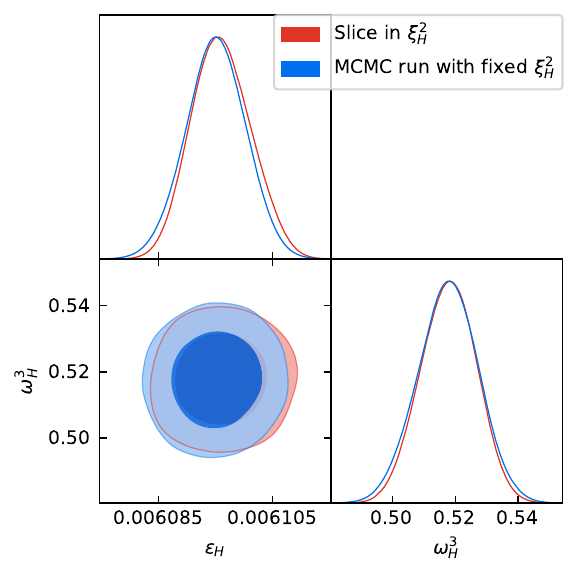}
	\includegraphics[width = 0.245\textwidth]{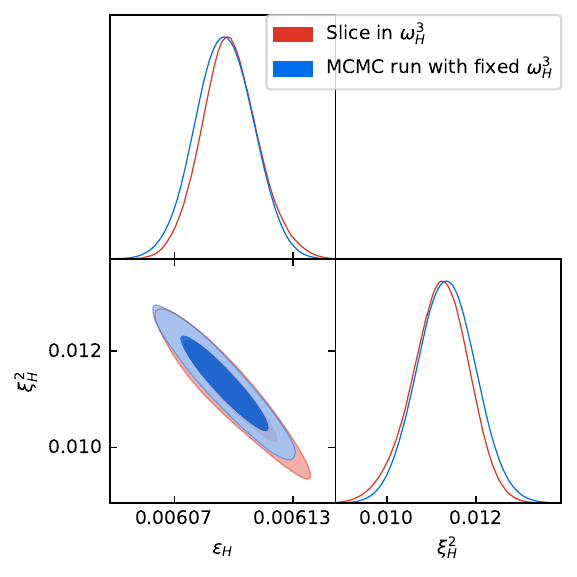} 
	\includegraphics[width = 0.245\textwidth]{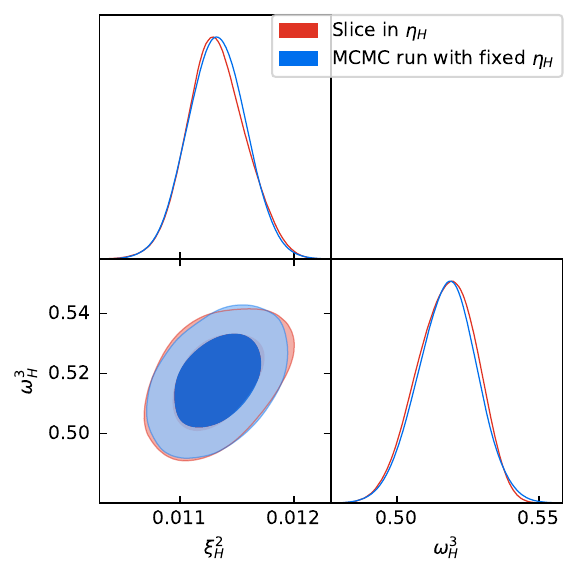} \\
	\includegraphics[width = 0.245\textwidth]{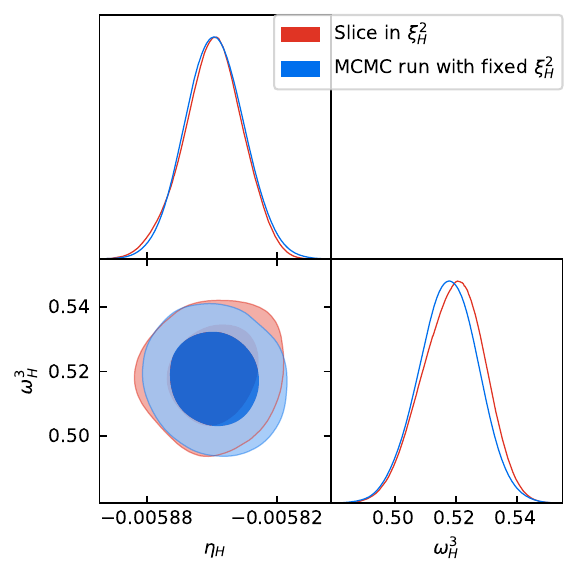}
	\includegraphics[width = 0.245\textwidth]{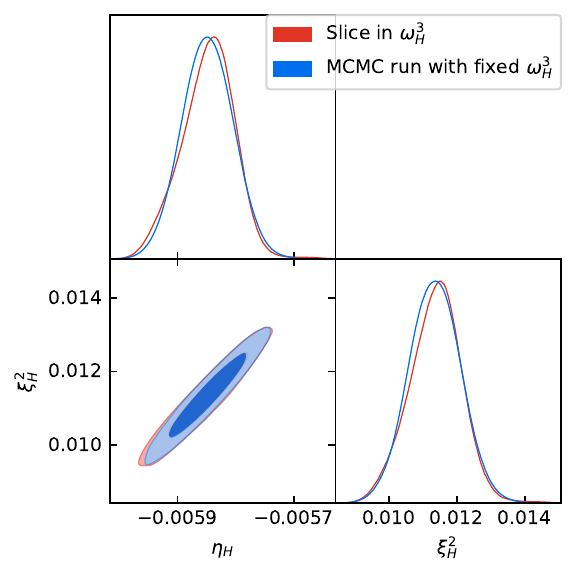} 
	\caption{Comparison between slices and 2D contours using the SKA projections. Slices are taken from the 3D chains in Fig. \ref{fig:SKA3D}.}
	\label{fig:SKA2D_slices2}
\end{figure}

The 3-dimensional Markov chains generated with the SKA likelihoods in
\ref{sec:ska_likelihood} are consistent with the two dimensional
Markov chains. Fig. \ref{fig:SKA2D_slices1} and
\ref{fig:SKA2D_slices2} show the comparison between fraction of the
three dimensional Markov chains and the two dimensional Markov
chains. To select appropriate points from the three dimensional Markov
chains one parameter is chosen and only those points within a small
region around its mean value found in \ref{tab:HSR_planck} are taken
into account to generate the figures. For each three dimensional
Markov chain three of these sliced chains are generated. The cut
chains exhibit similar mean parameter values and contours as the two
dimensional Markov chains.

\begin{figure}[t!]
	\includegraphics[width = 0.32\textwidth]{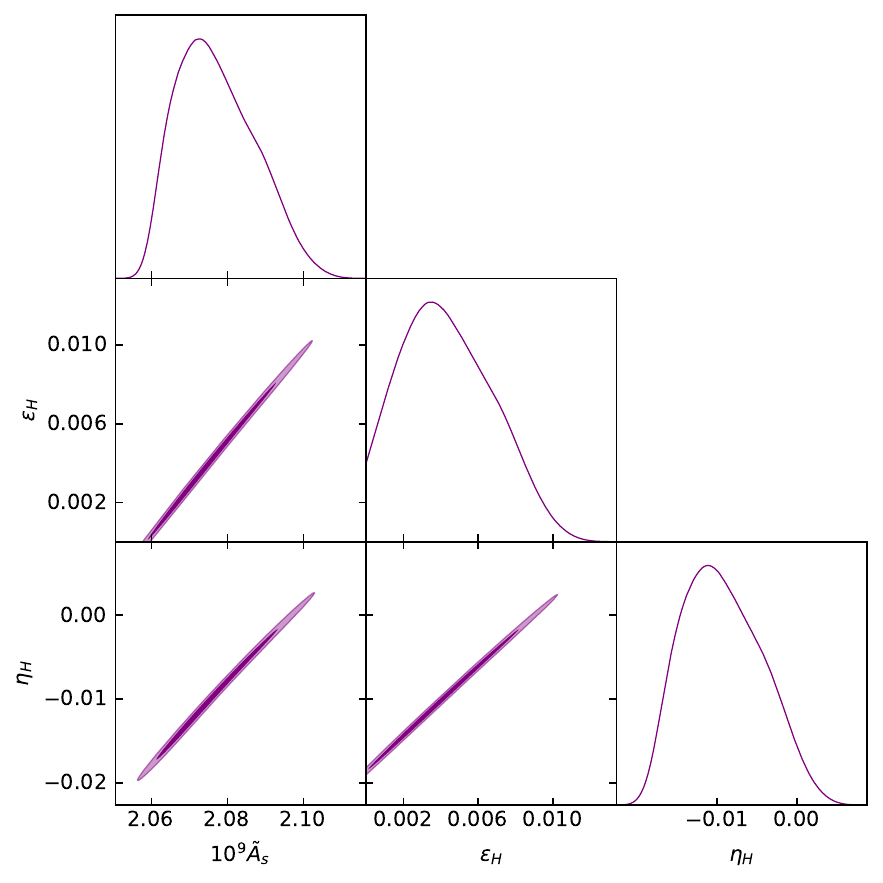}
	\includegraphics[width = 0.32\textwidth]{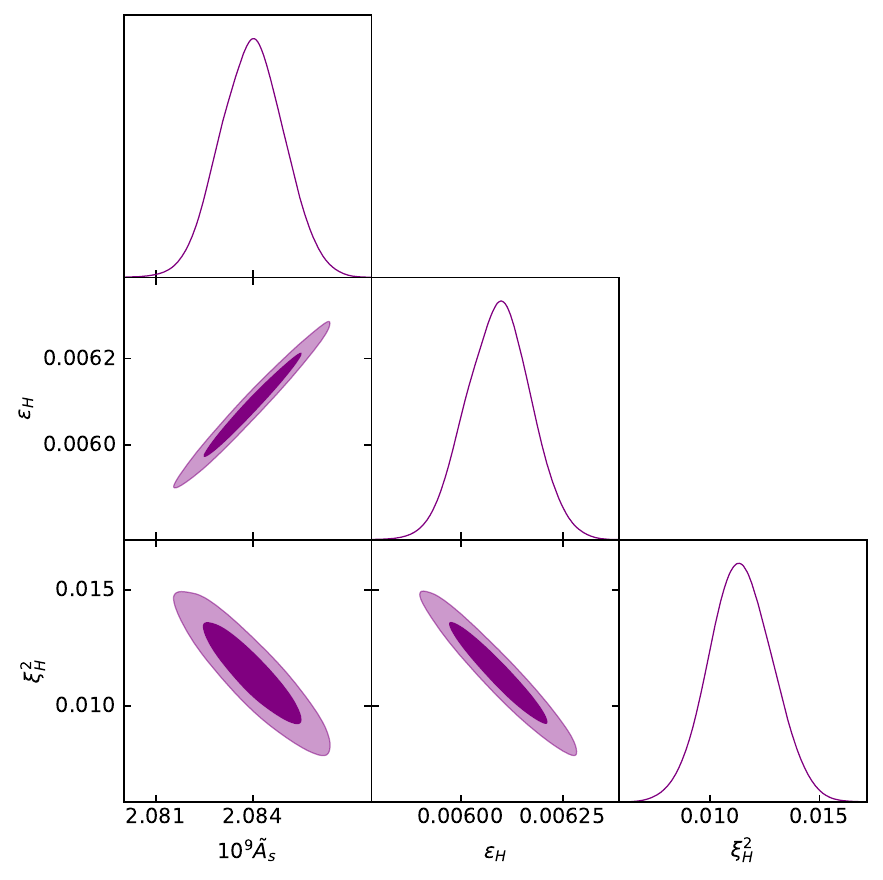}
	\includegraphics[width = 0.32\textwidth]{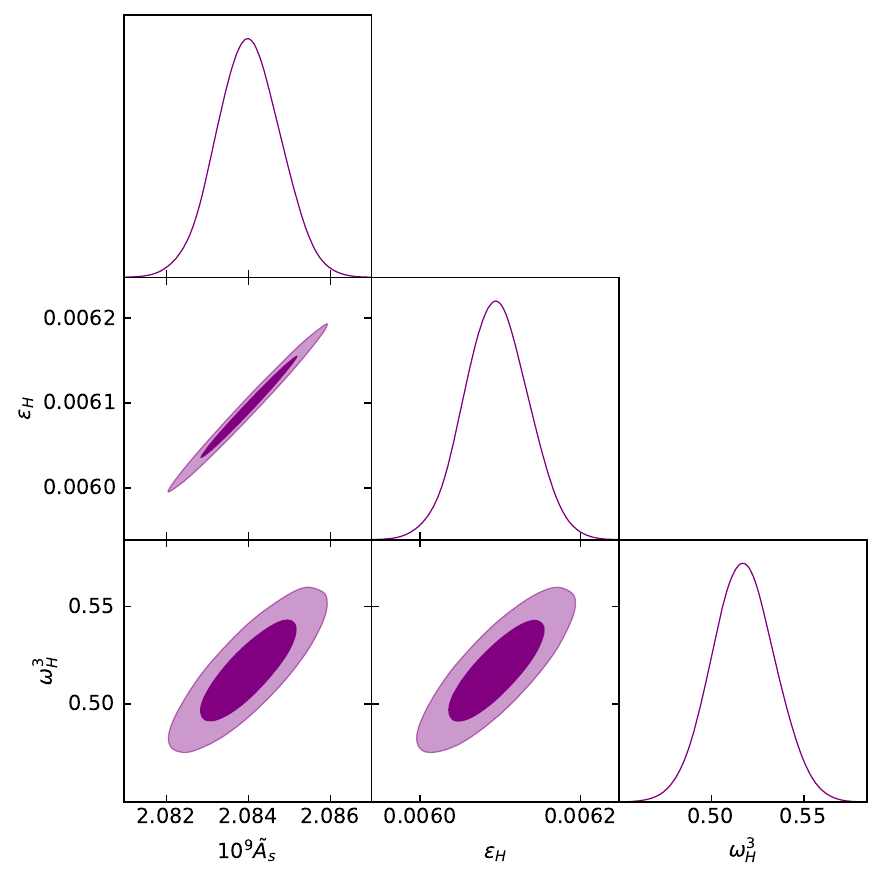} \\
	\includegraphics[width = 0.32\textwidth]{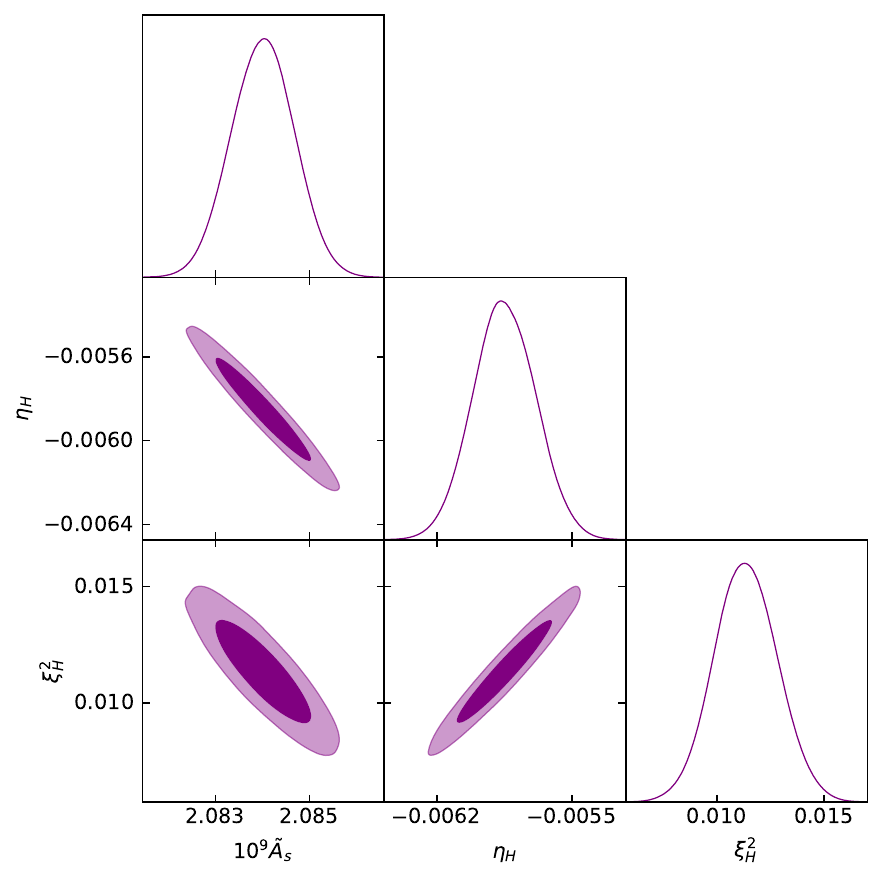}
	\includegraphics[width = 0.32\textwidth]{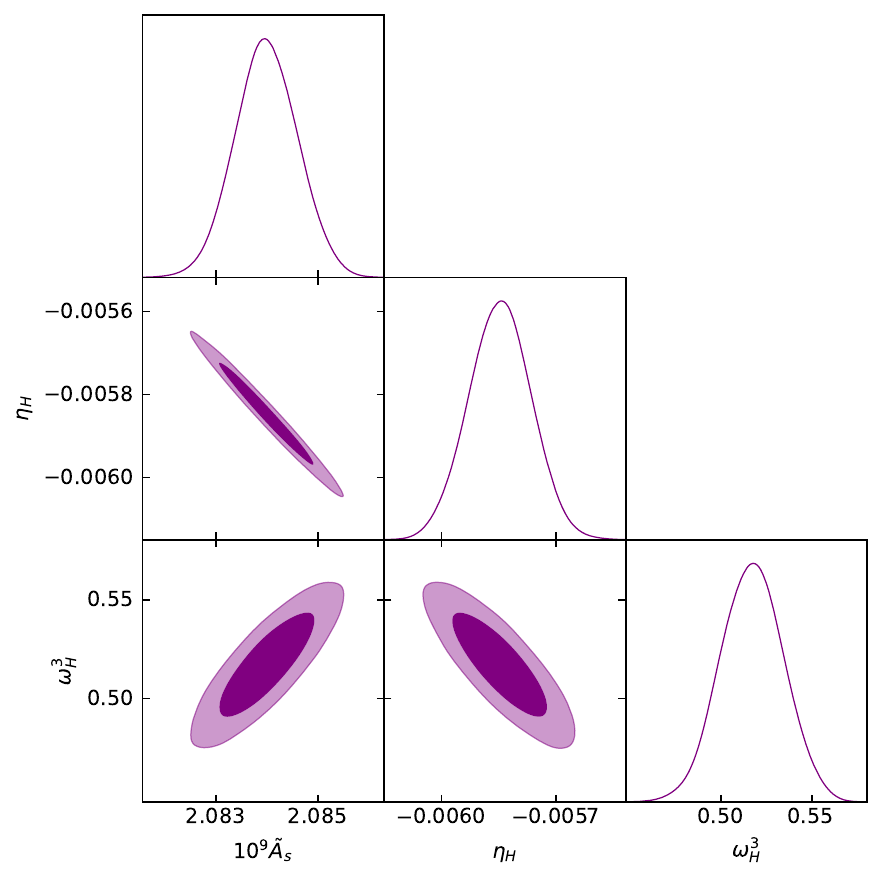}
	\includegraphics[width = 0.32\textwidth]{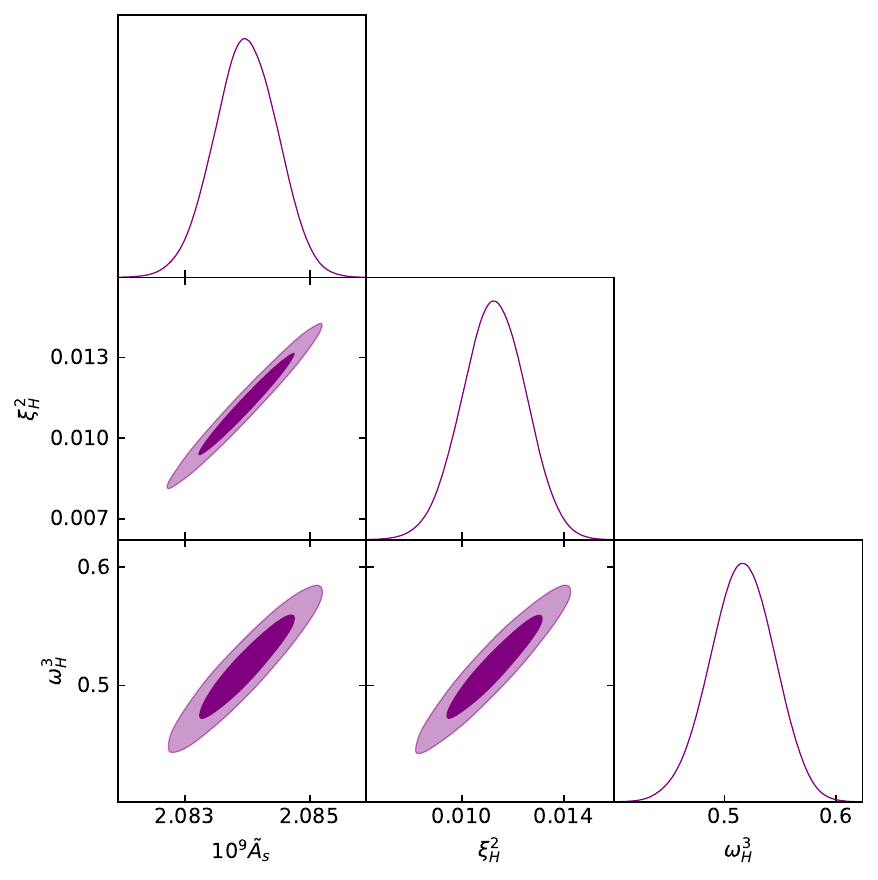} \\
	\includegraphics[width = 0.32\textwidth]{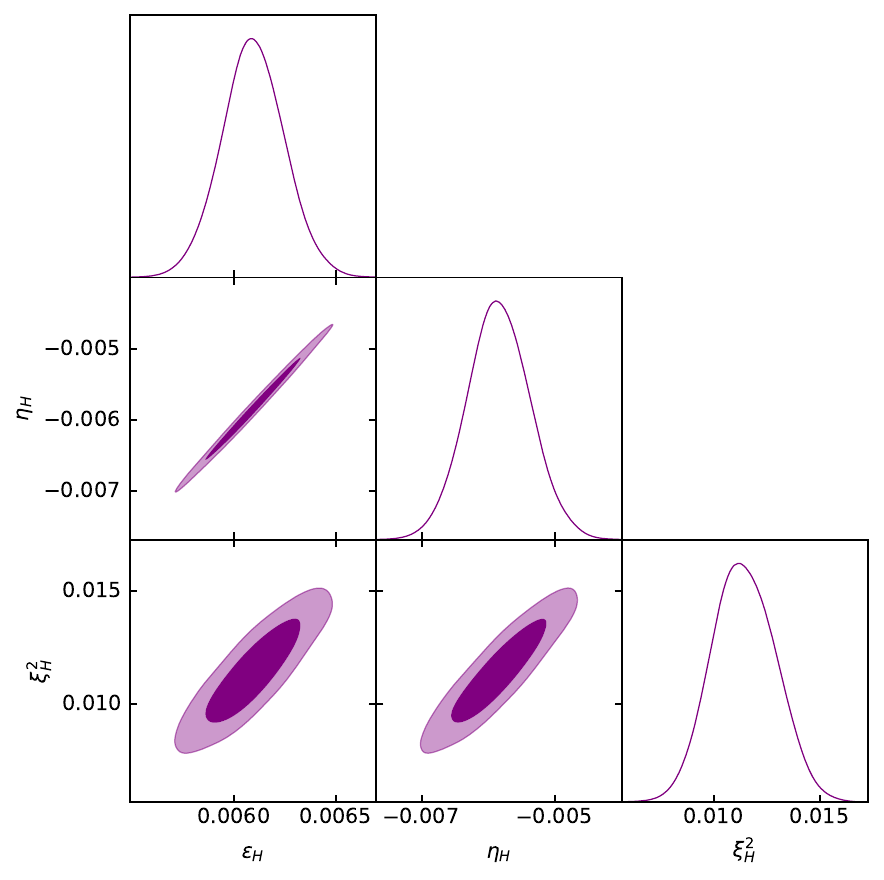}
	\includegraphics[width = 0.32\textwidth]{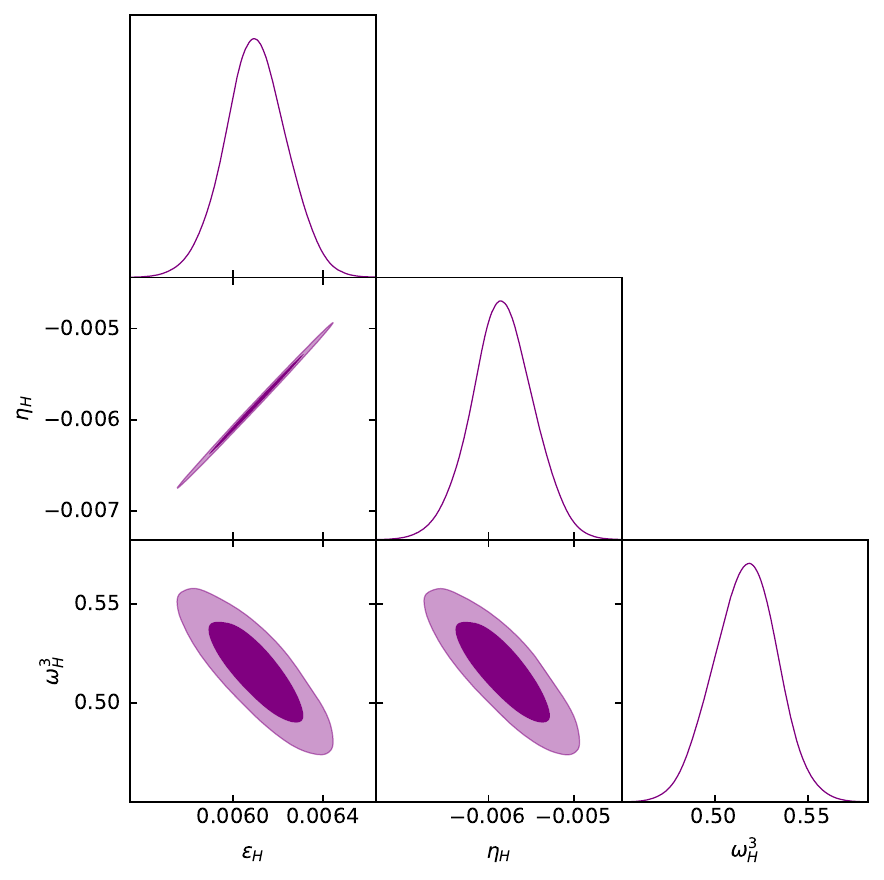}
	\includegraphics[width = 0.32\textwidth]{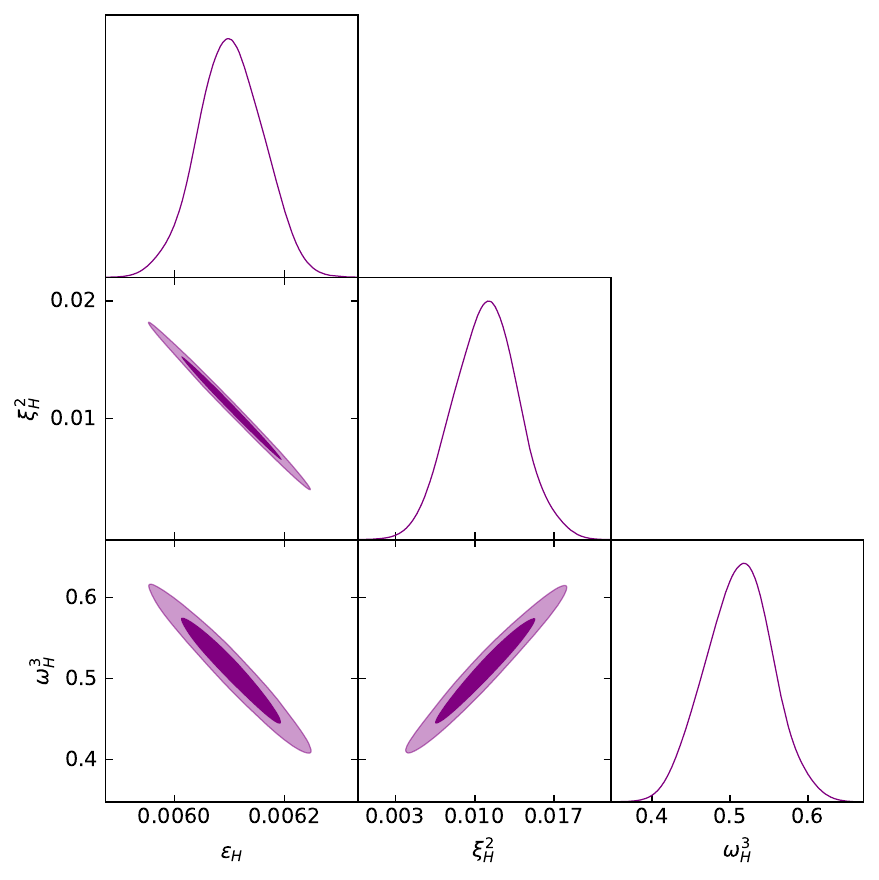} \\
	\includegraphics[width = 0.32\textwidth]{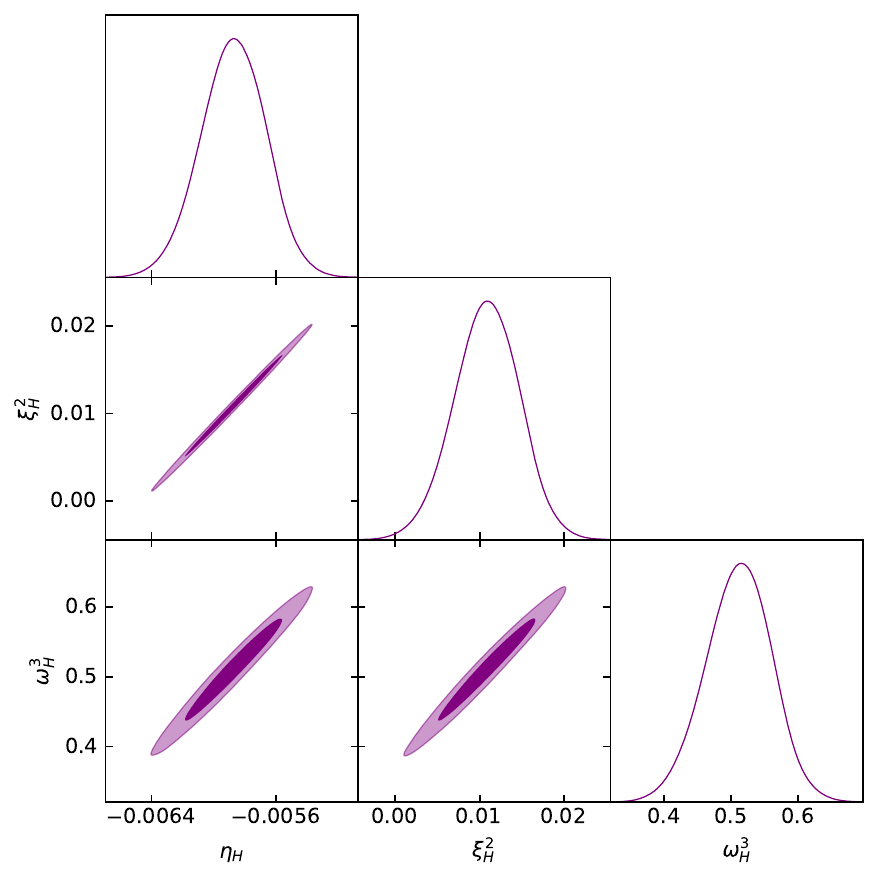}
	\caption{Sliced 3-dimensional likelihoods for the slow-roll
		parameters from a combination of Planck and SKA
		projections.}
	\label{fig:SKA+Planck3D}
\end{figure}

Adding the Planck likelihoods to the 3-dimensional
parameter estimation yields results very similar to the 3-dimensional
SKA limits alone. The marginalized likelihoods are shown in
Fig.~\ref{fig:SKA+Planck3D}. When constraining $\tilde{A}_s$,
$\epsilon_H$ and $\eta_H$ at the same time the marginalized
distributions for each of the parameters become less wide. The Planck
likelihood constrains the very edges of the strongly correlated
parameters, allowing for an easier numerical evaluation.

\subsubsection*{SKA with spectral index and running}

\begin{figure}[h!]
	\includegraphics[width = 1 \textwidth]{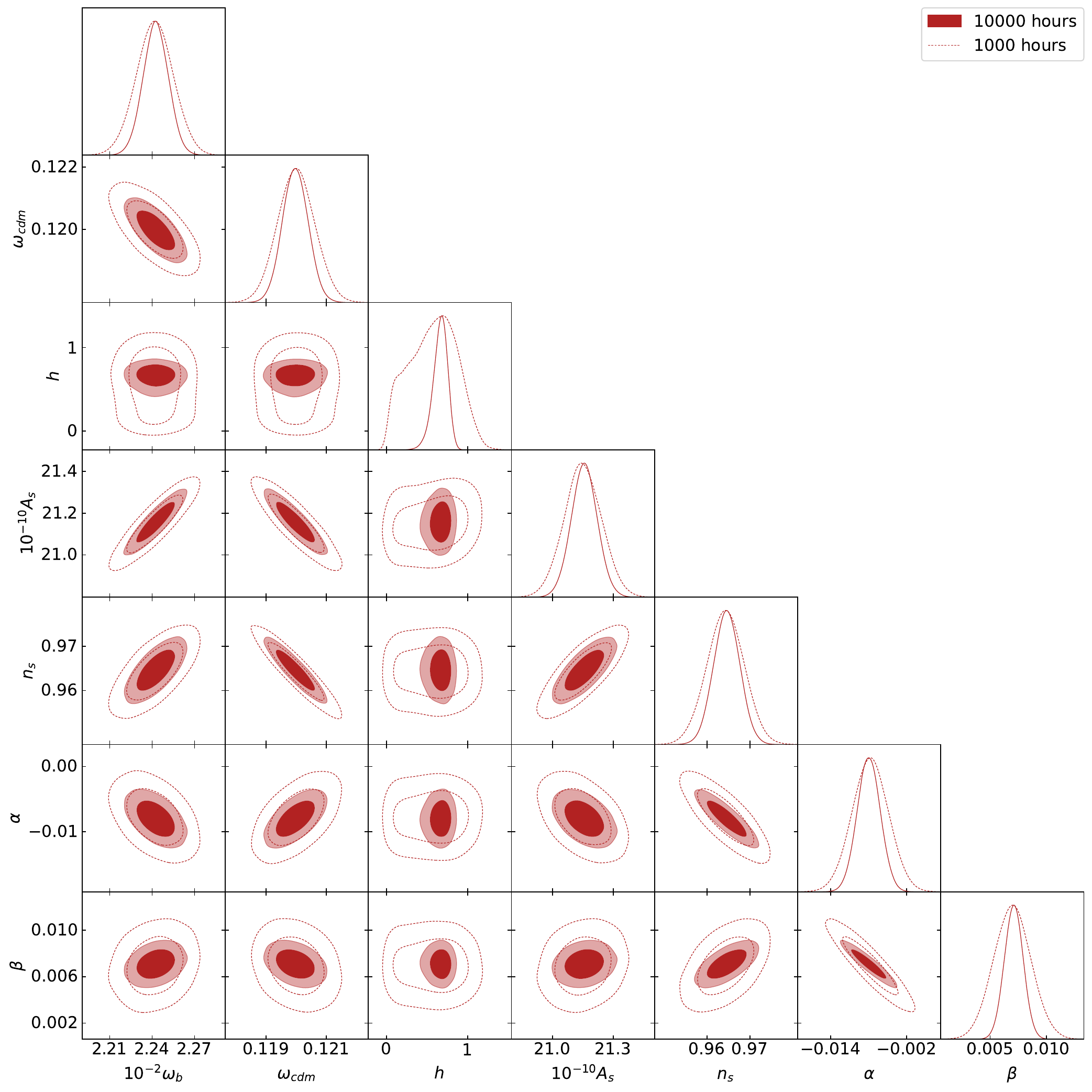}
	\caption{The contours for SKA likelihoods (red) and overlaid SKA with lower observation time of 1000 hours (red dashed lines).}
	\label{fig:SKA_alphabeta}
\end{figure}

The primordial power spectra $\mathcal{P}_\mathcal{R}(k)$ can also be
parameterized by the spectral index $n_s$ and its running $\alpha$ and
$\beta$, which can also be constrained by 21cm cosmology. We also
provide the constraints on these parameters from SKA alone in
Fig.~\ref{fig:SKA_alphabeta} for both 10000 (red solid) and 1000 hrs
(red dashed) for comparison.  The corresponding mean values and 68\%
error bars are given in Tab.~\ref{tab:SKA_alphabeta}. We find that our
constraints are a bit stronger than that found by
Ref~\cite{Munoz:2016owz} for 1000 hrs observation time.

\begin{table}[h!]
	\centering
	\begin{tabular}{l|ccc} 
		\toprule
		Parameter &  mean & 68\% CL & 68\% CL \\ 
		& & 10000 hrs & 1000 hrs \\ 
		\midrule
		$10^{10}A_s$      & $21.157$     &$ \pm0.064$          &$\pm0.092$\\
		
		$n_s$             & $0.9646$    &$\pm 0.0031$         &$\pm0.0043$\\
		
		$\alpha$          & $-0.0080$   &$\pm0.0018$          &$\pm0.0029$\\
		
		$\beta$           & $0.00710$   &$\pm0.00083$         &$\pm0.0016$\\
		
		\bottomrule
	\end{tabular} \\ 
	\caption{Best fit parameter values and error bars for SKA likelihoods constructed as in \ref{sec:ska_likelihood} for 
		primordial spectrum computation with spectral index and its runnings.} 
	\label{tab:SKA_alphabeta}
\end{table}

\bibliography{literature}

\providecommand{\href}[2]{#2}\begingroup\raggedright\begin{thebibliography}{10}

\bibitem{Starobinsky:1980te}
A.~A. Starobinsky, {\it {A New Type of Isotropic Cosmological Models Without
  Singularity}},  \href{http://dx.doi.org/10.1016/0370-2693(80)90670-X}{Phys.
  Lett. B {\bfseries 91} (1980)  99}.

\bibitem{Sato:1980yn}
K.~Sato, {\it {First Order Phase Transition of a Vacuum and Expansion of the
  Universe}},  Mon. Not. Roy. Astron. Soc. {\bfseries 195} (1981)  467.

\bibitem{Guth:1980zm}
A.~H. Guth, {\it {The Inflationary Universe: A Possible Solution to the Horizon
  and Flatness Problems}},
  \href{http://dx.doi.org/10.1103/PhysRevD.23.347}{Phys. Rev. D {\bfseries 23}
  (1981)  347}.

\bibitem{Mukhanov:1981xt}
V.~F. Mukhanov and G.~V. Chibisov, {\it {Quantum Fluctuations and a Nonsingular
  Universe}},  JETP Lett. {\bfseries 33} (1981)  532.

\bibitem{Starobinsky:1982ee}
A.~A. Starobinsky, {\it {Dynamics of Phase Transition in the New Inflationary
  Universe Scenario and Generation of Perturbations}},
  \href{http://dx.doi.org/10.1016/0370-2693(82)90541-X}{Phys. Lett. B
  {\bfseries 117} (1982)  175}.

\bibitem{Hawking:1982cz}
S.~W. Hawking, {\it {The Development of Irregularities in a Single Bubble
  Inflationary Universe}},
  \href{http://dx.doi.org/10.1016/0370-2693(82)90373-2}{Phys. Lett. B
  {\bfseries 115} (1982)  295}.

\bibitem{Guth:1982ec}
A.~H. Guth and S.~Y. Pi, {\it {Fluctuations in the New Inflationary Universe}},
   \href{http://dx.doi.org/10.1103/PhysRevLett.49.1110}{Phys. Rev. Lett.
  {\bfseries 49} (1982)  1110}.

\bibitem{Linde:1981mu}
A.~D. Linde, {\it {A New Inflationary Universe Scenario: A Possible Solution of
  the Horizon, Flatness, Homogeneity, Isotropy and Primordial Monopole
  Problems}},  \href{http://dx.doi.org/10.1016/0370-2693(82)91219-9}{Phys.
  Lett. B {\bfseries 108} (1982)  389}.

\bibitem{Albrecht:1982wi}
A.~Albrecht and P.~J. Steinhardt, {\it {Cosmology for Grand Unified Theories
  with Radiatively Induced Symmetry Breaking}},
  \href{http://dx.doi.org/10.1103/PhysRevLett.48.1220}{Phys. Rev. Lett.
  {\bfseries 48} (1982)  1220}.

\bibitem{Linde:1983gd}
A.~D. Linde, {\it {Chaotic Inflation}},
  \href{http://dx.doi.org/10.1016/0370-2693(83)90837-7}{Phys. Lett. B
  {\bfseries 129} (1983)  177}.

\bibitem{Planck:2018vyg}
Planck, N.~Aghanim {\em et al.}, {\it {Planck 2018 results. VI. Cosmological
  parameters}},  \href{http://dx.doi.org/10.1051/0004-6361/201833910}{Astron.
  Astrophys. {\bfseries 641} (2020)  A6},
  \href{http://arxiv.org/abs/1807.06209}{{arXiv:1807.06209 [astro-ph.CO]}}.
  [Erratum: Astron.Astrophys. 652, C4 (2021)].

\bibitem{Planck:2018jri}
Planck, Y.~Akrami {\em et al.}, {\it {Planck 2018 results. X. Constraints on
  inflation}},  \href{http://dx.doi.org/10.1051/0004-6361/201833887}{Astron.
  Astrophys. {\bfseries 641} (2020)  A10},
  \href{http://arxiv.org/abs/1807.06211}{{arXiv:1807.06211 [astro-ph.CO]}}.

\bibitem{Cosmology-SWG:2015tjb}
Cosmology-SWG, EoR/CD-SWG, J.~Pritchard {\em et al.}, {\it {Cosmology from
  EoR/Cosmic Dawn with the SKA}},
  \href{http://dx.doi.org/10.22323/1.215.0012}{PoS {\bfseries AASKA14} (2015)
  012}, \href{http://arxiv.org/abs/1501.04291}{{arXiv:1501.04291
  [astro-ph.CO]}}.

\bibitem{Furlanetto:2005ax}
S.~R. Furlanetto, M.~McQuinn, and L.~Hernquist, {\it {Characteristic scales
  during reionization}},
  \href{http://dx.doi.org/10.1111/j.1365-2966.2005.09687.x}{Mon. Not. Roy.
  Astron. Soc. {\bfseries 365} (2006)  115},
  \href{http://arxiv.org/abs/astro-ph/0507524}{{arXiv:astro-ph/0507524}}.

\bibitem{Furlanetto:2006jb}
S.~Furlanetto, S.~P. Oh, and F.~Briggs, {\it {Cosmology at Low Frequencies: The
  21 cm Transition and the High-Redshift Universe}},
  \href{http://dx.doi.org/10.1016/j.physrep.2006.08.002}{Phys. Rept. {\bfseries
  433} (2006)  181},
  \href{http://arxiv.org/abs/astro-ph/0608032}{{arXiv:astro-ph/0608032}}.

\bibitem{Loeb:2008dp}
A.~Loeb, {\it {Let there be Light: the Emergence of Structure out of the Dark
  Ages in the Early Universe}},
  \href{http://arxiv.org/abs/0804.2258}{{arXiv:0804.2258 [astro-ph]}}.

\bibitem{Mellema:2012ht}
G.~Mellema {\em et al.}, {\it {Reionization and the Cosmic Dawn with the Square
  Kilometre Array}},  \href{http://dx.doi.org/10.1007/s10686-013-9334-5}{Exper.
  Astron. {\bfseries 36} (2013)  235},
  \href{http://arxiv.org/abs/1210.0197}{{arXiv:1210.0197 [astro-ph.CO]}}.

\bibitem{Morales:2009gs}
M.~F. Morales and J.~S.~B. Wyithe, {\it {Reionization and Cosmology with 21 cm
  Fluctuations}},
  \href{http://dx.doi.org/10.1146/annurev-astro-081309-130936}{Ann. Rev.
  Astron. Astrophys. {\bfseries 48} (2010)  127},
  \href{http://arxiv.org/abs/0910.3010}{{arXiv:0910.3010 [astro-ph.CO]}}.

\bibitem{Natarajan:2014rra}
A.~Natarajan and N.~Yoshida, {\it {The Dark Ages of the Universe and Hydrogen
  Reionization}},  \href{http://dx.doi.org/10.1093/ptep/ptu067}{PTEP {\bfseries
  2014} (2014) 6, 06B112},
  \href{http://arxiv.org/abs/1404.7146}{{arXiv:1404.7146 [astro-ph.CO]}}.

\bibitem{Pritchard:2011xb}
J.~R. Pritchard and A.~Loeb, {\it {21-cm cosmology}},
  \href{http://dx.doi.org/10.1088/0034-4885/75/8/086901}{Rept. Prog. Phys.
  {\bfseries 75} (2012)  086901},
  \href{http://arxiv.org/abs/1109.6012}{{arXiv:1109.6012 [astro-ph.CO]}}.

\bibitem{Pritchard:2015fia}
Cosmology-SWG, EoR/CD-SWG, J.~Pritchard {\em et al.}, {\it {Cosmology from
  EoR/Cosmic Dawn with the SKA}},
  \href{http://dx.doi.org/10.22323/1.215.0012}{PoS {\bfseries AASKA14} (2015)
  012}, \href{http://arxiv.org/abs/1501.04291}{{arXiv:1501.04291
  [astro-ph.CO]}}.

\bibitem{Bull:2018lat}
A.~Weltman {\em et al.}, {\it {Fundamental physics with the Square Kilometre
  Array}},  \href{http://dx.doi.org/10.1017/pasa.2019.42}{Publ. Astron. Soc.
  Austral. {\bfseries 37} (2020)  e002},
  \href{http://arxiv.org/abs/1810.02680}{{arXiv:1810.02680 [astro-ph.CO]}}.

\bibitem{SKA:2018ckk}
SKA, D.~J. Bacon {\em et al.}, {\it {Cosmology with Phase 1 of the Square
  Kilometre Array: Red Book 2018: Technical specifications and performance
  forecasts}},  \href{http://dx.doi.org/10.1017/pasa.2019.51}{Publ. Astron.
  Soc. Austral. {\bfseries 37} (2020)  e007},
  \href{http://arxiv.org/abs/1811.02743}{{arXiv:1811.02743 [astro-ph.CO]}}.

\bibitem{Weltman:2018zrl}
A.~Weltman {\em et al.}, {\it {Fundamental physics with the Square Kilometre
  Array}},  \href{http://dx.doi.org/10.1017/pasa.2019.42}{Publ. Astron. Soc.
  Austral. {\bfseries 37} (2020)  e002},
  \href{http://arxiv.org/abs/1810.02680}{{arXiv:1810.02680 [astro-ph.CO]}}.

\bibitem{Zaroubi:2012in}
S.~Zaroubi, {\it {The Epoch of Reionization}},
  \href{http://arxiv.org/abs/1206.0267}{{arXiv:1206.0267 [astro-ph.CO]}}.

\bibitem{Furlanetto:2019wsj}
S.~R. Furlanetto, {\it {Physical Cosmology From the 21-cm Line}},
  \href{http://arxiv.org/abs/1909.12430}{{arXiv:1909.12430 [astro-ph.CO]}}.

\bibitem{Feix:2019lpo}
M.~Feix, J.~Frank, A.~Pargner, R.~Reischke, B.~M. Sch\"afer, and T.~Schwetz,
  {\it {Isocurvature bounds on axion-like particle dark matter in the
  post-inflationary scenario}},
  \href{http://dx.doi.org/10.1088/1475-7516/2019/05/021}{JCAP {\bfseries 05}
  (2019)  021}, \href{http://arxiv.org/abs/1903.06194}{{arXiv:1903.06194
  [astro-ph.CO]}}.

\bibitem{McQuinn:2005hk}
M.~McQuinn, O.~Zahn, M.~Zaldarriaga, L.~Hernquist, and S.~R. Furlanetto, {\it
  {Cosmological parameter estimation using 21 cm radiation from the epoch of
  reionization}},  \href{http://dx.doi.org/10.1086/505167}{Astrophys. J.
  {\bfseries 653} (2006)  815},
  \href{http://arxiv.org/abs/astro-ph/0512263}{{arXiv:astro-ph/0512263}}.

\bibitem{Oyama:2015gma}
Y.~Oyama, K.~Kohri, and M.~Hazumi, {\it {Constraints on the neutrino parameters
  by future cosmological 21 cm line and precise CMB polarization
  observations}},  \href{http://dx.doi.org/10.1088/1475-7516/2016/02/008}{JCAP
  {\bfseries 02} (2016)  008},
  \href{http://arxiv.org/abs/1510.03806}{{arXiv:1510.03806 [astro-ph.CO]}}.

\bibitem{Xu:2020uws}
Y.~Xu and X.~Zhang, {\it {Cosmological parameter measurement and neutral
  hydrogen 21 cm sky survey with the Square Kilometre Array}},
  \href{http://dx.doi.org/10.1007/s11433-020-1544-3}{Sci. China Phys. Mech.
  Astron. {\bfseries 63} (2020) 7, 270431},
  \href{http://arxiv.org/abs/2002.00572}{{arXiv:2002.00572 [astro-ph.CO]}}.

\bibitem{Shaw:2019qin}
A.~K. Shaw, S.~Bharadwaj, and R.~Mondal, {\it {The impact of non-Gaussianity on
  the error covariance for observations of the Epoch of Reionization 21-cm
  power spectrum}},  \href{http://dx.doi.org/10.1093/mnras/stz1561}{Mon. Not.
  Roy. Astron. Soc. {\bfseries 487} (2019) 4, 4951},
  \href{http://arxiv.org/abs/1902.08706}{{arXiv:1902.08706 [astro-ph.CO]}}.

\bibitem{Mondal:2016hmf}
R.~Mondal, S.~Bharadwaj, and S.~Majumdar, {\it {Statistics of the epoch of
  reionization (EoR) 21-cm signal \textendash{} II. The evolution of the
  power-spectrum error-covariance}},
  \href{http://dx.doi.org/10.1093/mnras/stw2599}{Mon. Not. Roy. Astron. Soc.
  {\bfseries 464} (2017) 3, 2992},
  \href{http://arxiv.org/abs/1606.03874}{{arXiv:1606.03874 [astro-ph.CO]}}.

\bibitem{Doussot:2019rdm}
A.~Doussot, E.~Eames, and B.~Semelin, {\it {Improved supervised learning
  methods for EoR parameters reconstruction}},
  \href{http://dx.doi.org/10.1093/mnras/stz2429}{Mon. Not. Roy. Astron. Soc.
  {\bfseries 490} (2019) 1, 371},
  \href{http://arxiv.org/abs/1904.04106}{{arXiv:1904.04106 [astro-ph.CO]}}.

\bibitem{Watkinson:2015vla}
C.~A. Watkinson and J.~R. Pritchard, {\it {The impact of spin temperature
  fluctuations on the 21-cm moments}},
  \href{http://dx.doi.org/10.1093/mnras/stv2010}{Mon. Not. Roy. Astron. Soc.
  {\bfseries 454} (2015) 2, 1416},
  \href{http://arxiv.org/abs/1505.07108}{{arXiv:1505.07108 [astro-ph.CO]}}.

\bibitem{Pacucci:2014wwa}
F.~Pacucci, A.~Mesinger, S.~Mineo, and A.~Ferrara, {\it {The X-ray spectra of
  the first galaxies: 21 cm signatures}},
  \href{http://dx.doi.org/10.1093/mnras/stu1240}{Mon. Not. Roy. Astron. Soc.
  {\bfseries 443} (2014) 1, 678},
  \href{http://arxiv.org/abs/1403.6125}{{arXiv:1403.6125 [astro-ph.CO]}}.

\bibitem{Pritchard:2006sq}
J.~R. Pritchard and S.~R. Furlanetto, {\it {21 cm fluctuations from
  inhomogeneous X-ray heating before reionization}},
  \href{http://dx.doi.org/10.1111/j.1365-2966.2007.11519.x}{Mon. Not. Roy.
  Astron. Soc. {\bfseries 376} (2007)  1680},
  \href{http://arxiv.org/abs/astro-ph/0607234}{{arXiv:astro-ph/0607234}}.

\bibitem{Warszawski:2008pz}
L.~Warszawski, P.~M. Geil, and S.~Wyithe, {\it {Modification of the 21-cm power
  spectrum by X-rays during the epoch of reionisation}},
  \href{http://dx.doi.org/10.1111/j.1365-2966.2009.14781.x}{Mon. Not. Roy.
  Astron. Soc. {\bfseries 396} (2009)  1106},
  \href{http://arxiv.org/abs/0809.1954}{{arXiv:0809.1954 [astro-ph]}}.

\bibitem{Ma:2018ltb}
Q.~Ma, B.~Ciardi, M.~B. Eide, and K.~Helgason, {\it {X-ray background and its
  correlation with the 21 cm signal}},
  \href{http://dx.doi.org/10.1093/mnras/sty1806}{Mon. Not. Roy. Astron. Soc.
  {\bfseries 480} (2018) 1, 26},
  \href{http://arxiv.org/abs/1807.01283}{{arXiv:1807.01283 [astro-ph.CO]}}.

\bibitem{Kim:2012jx}
H.-S. Kim, J.~S.~B. Wyithe, S.~Raskutti, and C.~G. Lacey, {\it {The Power
  spectrum of Redshifted 21cm Fluctuations in Hierarchical Galaxy Formation
  Models I: The Imprint of Supernova Feedback}},
  \href{http://dx.doi.org/10.1093/mnras/sts206}{Mon. Not. Roy. Astron. Soc.
  {\bfseries 428} (2013)  2467},
  \href{http://arxiv.org/abs/1203.3598}{{arXiv:1203.3598 [astro-ph.CO]}}.

\bibitem{Geil:2009ee}
P.~M. Geil and J.~S.~B. Wyithe, {\it {Modification of the 21-cm power spectrum
  by quasars during the epoch of reionisation}},
  \href{http://dx.doi.org/10.1111/j.1365-2966.2009.15451.x}{Mon. Not. Roy.
  Astron. Soc. {\bfseries 399} (2009)  1877},
  \href{http://arxiv.org/abs/0904.3163}{{arXiv:0904.3163 [astro-ph.CO]}}.

\bibitem{Barkana:2004zy}
R.~Barkana and A.~Loeb, {\it {A Method for separating the physics from the
  astrophysics of high-redshift 21 cm fluctuations}},
  \href{http://dx.doi.org/10.1086/430599}{Astrophys. J. Lett. {\bfseries 624}
  (2005)  L65},
  \href{http://arxiv.org/abs/astro-ph/0409572}{{arXiv:astro-ph/0409572}}.

\bibitem{Gnedin:2006uz}
N.~Y. Gnedin and X.-H. Fan, {\it {Cosmic Reionization Redux}},
  \href{http://dx.doi.org/10.1086/505790}{Astrophys. J. {\bfseries 648} (2006)
  1}, \href{http://arxiv.org/abs/astro-ph/0603794}{{arXiv:astro-ph/0603794}}.

\bibitem{Miralda-Escude:1998adl}
J.~Miralda-Escude, M.~Haehnelt, and M.~J. Rees, {\it {Reionization of the
  inhomogeneous universe}},  \href{http://dx.doi.org/10.1086/308330}{Astrophys.
  J. {\bfseries 530} (2000)  1},
  \href{http://arxiv.org/abs/astro-ph/9812306}{{arXiv:astro-ph/9812306}}.

\bibitem{Furlanetto:2004ha}
S.~Furlanetto, M.~Zaldarriaga, and L.~Hernquist, {\it {Statistical probes of
  reionization with 21 cm tomography}},
  \href{http://dx.doi.org/10.1086/423028}{Astrophys. J. {\bfseries 613} (2004)
  16}, \href{http://arxiv.org/abs/astro-ph/0404112}{{arXiv:astro-ph/0404112}}.

\bibitem{Iliev:2015aia}
I.~T. Iliev, M.~G. Santos, A.~Mesinger, S.~Majumdar, and G.~Mellema, {\it
  {Epoch of Reionization modelling and simulations for SKA}},
  \href{http://dx.doi.org/10.22323/1.215.0007}{PoS {\bfseries AASKA14} (2015)
  007}, \href{http://arxiv.org/abs/1501.04213}{{arXiv:1501.04213
  [astro-ph.CO]}}.

\bibitem{Trac:2009bt}
H.~Trac and N.~Y. Gnedin, {\it {Computer Simulations of Cosmic Reionization}},
  \href{http://dx.doi.org/10.1166/asl.2011.1214}{Adv. Sci. Lett. {\bfseries 4}
  (2011)  228}, \href{http://arxiv.org/abs/0906.4348}{{arXiv:0906.4348
  [astro-ph.CO]}}.

\bibitem{Shapiro:2008zf}
P.~R. Shapiro, I.~T. Iliev, G.~Mellema, U.-L. Pen, and H.~Merz, {\it {The
  Theory and Simulation of the 21-cm Background from the Epoch of
  Reionization}},  \href{http://dx.doi.org/10.1063/1.2973617}{AIP Conf. Proc.
  {\bfseries 1035} (2008) 1, 68},
  \href{http://arxiv.org/abs/0806.3091}{{arXiv:0806.3091 [astro-ph]}}.

\bibitem{Villanueva-Domingo:2020wpt}
P.~Villanueva-Domingo and F.~Villaescusa-Navarro, {\it {Removing Astrophysics
  in 21 cm maps with Neural Networks}},
  \href{http://dx.doi.org/10.3847/1538-4357/abd245}{Astrophys. J. {\bfseries
  907} (2021) 1, 44}, \href{http://arxiv.org/abs/2006.14305}{{arXiv:2006.14305
  [astro-ph.CO]}}.

\bibitem{Hassan:2019cal}
S.~Hassan, S.~Andrianomena, and C.~Doughty, {\it {Constraining the astrophysics
  and cosmology from 21 cm tomography using deep learning with the SKA}},
  \href{http://dx.doi.org/10.1093/mnras/staa1151}{Mon. Not. Roy. Astron. Soc.
  {\bfseries 494} (2020) 4, 5761},
  \href{http://arxiv.org/abs/1907.07787}{{arXiv:1907.07787 [astro-ph.CO]}}.

\bibitem{Mao:2008ug}
Y.~Mao, M.~Tegmark, M.~McQuinn, M.~Zaldarriaga, and O.~Zahn, {\it {How
  accurately can 21 cm tomography constrain cosmology?}},
  \href{http://dx.doi.org/10.1103/PhysRevD.78.023529}{Phys. Rev. D {\bfseries
  78} (2008)  023529}, \href{http://arxiv.org/abs/0802.1710}{{arXiv:0802.1710
  [astro-ph]}}.

\bibitem{Barger:2008ii}
V.~Barger, Y.~Gao, Y.~Mao, and D.~Marfatia, {\it {Inflationary Potential from
  21 cm Tomography and Planck}},
  \href{http://dx.doi.org/10.1016/j.physletb.2009.02.021}{Phys. Lett. B
  {\bfseries 673} (2009)  173},
  \href{http://arxiv.org/abs/0810.3337}{{arXiv:0810.3337 [astro-ph]}}.

\bibitem{Cooray:2008eb}
A.~Cooray, C.~Li, and A.~Melchiorri, {\it {The trispectrum of 21-cm background
  anisotropies as a probe of primordial non-Gaussianity}},
  \href{http://dx.doi.org/10.1103/PhysRevD.77.103506}{Phys. Rev. D {\bfseries
  77} (2008)  103506}, \href{http://arxiv.org/abs/0801.3463}{{arXiv:0801.3463
  [astro-ph]}}.

\bibitem{Kohri:2013mxa}
K.~Kohri, Y.~Oyama, T.~Sekiguchi, and T.~Takahashi, {\it {Precise Measurements
  of Primordial Power Spectrum with 21 cm Fluctuations}},
  \href{http://dx.doi.org/10.1088/1475-7516/2013/10/065}{JCAP {\bfseries 10}
  (2013)  065}, \href{http://arxiv.org/abs/1303.1688}{{arXiv:1303.1688
  [astro-ph.CO]}}.

\bibitem{Munoz:2016owz}
J.~B. Mu\~noz, E.~D. Kovetz, A.~Raccanelli, M.~Kamionkowski, and J.~Silk, {\it
  {Towards a measurement of the spectral runnings}},
  \href{http://dx.doi.org/10.1088/1475-7516/2017/05/032}{JCAP {\bfseries 05}
  (2017)  032}, \href{http://arxiv.org/abs/1611.05883}{{arXiv:1611.05883
  [astro-ph.CO]}}.

\bibitem{Pourtsidou:2016ctq}
A.~Pourtsidou, {\it {Synergistic tests of inflation}},
  \href{http://arxiv.org/abs/1612.05138}{{arXiv:1612.05138 [astro-ph.CO]}}.

\bibitem{Sekiguchi:2018kqe}
T.~Sekiguchi, T.~Takahashi, H.~Tashiro, and S.~Yokoyama, {\it {Probing
  primordial non-Gaussianity with 21 cm fluctuations from minihalos}},
  \href{http://dx.doi.org/10.1088/1475-7516/2019/02/033}{JCAP {\bfseries 02}
  (2019)  033}, \href{http://arxiv.org/abs/1807.02008}{{arXiv:1807.02008
  [astro-ph.CO]}}.

\bibitem{Liddle:2003py}
A.~R. Liddle, {\it {Inflationary flow equations}},
  \href{http://dx.doi.org/10.1103/PhysRevD.68.103504}{Phys. Rev. D {\bfseries
  68} (2003)  103504},
  \href{http://arxiv.org/abs/astro-ph/0307286}{{arXiv:astro-ph/0307286}}.

\bibitem{Brinckmann:2018cvx}
T.~Brinckmann and J.~Lesgourgues, {\it {MontePython 3: boosted MCMC sampler and
  other features}},  \href{http://dx.doi.org/10.1016/j.dark.2018.100260}{Phys.
  Dark Univ. {\bfseries 24} (2019)  100260},
  \href{http://arxiv.org/abs/1804.07261}{{arXiv:1804.07261 [astro-ph.CO]}}.

\bibitem{Audren:2012wb}
B.~Audren, J.~Lesgourgues, K.~Benabed, and S.~Prunet, {\it {Conservative
  Constraints on Early Cosmology: an illustration of the Monte Python
  cosmological parameter inference code}},
  \href{http://dx.doi.org/10.1088/1475-7516/2013/02/001}{JCAP {\bfseries 02}
  (2013)  001}, \href{http://arxiv.org/abs/1210.7183}{{arXiv:1210.7183
  [astro-ph.CO]}}.

\bibitem{Lesgourgues:2011rg}
J.~Lesgourgues, {\it {The Cosmic Linear Anisotropy Solving System (CLASS) III:
  Comparision with CAMB for LambdaCDM}},
  \href{http://arxiv.org/abs/1104.2934}{{arXiv:1104.2934 [astro-ph.CO]}}.

\bibitem{Blas:2011rf}
D.~Blas, J.~Lesgourgues, and T.~Tram, {\it {The Cosmic Linear Anisotropy
  Solving System (CLASS) II: Approximation schemes}},
  \href{http://dx.doi.org/10.1088/1475-7516/2011/07/034}{JCAP {\bfseries 07}
  (2011)  034}, \href{http://arxiv.org/abs/1104.2933}{{arXiv:1104.2933
  [astro-ph.CO]}}.

\bibitem{Lesgourgues:2007aa}
J.~Lesgourgues, A.~A. Starobinsky, and W.~Valkenburg, {\it {What do WMAP and
  SDSS really tell about inflation?}},
  \href{http://dx.doi.org/10.1088/1475-7516/2008/01/010}{JCAP {\bfseries 01}
  (2008)  010}, \href{http://arxiv.org/abs/0710.1630}{{arXiv:0710.1630
  [astro-ph]}}.

\bibitem{Planck:2015sxf}
Planck, P.~A.~R. Ade {\em et al.}, {\it {Planck 2015 results. XX. Constraints
  on inflation}},  \href{http://dx.doi.org/10.1051/0004-6361/201525898}{Astron.
  Astrophys. {\bfseries 594} (2016)  A20},
  \href{http://arxiv.org/abs/1502.02114}{{arXiv:1502.02114 [astro-ph.CO]}}.

\bibitem{Munoz:2015eqa}
J.~B. Mu\~noz, Y.~Ali-Ha\"\i{}moud, and M.~Kamionkowski, {\it {Primordial
  non-gaussianity from the bispectrum of 21-cm fluctuations in the dark ages}},
   \href{http://dx.doi.org/10.1103/PhysRevD.92.083508}{Phys. Rev. D {\bfseries
  92} (2015) 8, 083508},
  \href{http://arxiv.org/abs/1506.04152}{{arXiv:1506.04152 [astro-ph.CO]}}.

\bibitem{Bharadwaj:2004nr}
S.~Bharadwaj and S.~S. Ali, {\it {The CMBR fluctuations from HI perturbations
  prior to reionization}},
  \href{http://dx.doi.org/10.1111/j.1365-2966.2004.07907.x}{Mon. Not. Roy.
  Astron. Soc. {\bfseries 352} (2004)  142},
  \href{http://arxiv.org/abs/astro-ph/0401206}{{arXiv:astro-ph/0401206}}.

\bibitem{Gnedin:2004nj}
N.~Y. Gnedin, {\it {Reionization, SLOAN, and WMAP: Is the picture
  consistent?}},  \href{http://dx.doi.org/10.1086/421450}{Astrophys. J.
  {\bfseries 610} (2004)  9},
  \href{http://arxiv.org/abs/astro-ph/0403699}{{arXiv:astro-ph/0403699}}.

\bibitem{Choudhury:2016cyh}
T.~R. Choudhury, K.~Datta, S.~Majumdar, R.~Ghara, A.~Paranjape, R.~Mondal,
  S.~Bharadwaj, and S.~Samui, {\it {Modelling the 21 cm Signal From the Epoch
  of Reionization and Cosmic Dawn}},
  \href{http://dx.doi.org/10.1007/s12036-016-9403-z}{J. Astrophys. Astron.
  {\bfseries 37} (2016)  29},
  \href{http://arxiv.org/abs/1610.08179}{{arXiv:1610.08179 [astro-ph.CO]}}.

\bibitem{Gluscevic:2010iz}
V.~Gluscevic and R.~Barkana, {\it {Statistics of 21-cm Fluctuations in Cosmic
  Reionization Simulations: PDFs and Difference PDFs}},
  \href{http://dx.doi.org/10.1111/j.1365-2966.2010.17293.x}{Mon. Not. Roy.
  Astron. Soc. {\bfseries 408} (2010)  2373},
  \href{http://arxiv.org/abs/1005.3814}{{arXiv:1005.3814 [astro-ph.CO]}}.

\bibitem{Furlanetto:2006pg}
S.~Furlanetto and A.~Lidz, {\it {The Cross-Correlation of High-Redshift 21 cm
  and Galaxy Surveys}},  \href{http://dx.doi.org/10.1086/513009}{Astrophys. J.
  {\bfseries 660} (2007)  1030},
  \href{http://arxiv.org/abs/astro-ph/0611274}{{arXiv:astro-ph/0611274}}.

\bibitem{Fialkov:2019jcx}
A.~Fialkov, R.~Barkana, and M.~Jarvis, {\it {Extracting the global signal from
  21-cm fluctuations: the multitracer approach}},
  \href{http://dx.doi.org/10.1093/mnras/stz3208}{Mon. Not. Roy. Astron. Soc.
  {\bfseries 491} (2020) 3, 3108},
  \href{http://arxiv.org/abs/1904.10857}{{arXiv:1904.10857 [astro-ph.CO]}}.

\bibitem{Wang:2005my}
X.~Wang and W.~Hu, {\it {Redshift space 21 cm power spectra from
  reionization}},  \href{http://dx.doi.org/10.1086/503095}{Astrophys. J.
  {\bfseries 643} (2006)  585},
  \href{http://arxiv.org/abs/astro-ph/0511141}{{arXiv:astro-ph/0511141}}.

\bibitem{Schneider:2020xmf}
A.~Schneider, S.~K. Giri, and J.~Mirocha, {\it {Halo model approach for the
  21-cm power spectrum at cosmic dawn}},
  \href{http://dx.doi.org/10.1103/PhysRevD.103.083025}{Phys. Rev. D {\bfseries
  103} (2021) 8, 083025},
  \href{http://arxiv.org/abs/2011.12308}{{arXiv:2011.12308 [astro-ph.CO]}}.

\bibitem{Tegmark:2008au}
M.~Tegmark and M.~Zaldarriaga, {\it {The Fast Fourier Transform Telescope}},
  \href{http://dx.doi.org/10.1103/PhysRevD.79.083530}{Phys. Rev. D {\bfseries
  79} (2009)  083530}, \href{http://arxiv.org/abs/0805.4414}{{arXiv:0805.4414
  [astro-ph]}}.

\bibitem{Zaldarriaga:2003du}
M.~Zaldarriaga, S.~R. Furlanetto, and L.~Hernquist, {\it {21 Centimeter
  fluctuations from cosmic gas at high redshifts}},
  \href{http://dx.doi.org/10.1086/386327}{Astrophys. J. {\bfseries 608} (2004)
  622}, \href{http://arxiv.org/abs/astro-ph/0311514}{{arXiv:astro-ph/0311514
  [astro-ph]}}.

\bibitem{Greig:2015zra}
B.~Greig, A.~Mesinger, and L.~V.~E. Koopmans, {\it {Optimal core baseline
  design and observing strategy for probing the astrophysics of reionization
  with the SKA}},  \href{http://arxiv.org/abs/1509.03312}{{arXiv:1509.03312
  [astro-ph.CO]}}.

\bibitem{Munoz:2020itp}
J.~B. Mu\~noz and F.-Y. Cyr-Racine, {\it {Cosmic Variance of the 21-cm Global
  Signal}},  \href{http://dx.doi.org/10.1103/PhysRevD.103.023512}{Phys. Rev. D
  {\bfseries 103} (2021) 2, 023512},
  \href{http://arxiv.org/abs/2005.03664}{{arXiv:2005.03664 [astro-ph.CO]}}.

\bibitem{Sprenger:2018tdb}
T.~Sprenger, M.~Archidiacono, T.~Brinckmann, S.~Clesse, and J.~Lesgourgues,
  {\it {Cosmology in the era of Euclid and the Square Kilometre Array}},
  \href{http://dx.doi.org/10.1088/1475-7516/2019/02/047}{JCAP {\bfseries 02}
  (2019)  047}, \href{http://arxiv.org/abs/1801.08331}{{arXiv:1801.08331
  [astro-ph.CO]}}.

\bibitem{Pourtsidou:2014pra}
A.~Pourtsidou and R.~B. Metcalf, {\it {Gravitational lensing of cosmological 21
  cm emission}},  \href{http://dx.doi.org/10.1093/mnras/stv102}{Mon. Not. Roy.
  Astron. Soc. {\bfseries 448} (2015)  2368},
  \href{http://arxiv.org/abs/1410.2533}{{arXiv:1410.2533 [astro-ph.CO]}}.

\bibitem{Zahn:2005ap}
O.~Zahn and M.~Zaldarriaga, {\it {Lensing reconstruction using redshifted 21cm
  fluctuations}},  \href{http://dx.doi.org/10.1086/508916}{Astrophys. J.
  {\bfseries 653} (2006)  922},
  \href{http://arxiv.org/abs/astro-ph/0511547}{{arXiv:astro-ph/0511547}}.

\end{thebibliography}\endgroup
\end{document}